\documentclass[final,12pt]{colt2025} 

\usepackage{algorithm}
\usepackage[title]{appendix}
\usepackage{algpseudocode}
\usepackage{float}

\newcommand{\bv}{\boldsymbol{v}}
\newcommand{\bu}{\boldsymbol{u}}
\newcommand{\bx}{\boldsymbol{x}}
\newcommand{\by}{\boldsymbol{y}}
\newcommand{\bz}{\boldsymbol{z}}

\newcommand{\bb}{\boldsymbol{b}}

\newcommand{\F}{\mathbf{F}}

\newcommand{\mC}{\mathbf{C}}
\newcommand{\mY}{\mathbf{Y}}
\newcommand{\mX}{\mathbf{X}}
\newcommand{\mD}{\mathbf{D}}
\def\bx{{\bf x}}

\newcommand{\abs}[1]{|#1|}

\newcommand{\val}{\mathsf{val}}

\DeclareMathOperator*{\E}{\mathbb{E}}
\DeclareMathOperator*{\Var}{\mathsf{Var}}
\DeclareMathOperator*{\cov}{\mathsf{cov}}

\DeclareOldFontCommand{\bf}{\normalfont\bfseries}{\mathbf}
 \DeclareOldFontCommand{\it}{\normalfont\itshape}{\mathit}
 \DeclareOldFontCommand{\sf}{\normalfont\sffamily}{\mathsf}


\newcommand{\secret}{\mathbf{secret}}
\newcommand{\candidate}{\mathbf{candidate}}

\newenvironment{proofof}[1]{\bigskip \noindent {\bfseries \upshape Proof of #1.}\quad }
{\jmlrQED\par\vskip 4mm\par}

\makeatletter
\renewcommand*{\@fnsymbol}[1]{\ensuremath{\ifcase#1\or *\or \dagger\or \ddagger\or
   \mathsection\or \mathparagraph\or \|\or **\or \dagger\dagger
   \or \ddagger\ddagger \else\@ctrerr\fi}}
\makeatother

\title{Algorithms for Sparse LPN and LSPN Against Low-noise}
\usepackage{times}

\begin{document}
\coltauthor{%
 \Name{Xue Chen}\footnote{Accepted for presentation at the
Conference on Learning Theory (COLT) 2025} \Email{xuechen1989@ustc.edu.cn}\\
 \addr University of Science and Technology of China, Hefei 230026, China \& Hefei National Laboratory, Hefei 230088, China
 \AND
 \Name{Wenxuan Shu} \Email{wxshu@mail.ustc.edu.cn}\\
 \addr University of Science and Technology of China, Hefei 230026, China
 \AND
  \Name{Zhaienhe Zhou} \Email{zhaienhezhou@gmail.com}\\
 \addr University of Science and Technology of China, Hefei 230026, China
}
\maketitle

\begin{abstract}%
We consider sparse variants of the classical Learning Parities with random Noise (LPN) problem. Our main contribution is a new algorithmic framework that provides learning algorithms against low-noise for both Learning Sparse Parities (LSPN) problem and sparse LPN problem. Different from previous approaches for LSPN and sparse LPN \citep{Grig11,valiant2012finding,KKK18,RRS17,GKM}, this framework has a simple structure without fast matrix multiplication or tensor methods such that its algorithms are easy to implement and run in polynomial space. Let $n$ be the dimension, $k$ denote the sparsity, and $\eta$ be the noise rate such that each label gets flipped with probability $\eta$.

As a fundamental problem in computational learning theory \citep{Feldman09}, Learning Sparse Parities with Noise (LSPN) assumes the hidden parity is $k$-sparse instead of a potentially dense vector. While the simple enumeration algorithm takes ${n \choose k}=O(n/k)^k$ time, previously known results stills need at least ${n \choose k/2} = \Omega(n/k)^{k/2}$ time for any noise rate $\eta$ \citep{Grig11,valiant2012finding,KKK18}. Our framework provides a LSPN algorithm runs in time $O(\eta \cdot n/k)^k$ for any noise rate $\eta$, which improves the state-of-the-art of LSPN whenever $\eta \in (k/n,\sqrt{k/n})$.

The sparse LPN problem is closely related to the classical problem of refuting random $k$-CSP \citep{FKO06, RRS17, GKM} and has been widely used in cryptography as the hardness assumption \citep[e.g.,][]{Alekhnovich,ABW10,ADINZ18,DIJL_sparseLPN}. Different from the standard LPN that samples random vectors in $\mathbf{F}_2^n$, it samples random $k$-sparse vectors. Because the number of $k$-sparse vectors is ${n \choose k}<n^k$, sparse LPN has learning algorithms in polynomial time when $m>n^{k/2}$. However, much less is known about learning algorithms for a constant $k$ like 3 and  $m<n^{k/2}$ samples, except the Gaussian elimination algorithm and sum-of-squares algorithms \citep{BKS14,BM16,RRS17}. Our framework provides a learning algorithm in $e^{\tilde{O}(\eta \cdot n^{\frac{\delta+1}{2}})}$ time given $\delta \in (0,1)$ and $m=\max\{1,\frac{\eta \cdot n^{\frac{\delta+1}{2}}}{k^2}\} \cdot n^{1+(1-\delta)\cdot \frac{k-1}{2}}$ samples. This improves previous learning algorithms. For example, in the classical setting of $k=3$ and $m=n^{1.4}$ \citep{FKO06,ABW10}, our algorithm would be faster than previous approaches for any $\eta<n^{-0.7}$.
\end{abstract}

\acks{Xue Chen is supported by Innovation Program for Quantum Science and Technology 2021ZD0302901, NSFC 62372424, and CCF-HuaweiLK2023006.}

\section{Introduction}
Learning Parities with Noise (LPN) problem and its variants are ubiquitous in learning theory. It is equivalent to the famous problem of decoding random linear codes in coding
theory, and has been widely used in cryptography as the security assumption. In a dimension-$n$ LPN problem of noise rate $\eta$, the algorithm is trying to learn a hidden vector in $\mathbf{F}_2^n$, called $\secret$ in this work. However, this algorithm only has access to an oracle that generates random vectors $\bx_i$ with labels $\by_i$, where $\bx_i$ is uniformly distributed in $\mathbf{F}_2^n$ and $\by_i \in \mathbf{F}_2$ equals the inner product $\langle \secret, \bx_i \rangle$ with probability $1-\eta$. 

While there is a long line of research on this problem \citep[to name a few][]{BFKL93, blum2003noise, Lyubashevsky05, Regev09,Becker12, Raz18, bshouty2024}, we only have very limited algorithmic tools for LPN at this moment. In the high noise regime, say $\eta=\Omega(\frac{1}{\log n})$, the best known algorithm needs $2^{\Theta(n / \log n)}$ time and samples \citep{blum2003noise}. This algorithm also separates two classical learning models in computational learning theory --- the PAC model and statistical-query model.
In the low noise regime, say $\eta = o(\frac{1}{\log n})$, the best known algorithm needs roughly $e^{\eta n}$ time. This algorithm, called \emph{Gaussian elimination} in this work, repeatedly applies Gaussian elimination to solving linear equations of $n$ random samples until these samples are noiseless and linearly independent. Its expected running time is $n^{O(1)}/(1-\eta)^n \approx e^{\eta n}$. In cryptography, a common choice of the low-noise rate is $\eta=1/\sqrt{n}$ \citep{DMN12_PKC,KMP14_PKC,Yu19}; and it was further conjectured that LPN with noise rate $\eta$ as low as $\frac{\log^2 n}{n}$ is computationally hard \citep{Brakerski19}. 

In this work, we consider two sparse variants of LPN in the low-noise regime. This first variant, called Learning Sparse Parities with Noise (LSPN), assumes that the hidden vector $\secret$ in $\mathbf{F}_2^n$ is of sparsity $k$ and requires the algorithm to find $\secret$ given the same access of random examples $(\bx_i,\by_i)$ as the LPN problem. This problem plays an important role in learning --- \cite{Feldman09} have shown that other fundamental problems in PAC learning such as learning $k$-juntas and learning DNFs could be reduced to LSPN. Due to this connection, LSPN has been extensively studied in both learning theory \citep{Feldman07,Grig11,valiant2012finding,KKK18,KRT17} and cryptography \citep{Yu21,DSGKS21}. In the high-noise regime, say $\eta$ is a constant, Valaint's ingenious work and its improvement \citep{valiant2012finding,KKK18} provided algorithms in time $(\frac{n}{k})^{\frac{\omega}{3} \cdot k} \cdot (1-2\eta)^{O(1)} \le (\frac{n}{k})^{0.8k}$ given the matrix multiplication constant $\omega<2.238$ \citep{WXXZ_matrix_mutl24}. However, for any low-noise rate $\eta$, the running time is at least $(\frac{n}{k})^{k/2}$ \citep{Grig11} although the simple enumeration algorithm only needs $O(\frac{n}{k})^{k}$ time (given any $k\le n^{0.99}$).

The second sparse variant of LPN, called sparse LPN, changes the random oracle such that each vector $\bx_i$ is a random \emph{$k$-sparse} vector instead of a uniformly random vector in $\mathbf{F}_2^n$ and its label $\by_i$ still equals $\langle \secret, \bx_i \rangle$ with probability $1-\eta$. The goal is still to learn the hidden (dense) vector $\secret$. In crpytography, sparse LPN has been widely used as the hardness assumption \citep[e.g.,][]{Alekhnovich,ABW10,ADINZ18,CRR23,RRT23,DIJL_sparseLPN,RVV24}. 

In computational complexity, sparse LPN is closely related to the classical problem of refuting random $k$-XOR. Because there are only ${n \choose k}$ possible $k$-sparse vectors $\bx_i$, one would receive two random samples with the same $k$-sparse vector given $m = n^{k/2}$ samples. Hence, sparse LPN becomes computational easy when $m \ge n^{k/2}$ \citep{Applebaum16} such that it usually restrain the sample complexity $m < n^{k/2}$.
For $m<n^{k/2}$, A line of research \citep{FKO06, AOW_15, BM16, FPV18, RRS17, GKM, HKM_hyper_moore} has provided sub-exponential time algorithms that \emph{distinguish} between the case with totally random labels and a sparse LPN instance with noisy labels of $\langle \bx,\secret \rangle$. In this work, we call such an algorithm a \emph{distinguishing} algorithm and call an algorithm outputting $\secret$ a \emph{learning} algorithm. 
However, much less is known about \emph{learning} algorithms given $m<n^{k/2}$ random samples. While several works \citep{ABW10,Applebaum13,BSV2019} have provided reductions from distinguishing algorithms to learning algorithms, these reductions either needs cubicly many more samples \citep{ABW10,Applebaum13} or has a small success probability \citep{BSV2019}. To the best of our knowledge, the best known algorithms for the classical setting of $k=3$ and $m=n^{1.4}$ \citep{FKO06,Feige2008,ABW10} still needs $\min\{e^{\eta n},e^{\tilde{O}(n^{0.2})}\}$ time.

In this work, we study the time complexity of LSPN and sparse LPN  in the low-noise regime. Our main contribution is an algorithmic framework for both LSPN and sparse LPN. There are several advantages of our algorithmic framework. First of all, it improves the time complexity of LSPN and sparse LPN for a wide range of parameters. For LSPN, it provides the first algorithm faster than $\min\{(\frac{n}{k})^{k/2},e^{\eta n}\}$ when $\eta \in (k/n,\sqrt{k/n})$. For the classical setting of sparse LPN \citep{FKO06} with $k=3$ and $m=n^{1.4}$, our \emph{learning} algorithm would run faster than previous algorithms for any $\eta<n^{-0.7}$. Secondly, our framework is structurally simple such that its implementation is relatively easy. In particular, the space complexity of our algorithms is a polynomial of $n$ and $m$ while many previous algorithms need super-polynomial space (including the-state-of-the-art for LSPN \citep{Grig11,valiant2012finding,KKK18} and \emph{distinguishing} algorithms for sparse LPN \citep{RRS17,GKM,HKM_hyper_moore}). Also, our framework is very flexible that provides a non-trivial algorithm for LSPN against high-noise (see Section~\ref{sec:subset_BKW}). Last but not least, this framework shows a conceptual connection between these two sparse variants of LPN. To the best of our knowledge, this is the first algorithmic connection between these will-studied variants of LPN. Since the long line of research on these problems has discovered a variety of methods, we believe this connection could find more applications in LPN.


\subsection{Our Results}
The basic idea of our framework is to leverage the knowledge of sparsity (in LSPN and sparse LPN) into domain reductions. This is motivated by the domain reduction idea in the query-access model \citep[e.g.,][]{GoldreichLevin:89, akagolsaf03, HIKP12, ChenD20}. However, our reduction is different from theirs --- it samples a subset and reduces the original problem to a sub-problem by assuming that the sparse part of $\secret$ or $\bx$ is in this subset. While this reduction also works under high noise (see Section~\ref{sec:subset_BKW} for an example), we only have efficient algorithms to solve sub-problems in the low-noise regime (via Gaussian eliminations). 

Next we describe our main results against low-noise. Because this framework leads to algorithms of LSPN and sparse LPN in different contexts, we state them separately.

\paragraph{Learning Algorithm for LSPN.}
In a LSPN problem, $\secret$ is of sparsity $k$ and each vector $\bx_i$ is uniformly drawn from $\mathbf{F}_2^n$ such that each label $\by_i=\langle \secret, \bx_i \rangle + e_i$ for a random Bernoulli noise $e_i \in \mathbf{F}_2$ of rate $\eta$. Since a simple enumeration of $\secret$ takes time ${n \choose k} \approx (en/k)^k$, we will state the running time of our algorithm with respect to $(n/k)^k$. For convenience, we use $(k,\eta)$-LSPN to denote a learning sparse parity with noise instance whose $\secret$ is of sparsity $\le k$ and noise rate is $\eta$.

\begin{theorem}\label{thm:infor1}[Informal version of Theorem~\ref{thm:low_noise_main}]
    For any noise rate $\eta$ and sparsity $k$ satisfying $k/\eta<n$, there exists an algorithm to solve $(k,\eta)$-LSPN problems in $O(\frac{\eta n}{k})^{k}$ time and $O(\frac{k}{\eta}+ k \log \frac{n}{k})$ samples. 
\end{theorem}

When $\eta$ is small say $\eta<(k/n)^{1/2}$, this algorithm improves the state-of-the-art for any $k<n^{0.99}$ where the best known algorithm has time at least ${n \choose k/2} \ge (n/k)^{k/2}$ \citep{Grig11}. 
In fact, its precise running time is $e^{(1+O(\eta)) k} \cdot (\frac{\eta \cdot n}{k})^k \cdot (k/\eta)^{O(1)}$, where $(k/\eta)^{O(1)}$ is the time complexity of applying Gaussian eliminations to $k/\eta$ equations. This is always faster than the simple enumeration algorithm of time ${n \choose k} \approx (en/k)^k$ by a factor about $\eta^k$. Moreover, its sample complexity $O(\frac{k}{\eta}+ k \log \frac{n}{k})$ is always in $O(n)$ under the condition $n>\frac{k}{\eta}$.

Essentially, condition $n>\frac{k}{\eta}$ is to compare with the two trivial algorithms of running time $e^{\eta n}$ and ${n \choose k}$ separately. $\eta n = \tilde{\Omega}(k)$\footnote{In this work, we use $\tilde{}$ to hide terms like $\log^{O(1)} n$ and $n^{o(1)}$.} would imply the enumeration time ${n \choose k}<e^{\eta n}$. A similar condition has been proposed in previous work by \cite{Yu21}, whose algorithm has running time about $(\eta \cdot n)^{2k}$ in this range of parameters. Hence our algorithm provides an improvement upon \cite{Yu21}.

Moreover, we simplify Valiant's learning framework \citep{valiant2012finding} for LSPN in the high-noise regime and improve its time complexity and sample complexity to matching the state-of-the-art based on the light-bulb algorithm \citep{Grig11,KKK18}. Since Valiant's framework could use alternate methods to generate biased samples and enumerate parities of smaller sizes \citep[e.g.,][]{Yu21, DSGKS21}, it is \emph{more flexible} than the approach of light-bulb algorithms that match parities of size $k/2$. Our motivation here is to refine this flexible framework and provide more tools. 
Due to the space constraint, We defer more discussion and details of this part to Appendix~\ref{sec:valiant}.

\paragraph{Learning Algorithm for sparse LPN.} In a sparse LPN problem, each $\bx_i$ is a random $k$-sparse vector  and $\secret$ could have Hamming weight $\Omega(n)$  such that its label $y_i=\langle \secret, \bx_i \rangle + e_i$ with a random noise $e_i$ of rate $\eta$. Since $\bx_i$ only has ${n \choose k}<n^k$ possibilities, this problem becomes computational easy when $m > n^{k/2}$ by the birthday paradox (see Theorem 3.5 of \cite{Applebaum16}). Hence we use a parameter $\delta \in (0,1)$ to bound the time complexity for $m \approx n^{1 + (1-\delta) \cdot \frac{k-1}{2}}$ samples. This is similar to previous results about refuting $k$-XOR \citep{FKO06,RRS17,GKM}.

We use $(k,\eta)$-sparse LPN to denote a sparse LPN instance whose random vectors are $k$-sparse and noise rate is $\eta$ and state the result as follows.
\begin{theorem}\label{thm:infor_sparse_LPN}[Informal version of Theorem~\ref{thm:sparse_LPN}]
    For any $(k,\eta)$-sparse LPN problem, given any $\delta \in (0,1)$, there is a learning algorithm that takes $m:=\max \big\{1, \frac{\log n}{k}, \frac{\eta n^{\frac{1+\delta}{2}} \cdot \log n}{k^2} \big\} \cdot O(n)^{1+(1-\delta) \cdot \frac{k-1}{2}}$ samples to return $\secret$ in time $e^{\tilde{O}(\eta \cdot n^{\frac{1+\delta}{2}})} + n^{O(1)} \cdot m$.
\end{theorem}
There is an extra factor $\max \big\{1, \frac{\log n}{k}, \frac{\eta n^{\frac{1+\delta}{2}} \cdot \log n}{k^2} \big\}$ in our sample complexity $m$ besides the main term $n^{1+(1-\delta) \cdot \frac{k-1}{2}}$. The second term $\frac{\log n}{k}$ comes from the fact that it takes $\Theta(n) \cdot \max \{1, \frac{\log n}{k}\}$ random $k$-sparse vectors to span the whole space of dimension $n$ by the coupon collector argument. The last term comes from the fact that it takes more samples to verify a hypothesis $h$ in a sparse LPN problem. In particular, if $h$ and $\secret$ differ on 1 bit, it would take $\Theta(\frac{\eta n^2 }{k^2})$ samples to distinguish them under noise rate $\eta$.

In the classical setting of $k=3$ and $m=\tilde{O}(n^{1.4})$ \citep{FKO06, ABW10}, the best known \emph{learning} algorithm was Gaussian elimination of time $e^{\eta n}$ and sum-of-squares of time $e^{\tilde{O}(n^{0.2})}$ \citep{BKS14,BM16,RRS17}. 
Our learning algorithm is faster than them for any $\eta < n^{-0.7}$. Specifically, its running time is $n^{O(1)}$ when $\eta<n^{-0.8}$ (by setting $\delta=0.6$) and becomes $e^{\tilde{O}(n^{2c-0.4})}$ when $\eta=n^{-(1-c)}$ for any $c \in [0.2,0.4]$. When $k=5$ and $m=n^{1.8}$, our algorithm will be faster for any $\eta<n^{-0.45}$. For a super-constant $k$ and $m \approx n^{ (1-\delta) \cdot \frac{k}{2}}$, our algorithm has running time about $e^{\tilde{O}(\eta \cdot n^{\frac{1+\delta}{2}})}$, which is always faster.

\paragraph{Extensions.} We show several applications of this framework besides Theorem~\ref{thm:infor1} and Theorem~\ref{thm:infor_sparse_LPN}.  Firstly, our framework provides learning algorithms against \emph{adversarial} noise for LSPN and sparse LPN (see Remark~\ref{rm:adversarial_LSPN} and Appendix~\ref{sec:adv_sparse_LPN}). Secondly, we combine the classical \cite{blum2003noise} algorithm with our framework to obtain an efficient algorithm for LSPN against high noise in Section~\ref{sec:subset_BKW}. At last, we provide a \emph{distinguishing} algorithm for sparse LPN based on this framework in Appendix~\ref{sec:distinguish}, which extends the algorithm of \cite{DJ24}. Due to the space constraint, we defer the details to the corresponding parts.

\paragraph{Concurrent work.} \cite{DJ24} provided a distinguishing algorithm for sparse LPN, whose idea is similar to our idea behind Theorem~\ref{thm:infor_sparse_LPN}, independently and concurrently. Specifically, they show how to distinguish between a $(k,\eta)$-sparse LPN instance and an instance with totally random labels in polynomial time given $m=O(n)^{1+(1-\delta) \cdot \frac{k-1}{2}}$ and $\eta \le n^{-\frac{1+\delta}{2}}$ --- this is identical to our result in Theorem~\ref{thm:infor_sparse_LPN} restricted to $\eta \le n^{-\frac{1+\delta}{2}}$. However, our work is significantly different from theirs. On the first hand, they  consider $\eta \le n^{-\frac{1+\delta}{2}}$ for distinguishing sparse LPN, while our main results are \emph{learning} algorithms for both LSPN and sparse LPN\footnote{After realizing \cite{DJ24}, we show their \emph{distinguishing} algorithm would work for larger noise rate $\eta$ like Theorem~\ref{thm:infor_sparse_LPN} in Appendix~\ref{sec:distinguish}.}. Moreover, their result would not imply a learning algorithm like Theorem~\ref{thm:infor_sparse_LPN} after applying known reductions from distinguishing to learning. On the other hand, our main conceptual contribution is an algorithmic framework that supports both sparse LPN and LSPN. To the best of our knowledge, this framework provides the first connection between these two variants of LPN. 
The main conceptual contribution of \citep{DJ24} is a new LPN variant called Dense-Sparse LPN, which provides hardness in a low-noise regime where sparse LPN is easy with applications in lossy cryptography.

\subsection{Discussion}
In this work, we provide a new approach for LSPN and sparse LPN in the low-noise setting. Our work leaves several open questions, and we list some of them here.

For LSPN, since $O(k \log n)$ noiseless samples are enough to identify the hidden sparse vector $\secret$ statistically, an interesting question is to design efficient algorithms for $O(k \log n)$ \emph{noiseless} samples. It is equivalent to $\eta<\frac{1}{k \log n}$. While this problem is closely related to compressed sensing, convex optimization techniques there do not help with this problem in $\mathbf{F}_2$. To the best of our knowledge, there is no polynomial time algorithm for $k=\log n$ given $o(n)$ noiseless samples.

Moreover, given the reduction from LSPN to learning $k$-juntas and learning DNFs \citep{Feldman09}, it would be interesting to apply our framework to other problems in learning theory. 

For sparse LPN problems, previous works have proposed that the time complexity of $m=n^{1+(k/2-1)(1-\delta)}$ samples is $\text{exp}\big( \tilde{\Omega}_\eta(n^\delta) \big)$ for any $\delta \in (0,1)$ (e.g., Assumption 6 in \cite{ADINZ18} and Assumption 1 in \cite{CRR23}). The monomial $n^\delta$ in the exponent $\tilde{\Omega}_\eta(n^\delta)$ comes from the size of the smallest linearly dependent set among so many $k$-sparse samples, where recent works have proved Feige's conjecture \citep{Feige2008} and established an upper bound $\tilde{O}(n^{\delta})$ \citep{GKM, HKM_hyper_moore}. However, much less is known about the dependence on $\eta$. Since the hardness of low-noise sparse LPN problems has been widely used in cryptography \citep[e.g.,][]{Alekhnovich,BCGI22,DIJL_sparseLPN},  figuring out the time complexity of sparse LPN in the low-noise setting especially its dependence on the noise rate $\eta$ would be an important question. 

A key step towards efficient algorithms for sparse LPN is to find a small linearly dependent set among those $m$ random vectors \emph{efficiently}. While \cite{GKM, HKM_hyper_moore} have shown a upper bound $\tilde{O}(n^\delta)$ on the size of the smallest linear dependent set, their proofs only provide a brute-force search algorithm of time $e^{\tilde{O}(n^\delta)}$. On the other hand, our framework could be viewed as to find a larger linearly dependent set of size about $n^{\frac{1+\delta}{2}}$ in polynomial time. At this moment, we do not know any trade-off between the time complexity and the size of linearly dependent set. It would be interesting to have more algorithms for finding linearly dependent sets of small sizes.

\subsection{Related Work}
For the general LPN problem, learning algorithms and distinguishing algorithms are equivalent \citep{BFKL93}. So far the best known algorithm needs $2^{O(n/\log n)}$ time and samples from the seminal work of \cite{blum2003noise}. In this work, we call this seminal algorithm BKW for convenience. For polynomially many samples, the best known algorithm takes $2^{O(n/\log \log n)}$ time by \cite{Lyubashevsky05}. 

LPN problems with low-noise rate like $\eta=1/\sqrt{n}$ even $\eta=\frac{\log^2 n}{n}$ are believed to be computational hard in cryptography. These stronger assumptions has wide applications in the design of advanced cryptographic schemes, to name a few  \citep{DMN12_PKC,KMP14_PKC,Brakerski19,Yu19,JLS21,JLS22}. However, the best known algorithm is still of time $e^{O(\eta n)}$ for any $\eta<\frac{1}{\log n}$.

LSPN has been heavily studied in both learning theory and cryptography. While the trivial enumeration has running time ${n \choose k} \approx (n/k)^k$, the state-of-the-art only gets polynomial improvement on it. In the low-noise regime, the best algorithm \citep{Grig11} runs in time at least ${n \choose k/2} \approx (n/k)^{k/2}$. In the high-noise regime, \cite{valiant2012finding} gives the first algorithm faster than ${n \choose k}$ time. Later on, \cite{KKK18} obtained time complexity $(n/k)^{k \cdot \frac{\omega}{3}}$ and sample complexity $\tilde{O}(k)$ via matching parities of size $k/2$ (called the light-bulb problem). Recently, there are two improvements on \emph{special cases} of LSPN. When $\eta<n^{-1/8}$, \cite{Yu21} provided an algorithm whose time complexity and sample complexity are both $\approx (\eta n)^{2k}$. \cite{DSGKS21} showed two algorithms with time $(n/k)^{o(k)}$ for two specific ranges of parameters ($k=\frac{n}{\log^{1+o(1)} n}$ in the high-noise regime and $k/\eta \approx n$ in the low-noise regime).

Much less is known about learning algorithms for sparse LPN even though its distinguishing algorithms has been extensively studied as refuting random $k$-XOR. When $m>n^{k/2}$, both learning and distinguishing problems have polynomial time algorithms \citep{AOW_15,BM16,Applebaum16}. For $m<n^{k/2}$, say $m=n^{1+(\frac{k}{2}-1)(1-\delta)}$, \cite{RRS17} showed \emph{distinguishing} algorithms of time $e^{\tilde{O}(n^{\delta})}$ via tensor methods. Essentially, the exponent $\tilde{O}(n^{\delta})$ comes from the size of the smallest linearly dependent set among those $m$ random $k$-sparse vectors \citep{ Feige2008,GKM,HKM_hyper_moore}. For \emph{learning} algorithms with $m<n^{k/2}$, besides Gaussian elimination of time $e^{\eta n}$, another approach is to reduce those distinguishing algorithms to learning. However, known reductions either takes cubicly many more samples \citep{ABW10, Applebaum13} or has a small success probability like $e^{-\Omega(m^{6/k})}$ \citep{BSV2019}. To the best of our knowledge, the fastest learning algorithm for small constants $k$ was $e^{\eta n}$ given any noise rate $\eta$ and $m<n^{k/2}$.

Sparse LPN has been widely used in cryptography \citep[e.g.,][]{Alekhnovich,ABW10,ADINZ18,CRR23,RRT23,DIJL_sparseLPN,RVV24} since it has several advantages compared to the standard LPN. On the first hand, sparse LPN improves efficiency in practical cryptosystems. On the second hand, sparsity brings more gadgets to the design of crypto-systems. For example, \cite{ABW10} constructed a public-key encryption based on sparse LPN with sparsity $k=3$, $m=n^{1.4}$ samples, and noise rate $\eta = o(n^{-0.2}$). Compared to public-key protocols based on LPN with noise rate $\eta \approx n^{-0.5}$ \citep{DMN12_PKC,KMP14_PKC}, this construction tolerates a much higher noise rate. It was conjectured in cryptography that the time complexity of sparse LPN is $e^{\tilde{\Omega}_{\eta}(n^{\delta})}$ given $m=n^{1+(\frac{k}{2}-1)(1-\delta)}$ samples, although it is unclear what is the exact dependency on $\eta$.

Independently and concurrently, \cite{DJ24} showed a distinguishing algorithm for sparse LPN, whose idea is similar to Theorem~\ref{thm:infor_sparse_LPN}. However, this distinguishing algorithm only works for $\eta=O(n^{-\frac{1+\delta}{2}})$, unlike Theroem~\ref{thm:infor_sparse_LPN} that provides a learning algorithm for any $\eta$.

Various LPN problems have also been studied in the literature. Related to LSPN, hardness of non-uniform secrets has been investigated in \cite{ACPS09,Goldwasser10}. One strand of research \citep{AroraGe11,Bartusek19,golowich2024learning} considered structured noise where the errors across multiple samples are guaranteed to satisfy certain constraints. Finally we refer to the survey \citep{Pie12} for other variants and applications of LPN in cryptography.

\paragraph{Organization.} Section~\ref{sec:preli} provides
basic notations and definitions. In Section~\ref{sec:overview}, we provide an overview of our algorithms. Then we show the learning algorithms for LSPN and sparse LPN in Section~\ref{sec:low_noise} and Section~\ref{sec:sparse_sample} separately. Due to the space constraint, we review some probabilistic tools in Appendix~\ref{app:preli} and defer several proofs in the main body to Appendix~\ref{app:proof}.
Then we show our improvement upon Valiant's framework for LSPN against high-noise in Appendix~\ref{sec:valiant}. Next, for sparse LPN, we show the learning algorithms against adversarial noise in Appendix~\ref{sec:adv_sparse_LPN} and extend the distinguishing algorithm by \cite{DJ24} in Appendix~\ref{sec:distinguish}.

\section{Preliminaries}\label{sec:preli}
We always use $[n]$ to denote $\{1,2,\ldots,n\}$. For convenience, let ${[n] \choose k}$ denote the family of all $k$-subsets in $[n]$ and ${n \choose k}$ denote the binomial coefficient $\frac{n!}{k!(n-k)!}$. In this work, we extensively use the fact that ${n \choose k} \in \big[ (n/k)^k, (en/k)^k) \big]$ for any $k \le n$.

Since there is a 1-1 map between $\mathbf{F}_2^n$ and all subsets of $[n]$, we use $\vec{1}_S \in \mathbf{F}_2^n$ to denote the indicator vector of any $S \subset [n]$. Moreover, for a vector $\bx \in \mathbf{F}_2^n$, we use $supp(\bx)$ to denote its support set.

Let $|S|$ denote the number of elements in $S$, which corresponds to the Hamming weight of $\vec{1}_S$.
For two subsets $S$ and $T$, we use $S \Delta T$ to denote their set-difference, which is $(S \setminus T) \cup (T \setminus S)$. 

For a matrix $\mX$ of dimension $m \times n$, we use $\mX(S,T)$ to denote its sub-matrix in $\mathbb{R}^{S \times T}$. We use $\mX(i,)$ to denote row $i$ of the matrix $\mX$. Similarly, for a vector $\bx \in \mathbb{R}^n$, $\bx(S)$ denotes its sub-vector in $\mathbb{R}^S$ for any $S \subseteq [n]$. 



For a probability event $E$, we use $\mathbf{1}(E)$ to denote the indicator function of $E$. 

\paragraph{Learning Parity with Noise problem.} For $m$ random samples $(\bx_1,y_1),\ldots,(\bx_m,y_m)$ in $\mathbf{F}_2^n \times \mathbf{F}_2$, we call $\bx_1,\ldots,\bx_m$ random vectors and $y_1,\ldots,y_m$ their labels. In this work, we use $(\mX,\by)$, where $\mX$ is a matrix in $\mathbf{F}_2^{m \times n}$ and $\by$ is a vector in $\mathbf{F}_2^m$, to denote the input of $m$ random samples $(\bx_1,y_1),\ldots,(\bx_m,y_m)$. Then we know $\by=\mX \cdot \secret + \vec{e}$ with a noise vector $\vec{e} \in \mathbf{F}_2^m$.

Now we define LSPN formally. Different from LPN, there is an input parameter $k$ such that the hidden parity $\secret$ has sparsity at most $k$.

\begin{definition}
	In a $(k,\eta)$-LSPN problem, the hidden vector $\secret \in \mathbf{F}_2^n$ has sparsity $\le k$ such that each random sample is of the form $(\bx,y)$ where its vector $\bx \sim \mathbf{F}_2^n$ is uniformly random and its label $y=\langle \bx, \secret \rangle + e$ for a Bernoulli noise $e$ of rate $\eta$. The goal is to find $\secret$ efficiently.
\end{definition}

Finally, we define sparse LPN problems. In a $k$-sparse LPN problem, each $\bx_i$ is drawn uniformly from vectors in $\mathbf{F}_2^n$ of sparsity $k$. For convenience, we use $\bx_i \sim {[n] \choose k}$ to denote this random sampling. 
\begin{definition}
	In a $(k,\eta)$-sparse LPN problem, there is a hidden vector $\secret \in \mathbf{F}_2^n$ such that each random sample is of the form $(\bx,y)$ where its vector $\bx \sim {[n] \choose k}$ and its label $y=\langle \bx, \secret \rangle + e$ for a Bernoulli noise $e$ of rate $\eta$. The goal is to find $\secret$ efficiently.	
\end{definition}


 \section{Overview}\label{sec:overview}
We provide an overview of our learning algorithms in this section. The basic idea of our algorithms is to reduce the original dimension $n$ by leveraging the prior knowledge of sparsity. However, this leads to very different algorithms for sparse LPN and LSPN. Hence we describe them separately.

\paragraph{Algorithm for LSPN.} In a $(k,\eta)$-LSPN instance, $\secret$ is $k$-sparse and each sample has $\bx_i \sim \mathbf{F}_2^n$. To reduce the dimension $n$, the rough idea is to guess a support set $S$ for $\secret$. Intuitively, if a given subset $S$ satisfies $S \supseteq supp(\secret)$, it only needs $|S|$ linearly independent samples with noiseless labels to recover $\secret$. Specifically, one could take a subset of samples, say $T \subset [m]$ with $|S|+O(1)$ samples, and apply Gaussian elimination on $S \subset [n]$ to solve those linear equations indexed by $T$. In another word, it finds a candidate $\mathbf{ans} \in \mathbf{F}_2^S$ such that $\mX(T,S) \cdot \mathbf{ans}=\by(T)$.  When these samples are noiseless, $\mathbf{ans}$ would equal $\secret$. The latter one happens with probability $\approx e^{-\eta \cdot |S|}$.

While the algorithm wants $S$ to be as small as possible (o.w. $S=[n]$ is trivial), it has no clue about the support of $\secret$. Hence we sample a random subset $S$ of size $q$ in $[n]$ and projecting  all random samples from $[n]$ to $S$ as a domain reduction. It succeeds only if $\secret \subset S$ and the random samples in Gaussian elimination are noiseless. Since these two events are independent, the success probability is about $\frac{{n-k \choose q-k}}{{n \choose q}} \cdot e^{-\eta \cdot q} \approx (\frac{q}{n})^k \cdot e^{-\eta \cdot q}$. We choose $q:=k/\eta$ such that this probability is about $(\frac{k}{\eta n})^k \cdot e^{-k}$. 

To reduce the sample complexity, we prepare a pool of $O(q)$ samples and use $q+O(1)$ random samples in this pool for every round of Gaussian elimination instead of fresh samples. 

 
\paragraph{Learning algorithm for sparse LPN.} Different from LSPN, sparse LPN has a potentially dense vector $\secret \in \mathbf{F}_2^n$ and each $\bx_i \sim {[n] \choose k}$. To reduce the original problem of size $n$ to a sub-problem, we choose a subset $I \subset [n]$ and consider those samples $(\bx_i, y_i)_{i=1,\ldots,m}$ whose vectors $\bx_i$ have support $supp(\bx_i) \subset I$. The probability that a random $k$-sparse vector $\bx_i$ satisfies $supp(\bx_i) \subset I$ is about $(\frac{|I|}{n})^k$. Since this probability is relatively small given $m<n^{k/2}$ samples, we could only collect a limited number of such samples.
Now observe that this reduces to a sub-problem on $I$: given a limited number of samples $(\bx_i,y_i)$ such that $\bx_i \in \mathbf{F}_2^I$ and $y_i$ is a noisy label of $\langle \bx_i(I), \secret(I) \rangle$, how to figure out $\secret(I)$ the restriction of $\secret$ in $I$? 

Since Gaussian elimination is the only algorithm for sparse LPN, we consider how to apply it on $I$. This needs $|I|$ linearly independent vectors in $\mathbf{F}_2^I$; but how many $k$-sparse random vectors will span $\mathbf{F}_2^I$? 
It is known \citep{cooperRankRandomBinary2018} that $\Theta({|I|} \cdot \frac{\log |I|}{k})$ random samples suffice (when $k<\log |I|$). Hence, one could expect Gaussian elimination on $O({|I|} \cdot \frac{\log |I|}{k})$ samples of the subproblem will return $\secret(I)$ with probability $(1-\eta)^{|I|} \approx e^{-\eta \cdot |I|}$ (when these samples are noiseless).

However, there are two small issues. First of all, we want to recycle those random samples  to reduce the sample complexity to $<n^{k/2}$ instead of using fresh samples in $e^{\eta \cdot |I|}$ rounds of Gaussian eliminations. Similar to the LSPN algorithm, for each sub-problem $I$, we prepare a pool of samples and pick a subset of size $O({|I|} \cdot \frac{\log |I|}{k})$ in this pool. However, we can not guarantee the success probability is still $(1-\eta)^{|I|}$ for recycled samples and we use $e^{-\eta \cdot O({|I|} \cdot \frac{\log |I|}{k})}$ instead. Secondly, it takes more samples to test a hypothesis in sparse LPN. In particular, if the Hamming distance between the hypothesis and $\secret(I)$ is 1, it would take $\Theta(\frac{\eta \cdot |I|^2}{k^2})$ samples to distinguish between them under noise rate $\eta$. This is equivalent to distinguishing between two biased coins with expectations $\eta$ and $ \eta + \Theta(k/|I|)$. In fact, the previous calculation about the number of random $k$-sparse vectors spanning $\mathbf{F}_2^n$ is the number of testing samples with noise rate $\eta = 0$. Hence we provide a complete calculation to show that the verification needs $O(|I|) \cdot \max\{1,\frac{\log |I|}{k}, \frac{\eta |I| \cdot \log |I|}{k^2}\}$ random $k$-sparse vectors. 

Finally, we apply the above procedure several times for different subsets $I$ to decode $\secret \in \mathbf{F}_2^n$. The last point is that for disjoint subsets $I_1,\ldots,I_t$, the total number of samples is 
\[
\max_{j=1}^t O(|I_j|) \cdot \max\{1,\frac{\log |I_j|}{k}, \frac{\eta |I_j| \cdot \log |I_j|}{k^2}\} \cdot (n/|I_j|)^k,\]
not their summation over $I_1,\ldots,I_t$ (since those events that a random $\bx$ has $supp(\bx) \subset I_j$ are disjoint). 

\section{Learning Algorithm for LSPN}\label{sec:low_noise}
We state the formal version of Theorem~\ref{thm:infor1} for learning sparse parity against low noise and finish its proof in this section. In Section~\ref{sec:subset_BKW}, we combine our idea with the BKW algorithm to get another efficient algorithm against high-noise. For ease of exposition, we assume $\eta<0.05$ in this section.



\begin{theorem}
	\label{thm:low_noise_main}
	For any $\eta$ and $k$ with $n>k/\eta$, Algorithm~\ref{alg::LSPN} solves a given $(k,\eta)$-LSPN instance with probability $0.9$ in $O(\frac{\eta \cdot n}{k})^{k}$  time and $O(\frac{k}{\eta}+ k \log \frac{n}{k})$ samples.
\end{theorem}

\begin{remark}[Generalizations to adversarial noise]\label{rm:adversarial_LSPN}
    One could modify this algorithm to learn $\secret$ against adversarial noise of rate $\eta$ under the same number of samples and almost the same time. In the adversarial setting, there exists a distribution $D(\bx,y)$ on $\mathbf{F}_2^n \times \mathbf{F}_2$ whose marginal distribution of $\bx$ is uniform such that $\Pr_{(\bx,y)\sim D}[\langle \bx, \secret \rangle=y] \ge 1-\eta$. 
    
    The first approach is to apply the reduction from adversarial noise to random noise by \cite{Feldman09}. This reduction will increase the running time by a factor of $m=O(\frac{k}{\eta}+ k \log \frac{n}{k})$. The second approach is to modify the parameter $q'$ in Function \textsc{LearnInSubset} to $q+O(\eta \cdot q)$. By the same analysis, this will increase the running time by a factor of $e^{O(\eta \cdot k)}$.
\end{remark}

The idea of domain reduction behind Algorithm~\ref{alg::LSPN} is to assume $supp(\secret) \subseteq S$ for a random subset $S \subset [n]$ and solve the sub-problem in $S$. To do this, it applies Gaussian eliminations to $q':=|S|+O(1)$ random samples (say set $T \subset [m]$ of samples) in $\mathbf{F}_2^S$. Since $\bx_i \sim \mathbf{F}_2^n$, it is easy to test hypotheses generated by Gaussian eliminations in this case. The last point is to choose the size $q:=k/\eta$ to optimize the probability that $S \supseteq supp(\secret)$ and the success probability of Gaussian eliminations in $\mathbf{F}_2^S$.

\begin{algorithm}[h]

	\caption{Algorithm to Learn Sparse Secret}
	\label{alg::LSPN}
 \SetKwFunction{test}{Test} 
 \SetKwFunction{LearnInSubset}{LearnInSubset} 
  \SetKwFunction{Main}{Main} 
\SetKwProg{Fn}{Function}{:}{}
\setcounter{AlgoLine}{0}
\KwIn{$n,k,\eta$}    
\KwOut{$\mathbf{ans} \in \{0,1\}^n$ or $\bot$}        

\Fn{\test{$h$}}{

Compute $s \gets \sum_{i=1}^{m_2} \mathbf{1} \bigg( \big\langle \mX'(i,),h \big\rangle = y'(i) \bigg)$ \Comment{$\mX'$ and $y'$ are the samples for verification}

 \KwRet{$1$} if and only if $s \ge \frac{3}{4} \cdot m_2$ }

\textbf{EndFunction}
	
\Fn{\LearnInSubset{$S \subseteq [n], m_1,\mX \in \mathbf{F}_2^{m_1 \times n}, \by \in \mathbf{F}_2^{m_1}$}}{

            $q\gets |S|$ and $q' \gets q+25$ 
            
		\For {$j= 1,2,\cdots, O(1) \cdot e^{1.08 \cdot \eta q}$} {
		 Sample $T\sim \binom{[m_1]}{q'}$ \Comment{$T$ is the batch of samples in Gaussian elimination}
   
		 If $rank(\mX(T,S))<\abs{S}$, skip this round
   
    	Apply Gaussian elimination to find $\mathbf{ans} \in \mathbf{F}_2^S$ such that \[
			\mX(T,S) \cdot \mathbf{ans} = \by(T)
		\]
  
		 If $\mathbf{ans}$ does not exist or $|\mathbf{ans}|>k$, skip this round
   
            If \textsc{Test}($\mathbf{ans}$)=True, \KwRet{$\mathbf{ans}$}
  
		}
		 \KwRet{ $\bot$}
}
\textbf{EndFunction};

		\Fn{\Main{$n,k,\eta$} }  {
  \tcc{Main procedure of our algorithm}
  
    $m_1 \gets O(k/\eta)$ and $m_2 \gets O(k \log n/k)$
    
    Take $m_1$ random samples $(\bx_1,y_1),\ldots,(\bx_{m_1},y_{m_1})$ and store them as a matrix $\mX \in \mathbf{F}_2^{m_1 \times n}$ and $\by \in \mathbf{F}_2^{m_1}$           ~~~~~~~~~~~~~~~~~~~~~~~~~~\Comment{samples for Gaussian elimination}
    
    Take $m_2$ random samples $(\bx'_1,y'_1),\ldots,(\bx'_{m_2},y'_{m_2})$ and store them as a matrix $\mX' \in \mathbf{F}_2^{m_2 \times n}$ and $\by' \in\mathbf{F}_2^{m_2}$
            ~~~~~~~~~~~~~~~~\Comment{samples for verification}
            
             $q\gets \frac{k}{\eta}$   
             
		\For {$i= 1,2,\cdots, O(1) \cdot (n/q)^k \cdot e^{\eta k}$}{
		 Sample $S \sim {[n] \choose q}$
   
            $\mathbf{ans} \gets \textsc{LearnInSubset}(S,m_1,\mX, \by)$
            
		 If $\mathbf{ans}\neq \bot$, \KwRet{$\mathbf{ans}$}
		}
           \KwRet{$\bot$} }
\textbf{EndFunction};	
\end{algorithm}

We describe the main procedure and the Gaussian elimination procedure (called Function~\textsc{LearnInSubset}) in Algorithm~\ref{alg::LSPN}. Here are a few notations. In both procedures, the size of the random subset $S$ is $q:=k/\eta$ and the pool of random samples for applying Gaussian eliminations is of size $m_1 := 100 \cdot q + O(1)$, recorded as $\mX \in \mathbf{F}_2^{m_1 \times n}$ and $\by \in \mathbf{F}_2^{m_1}$. An extra point is to use another $m_2=O(k \log \frac{n}{k})$ random samples, recorded as $\mX' \in \mathbf{F}_2^{m_2 \times n}$ and $\by' \in \mathbf{F}_2^{m_2}$, to verify every solution returned by the Gaussian elimination.
Since the number of parities of size $\le k$ is at most ${n \choose 1} + \cdots + {n \choose k} \le k \cdot (en/k)^k$ and $\eta<0.05$, we assume that $\secret$ is the only parity whose agreements with $y_i$ are more than $\frac{3}{4} \cdot m_2$ given $m_2=O(k \log \frac{n}{k})$ random samples. Hence, the final sample complexity is $m=m_1+m_2$.

In the rest of this section, we sketch a proof for Theorem~\ref{thm:low_noise_main}. Our analysis relies on the following fact: Since $\mX$ is a random matrix, for any pair of subsets $(T, S)$, the rank of submatrix $\mX(T,S)$ equals $\min\{|S|,|T|\}$ with good probability.
\begin{proposition}
	\label{cla:T&S}
	For any $S \in {[n] \choose q}$ and $ T\in \binom{[m_1]}{q'}$ with $q'=q+25$, we have
	$$\Pr_{\mX}\bigg[ rank(\mX(T,S))=\abs S \bigg]\geq 0.99.$$
\end{proposition}

The following proposition guarantees the correctness of function \text{LearnInSubset} in Algorithm~\ref{alg::LSPN}.

\begin{proposition}\label{clm:SUBSETLEARN}
	If $supp(\secret) \subset S$ and $m_1 \ge 100q$, 
	Procedure \textsc{LearnInSubset($S,\mX,\by$)} finds $\secret$ in $q^{O(1)} \cdot e^{1.08  \eta \cdot q}$ time with probability $0.95$.
\end{proposition}

Since $q=k/\eta$, the running time of Proposition~\ref{clm:SUBSETLEARN} would be $(k/\eta)^{O(1)} \cdot e^{1.08k}$ eventually. For completeness, we prove Proposition~\ref{cla:T&S} in Appendix~\ref{sec:T&S} and prove Proposition~\ref{clm:SUBSETLEARN} in Appendix~\ref{sec:proof_of_less_main} separately. Now, we are ready to prove Theorem~\ref{thm:low_noise_main}.

\begin{proofof}{Theorem~\ref{thm:low_noise_main}}

	Firstly, we bound the time complexity. Since the running time of Proposition~\ref{clm:SUBSETLEARN} is $(k/\eta)^{O(1)} \cdot e^{1.08k}$ given $q=k/\eta$, the total time complexity is 
 \[
 (k/\eta)^{O(1)} \cdot e^{1.08k} \cdot (n/q)^k \cdot e^{\eta k} \cdot m_2 = (k/\eta)^{O(1)} \cdot O(\eta n/k)^k.
 \]

	Next, we show the success probability $\ge 0.9$. Because $\Pr[\secret \subset S]={{n-k}\choose {q-k}}/{n\choose q}$, we show ${n\choose q}/{{n-k}\choose {q-k}} = O(1) \cdot (n/q)^k \cdot e^{\eta k}$ to guarantee $\secret \subset S$ occurs at least once with probability $0.99$ after taking enough samplings of $S$. 
	\begin{align*}
		 {{n\choose q}}/{{{n-k}\choose {q-k}}}		 & =O(1) \cdot \frac{n^{n+0.5}}{q^{q+0.5}(n-q)^{n-q+0.5}}\cdot \frac{(q-k)^{q-k+0.5}(n-q)^{n-q+0.5}}{(n-k)^{n-k+0.5}} \tag{\text{apply the Stirling's Formula}}                             \\
		 & =O(1) \cdot \frac{n^{n}}{(n-k)^{n-k}}\cdot \frac{(q-k)^{q-k}}{q^{q}}                                               \\
		 & = O(1) \cdot \left(\frac{n}{q}\right)^k\cdot \left(1+\frac{k}{n-k}\right)^{n-k}\cdot\left(1-\frac kq\right)^{q-k}  \\
		 & \le O(1) \cdot \left(\frac{n}{q}\right)^k \cdot e^{\frac{k}{n-k}\cdot(n-k)} \cdot e^{-\frac{k}{q}\cdot(q-k)}  = O(1) \cdot \left(\frac{n}{q}\right)^k\cdot e^{\frac{k^2}{q}}.
	\end{align*}
        Since $q=k/\eta$, the last term $e^{\frac{k^2}{q}}=e^{\eta k}$.

Then by Proposition~\ref{clm:SUBSETLEARN}, function \textsc{LearnInSubset}($S,m_1,m_2,\mX,\by$) will return $\mathbf{ans}$ with probability at least $0.95$ under the condition of $\secret\subset S$. From all discussion above, we conclude that, with probability 0.9, Algorithm~\ref{alg::LSPN} returns $\secret$.        

\end{proofof}

\subsection{Subset Learning with BKW}\label{sec:subset_BKW}

In this section, we show an algorithm by combining the BKW algorithm with our sampling idea.
\begin{theorem}
	\label{thm:bkw_main}
	For any constant $\eta<0.5$ and $k<n/\log n$, there exists an algorithm in $O(\frac{n}{k\log k})^{k}$ time and $m=O(2^{O(k)}+k\log \frac{n}{k})$ samples to solve $(k,\eta)$-LSPN problem.
\end{theorem}

\begin{proofof}{Theorem~\ref{thm:bkw_main}}
	We modify Algorithm~\ref{alg::LSPN} in two aspects: (a) reset $q=k \log k$; (b) replace calling function~\textsc{LearnInSubset} by applying the classical BKW algorithm to $S$ (i.e., we only retain the bits in each sample vector where the subscript belongs to $S$, and run $\mathrm{BKW}$ on these new samples). The time complexity is $2^{O(q/\log q)} \cdot 10 \cdot \frac{{n \choose q}}{{n-k \choose q-k}}=O(\frac{n}{k \log k})^k$.

	Note that we anticipate the success of $\mathrm{BKW(S)}$ only at the precise run when $\secret\subseteq S$. Therefore, there is no need to use new samples every time when we invoke BKW on $S$. And $2^{O(\frac{q}{\log q})}=2^{O(k)}$ samples will suffice.

\end{proofof}

One remark is that if $k=\frac{n}{r \cdot \log n}$ for $r \in  [\Omega(1),\log^{o(1)} n]$, the running time of Theorem~\ref{thm:bkw_main} is $O(r)^{k}=o(n/k)^{k}=(n/k)^{o(k)}$ faster than BKW and the naive algorithm of running time $(n/k)^k=(r \log n)^k$.

\section{Learning Algorithm for sparse LPN}\label{sec:sparse_sample}
We show the learning algorithm of sparse LPN in this section. Let $k$ denote the sparsity of those random vectors. For ease of exposition, we assume that $k$ is odd such that the answer is unique; otherwise both $\secret$ and its flip $\overline{\secret}$ are valid when $k$ is even. 

\begin{theorem}\label{thm:sparse_LPN}
    For any $(k,\eta)$-sparse LPN problem, given any $\delta>0$ satisfying $k=o(n^{\frac{1+\delta}{2}})$, Function \textsc{Main} in Algorithm~\ref{alg::sparseLPN_main} takes $m:=\max\{1, \frac{\log n}{k}, \frac{\eta n^{\frac{1+\delta}{2}} \cdot \log n}{k^2}\} \cdot O(n)^{1+(1-\delta) \cdot \frac{k-1}{2}}$ samples to return $\secret$ in time $n^{O(1)} \cdot \big( e^{O(\eta \cdot n^{\frac{1+\delta}{2}} \cdot \max\{1, \frac{\log n}{k}\})} + m \big)$ with probability $1-1/n$.
\end{theorem}

\begin{algorithm}[H]
\SetAlgoLined
    \caption{Algorithm for sparse LPN}
    \label{alg::sparseLPN_main}
    \SetKwFunction{Main}{Main} 
\SetKwProg{Fn}{Function}{:}{}
\setcounter{AlgoLine}{0}    
    \KwIn{$k$, $\eta$, and a parameter $\delta \in (0,1)$}    
    \KwOut{$\mathbf{ans} \in \mathbf{F}_2^n$ or $\bot$}   
        \Fn{\Main{} }{           
       
         $t \gets n^{\frac{1-\delta}{2}}$ and $q \gets n/t$ 

         $m \gets O(q) \cdot \max\{1,\frac{\log q}{k}, \frac{\eta q \log q}{k^2}\} \cdot (et)^k$ \Comment{Number of samples}

         define $t$ sub-problems $I_1,\ldots, I_t \subset [n]$: $I_j \gets \{(j-1)\cdot q + 1, \ldots, j \cdot q\}$ for each $j \in [t]$
        
         initialize $m_j =0$ and  matrix $\mX_j \in \mathbf{F}_2^{m_j \times n}$ and $\by_j \in \mathbf{F}_2^{m_j}$ for each $j \in [t]$ \Comment{$\mX_j$ records random samples for sub-problem $I_j$}
        
        \For {$i= 1,2,\ldots, m$}{   \tcc{Allocate $m$ random samples into $t$ sub-problems}
        
             Take a fresh sample $(\bx,y)$
             
            \If{$supp(\bx) \subset I_j$ for some $I_j$}{
                 $m_j \gets m_j +1$
                 
                 Store $\bx$ into a new row: $\mX_j(m_j,) \gets \bx$ and $\by_j(m_j)\gets y$
            }
        }
        \For {$j=1,\ldots,t$}{
             $\mathbf{ans}(I_j) \gets \textsc{PartialLearn}(I_j,m_j,\mX_j, \by_j)$
        }
       \KwRet{$\mathbf{ans}$}                
       }
\textbf{EndFunction}
\end{algorithm}

While Algorithm~\ref{alg::sparseLPN_main} shares the same idea as Algorithm~\ref{alg::LSPN} for LSPN problems of applying domain reductions, there are several differences. First of all, the hidden $\secret$ is no longer sparse such that solving one sub-problem is not enough to learn $\secret$ here. So we will split $n$ entries  into $t$ subsets $I_1,\ldots,I_t$ 
and decode $\secret$ in these sub-problems of $\secret(I_1),\ldots,\secret(I_n)$ separately. The key step is Function~\textsc{PartialLearn} in Algorithm~\ref{alg::sparseLPN} which learns $\secret(I)$ given sufficient samples with $supp(\bx_i)\subset I$ (stated in Proposition~\ref{clm:decode_subset}). Moreover, the number of random samples will be the main parameter in the design of Algorithm~\ref{alg::sparseLPN} since different sub-problems need different random samples. In particular, more random samples would provide smaller sub-problems and faster running time in Algorithm~\ref{alg::sparseLPN_main}. At last, it takes more random samples to validate the correctness of a hypothesis $\mathbf{h} \in \mathbf{F}_2^n$. For example, if $\mathbf{h}$ and $\secret$ differ on only 1 bit, it takes $\Theta(n/k)$ random samples to distinguish them even in the noiseless setting. 

\begin{algorithm}[h]
\SetAlgoLined
    \caption{Solve the sub-problem in $I$}
    \label{alg::sparseLPN}
     \SetKwFunction{Test}{Test} 
 \SetKwFunction{PartialLearn}{PartialLearn} 
\SetKwProg{Fn}{Function}{:}{}
\setcounter{AlgoLine}{0}
\KwIn{$I\subset [n],m,\mX \in \mathbf{F}_2^{m \times I},\by \in \mathbf{F}_2^m$}    
\KwOut{$ans \in \mathbf{F}_2^I$ or $\bot$}   
\Fn{\Test{$\mathbf{h}$}}{ \tcc{sub-function of \textsc{PartialLearn} to verify that $\mathbf{h} \in \mathbf{F}_2^I$ equals $\secret(I)$ given $m,m',\mX$, and $\by$ in \textsc{PartialLearn}}
        
$s \gets \sum_{i=m'+1}^m \mathbf{1}\big( \langle \bx_i, \mathbf{h} \rangle \neq y_i \big)$ 
                 
\If{$s \le (m-m')\cdot (\eta + c_0 \cdot \frac{k}{q})$} {            \tcc{$c_0$ is a small constant from Corollary~\ref{cor:verification}}
       \KwRet{True}
    } 
}
\textbf{EndFunction}
        
        \Fn{\PartialLearn{$I,m,\mX,\by$}} { \tcc{the goal is to learn $\secret(I)$ given $m$ samples $(\bx_i,y_i)$ in $(\mX,\by)$ with $supp(\bx_i) \subset I$}
        
         $q \gets |I|$
        
         $C \gets $ a sufficiently large constant
        
         If {$m<C \cdot q \cdot \max\{1,\frac{\log q}{k}, \frac{\eta q \log q}{k^2}\}$}, return $\bot$
        
      $d \gets c_1 \cdot q \cdot \max\{1,\frac{\log q}{k}\}$ is the number of required samples from Proposition~\ref{lem:ksparse_span} 
        
        $m' \gets 100 \cdot d$ \qquad \Comment{Use the first $m'$ samples in $(\mX,\by)$ to try Gaussian eliminations and the rest $(m-m')$ samples for verification}

        $L \gets 10 q^2 + 10 \log q \cdot e^{\frac{200c_1}{99} \cdot \eta q \max\{1,\frac{\log q}{k}\}}$ \Comment{Rounds of Gaussian eliminations}
        
        \For{$i=1,\ldots,L$}{
            Pick a random set of samples: $T \sim {[m'] \choose d}$ 
            
            \If{rank$\{\mX(T,I)\}=|I|$}{ 
            
               Apply Gaussian elimination to find $\mathbf{ans}\in \mathbf{F}_2^I$ such that                $
               \mX(T,I) \cdot \mathbf{ans} = \by(T)
               $
                    
                If $\textsc{Test}(\mathbf{ans})=True$, \KwRet{$\mathbf{ans}$}
            }
        }
          \KwRet{$\bot$}     }   
        \textbf{EndFunction}
      \end{algorithm}

The idea of Function \textsc{PartialLearn} is similar to the idea of Function \textsc{LearnInSubset} in the LSPN Algorithm~\ref{alg::LSPN}. We prepare a pool of random samples to try Gaussian elimination, say the first $m'$ random samples, and apply the rest to validate each hypothesis $h$ (in Function~\textsc{Test}). Then each Gaussian elimination would use $d:=\tilde{O}(|I|)$ random samples in the first $m'$ random samples to find a hypothesis $h$. The correctness of Function \textsc{PartialLearn} is stated in Proposition~\ref{clm:decode_subset}, whose proof is deferred to Section~\ref{sec:proof_decode_subset}. 

\begin{proposition}\label{clm:decode_subset}
    Given a subset $I \subset [n]$ and $m \ge C \cdot \max\{1, \frac{\log |I|}{k}, \frac{\eta |I| \cdot \log |I|}{k^2}\} \cdot |I|$ random samples (for some universal constant $C$), Function \textsc{PartialLearn} in Algorithm~\ref{alg::sparseLPN} outputs $\secret(I)$ with probability $1- \frac{O(1)}{|I|^4}$ in time $(m+|I|^{O(1)}) \cdot e^{O(\eta \cdot |I| \cdot \max\{1, \frac{\log |I|}{k}\})}$.
\end{proposition}

The proof of  Proposition~\ref{clm:decode_subset} relies on the following two calculations about the sample complexity. The first one is to bound the number $d$ of random samples in each Gaussian elimination. This is equivalent to the number of random sparse vectors spanning the whole space. 

\begin{proposition}\label{lem:ksparse_span}
     For any odd number $k<n/\log n$, $t:=c_1 \cdot n \cdot \max\{1,\frac{\log n}{k}\}$ random $k$-sparse vectors in $\mathbf{F}_2^n$ span the whole space $\mathbf{F}_2^{n}$ with probability $1-1/n^5$ (for some constant $c_1$).
\end{proposition}

While this was known before \citep{cooperRankRandomBinary2018}, we provide a proof of Proposition~\ref{lem:ksparse_span} in Section~\ref{sec:proof_sparse_span} for completeness. The second calculation is an extension of Proposition~\ref{lem:ksparse_span} that bounds the sample complexity of Function \textsc{Test} in Algorithm~\ref{alg::sparseLPN}.

\begin{corollary}{\label{cor:verification}}
    For the $(k,\eta)$-Sparse LPN problem of dimension $n$, $m=O(n) \cdot \max\{1,\frac{\log n}{k}, \frac{\eta n \log n}{k^2}\}$ random samples guarantee that with probability $1-1/n^5$, $\secret$ is the only vector $\bv$ in $\mathbf{F}_2^n$ satisfying
    $\sum_{i=1}^m \mathbf{1}(\langle \bx_i,\bv\rangle \neq y_i) \le (\eta + c_0 \cdot \frac{k}{n}) \cdot m$ (for some constant $c_0$).
\end{corollary}

Essentially, the difference between Corollary~\ref{cor:verification} and Proposition~\ref{lem:ksparse_span}  is that Corollary~\ref{cor:verification} bounds the number of samples under noise rate $\eta$ and Proposition~\ref{lem:ksparse_span} bounds the number of samples in the noiseless setting. Due to the space constraint, we defer their proofs (including Theorem~\ref{thm:sparse_LPN}) to Appendix~\ref{app:proof}.

\bibliography{cit}

\begin{thebibliography}{55}
\providecommand{\natexlab}[1]{#1}
\providecommand{\url}[1]{\texttt{#1}}
\expandafter\ifx\csname urlstyle\endcsname\relax
  \providecommand{\doi}[1]{doi: #1}\else
  \providecommand{\doi}{doi: \begingroup \urlstyle{rm}\Url}\fi

\bibitem[Akavia et~al.(2003)Akavia, Goldwasser, and Safra]{akagolsaf03}
Adi Akavia, Shafi Goldwasser, and Samuel Safra.
\newblock Proving hard-core predicates using list decoding.
\newblock pages 146--159, 2003.

\bibitem[Alekhnovich(2003)]{Alekhnovich}
Michael Alekhnovich.
\newblock More on average case vs approximation complexity.
\newblock FOCS '03, page 298. IEEE Computer Society, 2003.
\newblock ISBN 0769520405.

\bibitem[Allen et~al.(2015)Allen, ODonnell, and Witmer]{AOW_15}
Sarah~R. Allen, Ryan ODonnell, and David Witmer.
\newblock How to refute a random csp.
\newblock In \emph{Proceedings of the 2015 IEEE 56th Annual Symposium on Foundations of Computer Science}, FOCS '15, page 689–708, USA, 2015. IEEE.
\newblock ISBN 9781467381918.
\newblock \doi{10.1109/FOCS.2015.48}.

\bibitem[Applebaum(2013)]{Applebaum13}
Benny Applebaum.
\newblock Pseudorandom generators with long stretch and low locality from random local one-way functions.
\newblock \emph{SIAM Journal on Computing}, 42\penalty0 (5):\penalty0 2008--2037, 2013.
\newblock \doi{10.1137/120884857}.

\bibitem[Applebaum(2016)]{Applebaum16}
Benny Applebaum.
\newblock Cryptographic hardness of random local functions.
\newblock \emph{Computational complexity}, 25:\penalty0 667--722, 2016.
\newblock \doi{10.1007/s00037-015-0121-8}.

\bibitem[Applebaum et~al.(2009)Applebaum, Cash, Peikert, and Sahai]{ACPS09}
Benny Applebaum, David Cash, Chris Peikert, and Amit Sahai.
\newblock Fast cryptographic primitives and circular-secure encryption based on hard learning problems.
\newblock In \emph{Proceedings of the 29th Annual International Cryptology Conference on Advances in Cryptology}, CRYPTO '09, page 595–618. Springer-Verlag, 2009.
\newblock ISBN 9783642033551.
\newblock \doi{10.1007/978-3-642-03356-8_35}.

\bibitem[Applebaum et~al.(2010)Applebaum, Barak, and Wigderson]{ABW10}
Benny Applebaum, Boaz Barak, and Avi Wigderson.
\newblock Public-key cryptography from different assumptions.
\newblock In \emph{Proceedings of the Forty-Second ACM Symposium on Theory of Computing}, STOC '10, page 171–180. Association for Computing Machinery, 2010.
\newblock ISBN 9781450300506.
\newblock \doi{10.1145/1806689.1806715}.

\bibitem[Applebaum et~al.(2017)Applebaum, Damg{\aa}rd, Ishai, Nielsen, and Zichron]{ADINZ18}
Benny Applebaum, Ivan Damg{\aa}rd, Yuval Ishai, Michael Nielsen, and Lior Zichron.
\newblock Secure arithmetic computation with constant computational overhead.
\newblock In \emph{Advances in Cryptology -- CRYPTO 2017}, pages 223--254. Springer International Publishing, 2017.
\newblock ISBN 978-3-319-63688-7.

\bibitem[Arora and Ge(2011)]{AroraGe11}
Sanjeev Arora and Rong Ge.
\newblock New algorithms for learning in presence of errors.
\newblock In \emph{Proceedings of the 38th International Colloquim Conference on Automata, Languages and Programming - Volume Part I}, ICALP'11, page 403–415. Springer-Verlag, 2011.
\newblock ISBN 9783642220050.

\bibitem[Barak and Moitra(2022)]{BM16}
Boaz Barak and Ankur Moitra.
\newblock Noisy tensor completion via the sum-of-squares hierarchy.
\newblock \emph{Math. Program.}, 193\penalty0 (2):\penalty0 513–548, June 2022.
\newblock ISSN 0025-5610.
\newblock \doi{10.1007/s10107-022-01793-9}.
\newblock URL \url{https://doi.org/10.1007/s10107-022-01793-9}.

\bibitem[Barak et~al.(2014)Barak, Kelner, and Steurer]{BKS14}
Boaz Barak, Jonathan~A. Kelner, and David Steurer.
\newblock Rounding sum-of-squares relaxations.
\newblock In \emph{Proceedings of the Forty-Sixth Annual ACM Symposium on Theory of Computing}, STOC '14, page 31–40, New York, NY, USA, 2014. Association for Computing Machinery.
\newblock ISBN 9781450327107.
\newblock \doi{10.1145/2591796.2591886}.
\newblock URL \url{https://doi.org/10.1145/2591796.2591886}.

\bibitem[Bartusek et~al.(2019)Bartusek, Lepoint, Ma, and Zhandry]{Bartusek19}
James Bartusek, Tancr\`{e}de Lepoint, Fermi Ma, and Mark Zhandry.
\newblock New techniques for obfuscating conjunctions.
\newblock In \emph{Advances in Cryptology – EUROCRYPT 2019}, page 636–666. Springer-Verlag, 2019.
\newblock ISBN 978-3-030-17658-7.
\newblock \doi{10.1007/978-3-030-17659-4_22}.

\bibitem[Becker et~al.(2012)Becker, Joux, May, and Meurer]{Becker12}
Anja Becker, Antoine Joux, Alexander May, and Alexander Meurer.
\newblock Decoding random binary linear codes in 2n/20: how 1 + 1 = 0 improves information set decoding.
\newblock In \emph{Proceedings of the 31st Annual International Conference on Theory and Applications of Cryptographic Techniques}, EUROCRYPT'12, page 520–536. Springer-Verlag, 2012.
\newblock ISBN 9783642290107.
\newblock \doi{10.1007/978-3-642-29011-4_31}.

\bibitem[Blum et~al.(1994)Blum, Furst, Kearns, and Lipton]{BFKL93}
Avrim Blum, Merrick Furst, Michael Kearns, and Richard~J. Lipton.
\newblock Cryptographic primitives based on hard learning problems.
\newblock In \emph{Advances in Cryptology --- CRYPTO' 93}, pages 278--291. Springer Berlin Heidelberg, 1994.
\newblock ISBN 978-3-540-48329-8.

\bibitem[Blum et~al.(2003)Blum, Kalai, and Wasserman]{blum2003noise}
Avrim Blum, Adam Kalai, and Hal Wasserman.
\newblock Noise-tolerant learning, the parity problem, and the statistical query model.
\newblock \emph{Journal of the ACM (JACM)}, 50\penalty0 (4):\penalty0 506--519, 2003.

\bibitem[Bogdanov et~al.(2019)Bogdanov, Sabin, and Vasudevan]{BSV2019}
Andrej Bogdanov, Manuel Sabin, and Prashant~Nalini Vasudevan.
\newblock {XOR} codes and sparse learning parity with noise.
\newblock In \emph{Proceedings of the Thirtieth Annual ACM-SIAM Symposium on Discrete Algorithms}, SODA '19, page 986–1004. SIAM, 2019.

\bibitem[Boyle et~al.(2022)Boyle, Couteau, Gilboa, Ishai, Kohl, Resch, and Scholl]{BCGI22}
Elette Boyle, Geoffroy Couteau, Niv Gilboa, Yuval Ishai, Lisa Kohl, Nicolas Resch, and Peter Scholl.
\newblock Correlated pseudorandomness from expand-accumulate codes.
\newblock In \emph{Advances in Cryptology -- CRYPTO 2022}, pages 603--633. Springer Nature Switzerland, 2022.
\newblock ISBN 978-3-031-15979-4.

\bibitem[Brakerski et~al.(2019)Brakerski, Lyubashevsky, Vaikuntanathan, and Wichs]{Brakerski19}
Zvika Brakerski, Vadim Lyubashevsky, Vinod Vaikuntanathan, and Daniel Wichs.
\newblock Worst-case hardness for {LPN} and cryptographic hashing via code smoothing.
\newblock In \emph{Advances in Cryptology – EUROCRYPT 2019}, page 619–635. Springer-Verlag, 2019.
\newblock ISBN 978-3-030-17658-7.
\newblock \doi{10.1007/978-3-030-17659-4_21}.

\bibitem[Bshouty and Haddad(2024)]{bshouty2024}
Nader~H. Bshouty and George Haddad.
\newblock {Approximating the Number of Relevant Variables in a Parity Implies Proper Learning}.
\newblock In \emph{APPROX/RANDOM 2024}, volume 317 of \emph{Leibniz International Proceedings in Informatics (LIPIcs)}, pages 38:1--38:15. Schloss Dagstuhl -- Leibniz-Zentrum f{\"u}r Informatik, 2024.
\newblock \doi{10.4230/LIPIcs.APPROX/RANDOM.2024.38}.

\bibitem[Chen and De(2020)]{ChenD20}
Xue Chen and Anindya De.
\newblock Reconstruction under outliers for fourier-sparse functions.
\newblock In \emph{Proceedings of the 2020 {ACM-SIAM} Symposium on Discrete Algorithms, {SODA} 2020}, pages 2010--2029. {SIAM}, 2020.
\newblock \doi{10.1137/1.9781611975994.124}.
\newblock URL \url{https://doi.org/10.1137/1.9781611975994.124}.

\bibitem[Cooper et~al.(2019)Cooper, Frieze, and Pegden]{cooperRankRandomBinary2018}
Colin Cooper, Alan Frieze, and Wesley Pegden.
\newblock On the rank of a random binary matrix.
\newblock In \emph{Proceedings of the Thirtieth Annual ACM-SIAM Symposium on Discrete Algorithms}, pages 946--955. SIAM, 2019.

\bibitem[Couteau et~al.(2021)Couteau, Rindal, and Raghuraman]{CRR23}
Geoffroy Couteau, Peter Rindal, and Srinivasan Raghuraman.
\newblock Silver: Silent vole and oblivious transfer from hardness of decoding structured {LDPC} codes.
\newblock In \emph{Advances in Cryptology – CRYPTO 2021: 41st Annual International Cryptology Conference, CRYPTO 2021}, page 502–534. Springer-Verlag, 2021.
\newblock ISBN 978-3-030-84251-2.
\newblock \doi{10.1007/978-3-030-84252-9_17}.

\bibitem[Dachman-Soled et~al.(2021)Dachman-Soled, Gong, Kippen, and Shahverdi]{DSGKS21}
Dana Dachman-Soled, Huijing Gong, Hunter Kippen, and Aria Shahverdi.
\newblock {BKW} meets fourier new algorithms for {LPN} with sparse parities.
\newblock In \emph{Theory of Cryptography: 19th International Conference, TCC 2021}, page 658–688. Springer-Verlag, 2021.
\newblock ISBN 978-3-030-90452-4.
\newblock \doi{10.1007/978-3-030-90453-1_23}.

\bibitem[Dao and Jain(2024)]{DJ24}
Quang Dao and Aayush Jain.
\newblock Lossy cryptography from code-based assumptions.
\newblock In \emph{Advances in Cryptology – CRYPTO 2024: 44th Annual International Cryptology Conference}, page 34–75, Berlin, Heidelberg, 2024. Springer-Verlag.
\newblock ISBN 978-3-031-68381-7.
\newblock \doi{10.1007/978-3-031-68382-4_2}.

\bibitem[Dao et~al.(2023)Dao, Ishai, Jain, and Lin]{DIJL_sparseLPN}
Quang Dao, Yuval Ishai, Aayush Jain, and Huijia Lin.
\newblock Multi-party homomorphic secret sharing and sublinear mpc from sparse {LPN}.
\newblock In \emph{Advances in Cryptology – CRYPTO 2023: 43rd Annual International Cryptology Conference, CRYPTO 2023}, page 315–348. Springer-Verlag, 2023.
\newblock ISBN 978-3-031-38544-5.
\newblock \doi{10.1007/978-3-031-38545-2_11}.

\bibitem[D\"{o}ttling et~al.(2012)D\"{o}ttling, M\"{u}ller-Quade, and Nascimento]{DMN12_PKC}
Nico D\"{o}ttling, J\"{o}rn M\"{u}ller-Quade, and Anderson C.~A. Nascimento.
\newblock {IND-CCA} secure cryptography based on a variant of the {LPN} problem.
\newblock In \emph{Proceedings of the 18th International Conference on The Theory and Application of Cryptology and Information Security}, ASIACRYPT'12, page 485–503. Springer-Verlag, 2012.
\newblock ISBN 9783642349607.
\newblock \doi{10.1007/978-3-642-34961-4_30}.

\bibitem[Feige(2008)]{Feige2008}
Uriel Feige.
\newblock \emph{Small Linear Dependencies for Binary Vectors of Low Weight}, pages 283--307.
\newblock Springer Berlin Heidelberg, 2008.
\newblock ISBN 978-3-540-85221-6.
\newblock \doi{10.1007/978-3-540-85221-6_9}.

\bibitem[Feige et~al.(2006)Feige, Kim, and Ofek]{FKO06}
Uriel Feige, Jeong~Han Kim, and Eran Ofek.
\newblock Witnesses for non-satisfiability of dense random 3{CNF} formulas.
\newblock In \emph{2006 47th Annual IEEE Symposium on Foundations of Computer Science (FOCS'06)}, pages 497--508, 2006.
\newblock \doi{10.1109/FOCS.2006.78}.

\bibitem[Feldman(2007)]{Feldman07}
Vitaly Feldman.
\newblock Attribute-efficient and non-adaptive learning of parities and dnf expressions.
\newblock \emph{J. Mach. Learn. Res.}, 8:\penalty0 1431–1460, December 2007.
\newblock ISSN 1532-4435.

\bibitem[Feldman et~al.(2009)Feldman, Gopalan, Khot, and Ponnuswami]{Feldman09}
Vitaly Feldman, Parikshit Gopalan, Subhash Khot, and Ashok~Kumar Ponnuswami.
\newblock On agnostic learning of parities, monomials, and halfspaces.
\newblock \emph{SIAM J. Comput.}, 39\penalty0 (2):\penalty0 606–645, jul 2009.
\newblock ISSN 0097-5397.
\newblock \doi{10.1137/070684914}.

\bibitem[Feldman et~al.(2018)Feldman, Perkins, and Vempala]{FPV18}
Vitaly Feldman, Will Perkins, and Santosh Vempala.
\newblock On the complexity of random satisfiability problems with planted solutions.
\newblock \emph{SIAM Journal on Computing}, 47\penalty0 (4):\penalty0 1294--1338, 2018.
\newblock \doi{10.1137/16M1078471}.
\newblock URL \url{https://doi.org/10.1137/16M1078471}.

\bibitem[Goldreich and Levin(1989)]{GoldreichLevin:89}
O.~Goldreich and L.~Levin.
\newblock A hard-core predicate for all one-way functions.
\newblock In \emph{Proceedings of the Twenty-First Annual Symposium on Theory of Computing}, pages 25--32, 1989.

\bibitem[Goldwasser et~al.(2010)Goldwasser, Kalai, Peikert, and Vaikuntanathan]{Goldwasser10}
Shafi Goldwasser, Yael~Tauman Kalai, Chris Peikert, and Vinod Vaikuntanathan.
\newblock Robustness of the learning with errors assumption.
\newblock In \emph{Innovations in Computer Science - {ICS} 2010}, pages 230--240. Tsinghua University Press, 2010.

\bibitem[Golowich et~al.(2024)Golowich, Moitra, and Rohatgi]{golowich2024learning}
Noah Golowich, Ankur Moitra, and Dhruv Rohatgi.
\newblock On learning parities with dependent noise.
\newblock \emph{arXiv preprint arXiv:2404.11325}, 2024.

\bibitem[Greene and Wellner(2017)]{hypergeometric_GW}
Evan Greene and Jon~A Wellner.
\newblock Exponential bounds for the hypergeometric distribution.
\newblock \emph{Bernoulli: official journal of the Bernoulli Society for Mathematical Statistics and Probability}, 23\penalty0 (3):\penalty0 1911, 2017.

\bibitem[Grigorescu et~al.(2011)Grigorescu, Reyzin, and Vempala]{Grig11}
Elena Grigorescu, Lev Reyzin, and Santosh Vempala.
\newblock On noise-tolerant learning of sparse parities and related problems.
\newblock In \emph{Proceedings of the 22nd International Conference on Algorithmic Learning Theory}, ALT'11, page 413–424, 2011.
\newblock ISBN 9783642244117.

\bibitem[Guruswami et~al.(2022)Guruswami, Kothari, and Manohar]{GKM}
Venkatesan Guruswami, Pravesh~K. Kothari, and Peter Manohar.
\newblock Algorithms and certificates for boolean {CSP} refutation: Smoothed is no harder than random.
\newblock \emph{SIAM Journal on Computing}, 0\penalty0 (0):\penalty0 STOC22--282--STOC22--337, 2022.
\newblock \doi{10.1137/22M1537771}.

\bibitem[Hassanieh et~al.(2012)Hassanieh, Indyk, Katabi, and Price]{HIKP12}
Haitham Hassanieh, Piotr Indyk, Dina Katabi, and Eric Price.
\newblock Nearly optimal sparse fourier transform.
\newblock In \emph{Proceedings of the 44th Symposium on Theory of Computing Conference, {STOC}}, pages 563--578. {ACM}, 2012.
\newblock \doi{10.1145/2213977.2214029}.
\newblock URL \url{https://doi.org/10.1145/2213977.2214029}.

\bibitem[Hsieh et~al.(2023)Hsieh, Kothari, and Mohanty]{HKM_hyper_moore}
Jun{-}Ting Hsieh, Pravesh~K. Kothari, and Sidhanth Mohanty.
\newblock A simple and sharper proof of the hypergraph moore bound.
\newblock In \emph{Proceedings of the 2023 {ACM-SIAM} Symposium on Discrete Algorithms, {SODA} 2023}, pages 2324--2344. {SIAM}, 2023.
\newblock \doi{10.1137/1.9781611977554.CH89}.

\bibitem[Jain et~al.(2021)Jain, Lin, and Sahai]{JLS21}
Aayush Jain, Huijia Lin, and Amit Sahai.
\newblock Indistinguishability obfuscation from well-founded assumptions.
\newblock In \emph{Proceedings of the 53rd Annual ACM SIGACT Symposium on Theory of Computing}, STOC 2021, page 60–73. ACM, 2021.
\newblock ISBN 9781450380539.
\newblock \doi{10.1145/3406325.3451093}.

\bibitem[Jain et~al.(2022)Jain, Lin, and Sahai]{JLS22}
Aayush Jain, Huijia Lin, and Amit Sahai.
\newblock Indistinguishability obfuscation from {LPN} over {$F_p$}, {DLIN}, and {PRG}s in {$NC_0$}.
\newblock In \emph{Advances in Cryptology – EUROCRYPT 2022: 41st Annual International Conference on the Theory and Applications of Cryptographic Techniques}, page 670–699. Springer-Verlag, 2022.
\newblock ISBN 978-3-031-06943-7.
\newblock \doi{10.1007/978-3-031-06944-4_23}.

\bibitem[Karppa et~al.(2018)Karppa, Kaski, and Kohonen]{KKK18}
Matti Karppa, Petteri Kaski, and Jukka Kohonen.
\newblock A faster subquadratic algorithm for finding outlier correlations.
\newblock \emph{ACM Trans. Algorithms}, 14\penalty0 (3), jun 2018.
\newblock ISSN 1549-6325.
\newblock \doi{10.1145/3174804}.

\bibitem[Kiltz et~al.(2014)Kiltz, Masny, and Pietrzak]{KMP14_PKC}
Eike Kiltz, Daniel Masny, and Krzysztof Pietrzak.
\newblock Simple chosen-ciphertext security from low-noise {LPN}.
\newblock In \emph{Proceedings of the 17th International Conference on Public-Key Cryptography --- PKC 2014 - Volume 8383}, page 1–18. Springer-Verlag, 2014.
\newblock ISBN 9783642546303.
\newblock \doi{10.1007/978-3-642-54631-0_1}.

\bibitem[Kol et~al.(2017)Kol, Raz, and Tal]{KRT17}
Gillat Kol, Ran Raz, and Avishay Tal.
\newblock Time-space hardness of learning sparse parities.
\newblock In \emph{Proceedings of the 49th Annual ACM SIGACT Symposium on Theory of Computing}, STOC 2017, page 1067–1080. ACM, 2017.
\newblock ISBN 9781450345286.
\newblock \doi{10.1145/3055399.3055430}.

\bibitem[Lyubashevsky(2005)]{Lyubashevsky05}
Vadim Lyubashevsky.
\newblock The parity problem in the presence of noise, decoding random linear codes, and the subset sum problem.
\newblock In \emph{Approximation, Randomization and Combinatorial Optimization. Algorithms and Techniques}, pages 378--389. Springer Berlin Heidelberg, 2005.

\bibitem[Pietrzak(2012)]{Pie12}
Krzysztof Pietrzak.
\newblock Cryptography from learning parity with noise.
\newblock In \emph{SOFSEM 2012: Theory and Practice of Computer Science}, pages 99--114. Springer Berlin Heidelberg, 2012.
\newblock ISBN 978-3-642-27660-6.

\bibitem[Ragavan et~al.(2024)Ragavan, Vafa, and Vaikuntanathan]{RVV24}
Seyoon Ragavan, Neekon Vafa, and Vinod Vaikuntanathan.
\newblock Indistinguishability obfuscation from bilinear maps and lpn variants.
\newblock In \emph{Theory of Cryptography Conference}, pages 3--36. Springer, 2024.

\bibitem[Raghavendra et~al.(2017)Raghavendra, Rao, and Schramm]{RRS17}
Prasad Raghavendra, Satish Rao, and Tselil Schramm.
\newblock Strongly refuting random {CSP}s below the spectral threshold.
\newblock In \emph{Proceedings of the 49th Annual ACM SIGACT Symposium on Theory of Computing}, STOC 2017, page 121–131. ACM, 2017.
\newblock ISBN 9781450345286.
\newblock \doi{10.1145/3055399.3055417}.

\bibitem[Raghuraman et~al.(2023)Raghuraman, Rindal, and Tanguy]{RRT23}
Srinivasan Raghuraman, Peter Rindal, and Titouan Tanguy.
\newblock Expand-convolute codes for pseudorandom correlation generators from lpn.
\newblock In \emph{Advances in Cryptology -- CRYPTO 2023}, pages 602--632. Springer Nature Switzerland, 2023.
\newblock ISBN 978-3-031-38551-3.

\bibitem[Raz(2018)]{Raz18}
Ran Raz.
\newblock Fast learning requires good memory: A time-space lower bound for parity learning.
\newblock \emph{J. ACM}, 66\penalty0 (1), December 2018.
\newblock ISSN 0004-5411.
\newblock \doi{10.1145/3186563}.
\newblock URL \url{https://doi.org/10.1145/3186563}.

\bibitem[Regev(2009)]{Regev09}
Oded Regev.
\newblock On lattices, learning with errors, random linear codes, and cryptography.
\newblock \emph{J. ACM}, 56\penalty0 (6), sep 2009.
\newblock ISSN 0004-5411.
\newblock \doi{10.1145/1568318.1568324}.

\bibitem[Valiant(2015)]{valiant2012finding}
Gregory Valiant.
\newblock Finding correlations in subquadratic time, with applications to learning parities and the closest pair problem.
\newblock \emph{J. ACM}, 62\penalty0 (2), may 2015.
\newblock ISSN 0004-5411.
\newblock \doi{10.1145/2728167}.

\bibitem[Williams et~al.(2024)Williams, Xu, Xu, and Zhou]{WXXZ_matrix_mutl24}
Virginia~Vassilevska Williams, Yinzhan Xu, Zixuan Xu, and Renfei Zhou.
\newblock New bounds for matrix multiplication: from alpha to omega.
\newblock In \emph{Proceedings of the 2024 {ACM-SIAM} Symposium on Discrete Algorithms, {SODA} 2024}, pages 3792--3835. {SIAM}, 2024.
\newblock \doi{10.1137/1.9781611977912.134}.

\bibitem[Yan et~al.(2021)Yan, Yu, Liu, Zhao, and Zhang]{Yu21}
Di~Yan, Yu~Yu, Hanlin Liu, Shuoyao Zhao, and Jiang Zhang.
\newblock An improved algorithm for learning sparse parities in the presence of noise.
\newblock \emph{Theoretical Computer Science}, 873:\penalty0 76--86, 2021.
\newblock ISSN 0304-3975.
\newblock \doi{https://doi.org/10.1016/j.tcs.2021.04.026}.

\bibitem[Yu et~al.(2019)Yu, Zhang, Weng, Guo, and Li]{Yu19}
Yu~Yu, Jiang Zhang, Jian Weng, Chun Guo, and Xiangxue Li.
\newblock Collision resistant hashing from sub-exponential learning parity with noise.
\newblock In \emph{Advances in Cryptology – ASIACRYPT 2019}, page 3–24. Springer-Verlag, 2019.
\newblock ISBN 978-3-030-34620-1.
\newblock \doi{10.1007/978-3-030-34621-8_1}.

\end{thebibliography}



\clearpage


\begin{subappendices}
    \renewcommand{\thesection}{\Alph{section}}

\section{Probabilistic Tools}\label{app:preli}
We state the classical Chernoff bound and Chebyshev's bound.
\begin{theorem}[Chernoff bound]\label{THM:Chernoff}
    Let $X_1,X_2,...,X_n$ be independent random variables. Let $X=\sum_{i=1}^n X_i$, we have the following two kinds of inequalities:
    \begin{itemize}
        \item If $X_i\in [l,r]$, then for all $a>0$, $\Pr[X\geq\E[X]+a]\leq e^{-\frac{2a^2}{n(r-l)^2}}$ and $\Pr[X\leq\E[X]-a]\leq e^{-\frac{2a^2}{n(r-l)^2}}$.

        \item If $X_i\in [0,1]$, then for any $0<\epsilon$, $\Pr[X\geq(1+\epsilon)\E[X]]\leq e^{-\frac{\E[X]\epsilon^2}{2+\epsilon}}$, and for any $0<\epsilon<1$ $\Pr[X\leq (1-\epsilon)\E[X]]\leq e^{-\frac{\E[X]\epsilon^2}{2}}$.
    \end{itemize}
\end{theorem}

\begin{theorem}[Chebyshev's inequality]
\label{thm:chebyshev}
$\Pr \big[ |X-\E[X]|\geq k\sqrt{\Var(X)} \big] \leq \frac{1}{k^2}$ for any $X$.
\end{theorem}

In this work, we always use the total variation distance to compare two distributions.
\begin{definition}[Total variation distance]
    For two distributions $W$ and $V$ defined on the same support set $E$, the total variation distance between them is defined as 
    $$\lVert W-V \rVert_{TV}=\sup_{S\subset E} \left| \Pr_W[S]-\Pr_V[S] \right|,$$
    where $\Pr_W[S]$ denotes the probability of event $S$ under distribution $W$.
\end{definition}

For convenience, we say two distributions $W$ and $V$ are $\epsilon$-close only if 
$\|W-V\|_{TV}\leq \epsilon.$ Moreover, two random variables (or vectors, matrices) are $\epsilon$-close, if their distributions are $\epsilon$-close.
When $E$ is a finite or countable set, we have $$\lVert W-V \rVert_{TV}=\frac{1}{2}\sum_{x\in E}|W(x)-V(x)|=\sum_{x,W(x)\geq V(x)}(W(x)-V(x)).$$

We use $W^{\otimes k}$ to denote the distribution of $(X_1,X_2,...,X_k)$, where $X_i$ for $i\in [k]$ are i.i.d. random variables (or vectors) following the distribution $W$. The following two properties of total variation distance will be useful in our analysis.


\begin{proposition}
\label{ca::tvd}
\begin{enumerate}
    \item     Let $X,Y$ be two random variables( or vectors) with the same support set $E$ under distributions $W$ and $V$ separately. For any function $f: E\rightarrow \{0,1\},$
    $$\left| \Pr[f(X)=1]-\Pr[f(Y)=1] \right|\leq \lVert W-V \rVert_{TV}. $$
    \item     Given two distribution $W$ and $V$ with the same support, then $\|W^{\otimes k}-V^{\otimes k}\|_{TV} \le k \cdot \|W-V\|_{TV}$.
\end{enumerate}
\end{proposition}

By Proposition~\ref{ca::tvd} we have the following result.
\begin{corollary}
\label{cor::stat_dist}
    Let $\mX,\mY\in \{\pm 1\}^{k\times n}$ be two random matrices with each row sampled independently from the distribution $W$ and $V$ supported on $E=\{-1,1\}^n$ separately. For any function $f:\{\pm 1\}^{k\times n}\rightarrow \{0,1\}$, we have
    $$\left| \Pr[f(\mX)=1]-\Pr[f(\mY)=1] \right|\leq k \cdot \lVert W-V \rVert_{TV}. $$
\end{corollary}

\section{Omitted proofs}\label{app:proof}
We supplement two proofs for the learning algorithm of LSPN in Appendix~\ref{sec:T&S} and Appendix~\ref{sec:proof_of_less_main}. Then we finish the analysis of the learning algorithm of sparse LPN in Appendix~\ref{sec:proof_sparse_LPN}.

\subsection{Proof of Proposition~\ref{cla:T&S}}\label{sec:T&S}
We finish the proof of Proposition~\ref{cla:T&S} in this section. Recall $q=|S|$ and $q'=q+25$.

Since $\mX(T,S)$ is a random matrix in $\mathbf{F}_2^{|T| \times |S|}$, we show that $|T|$ random vectors in $\mathbf{F}_2^{S}$ is of rank $q$ with high probability.


We consider the following random process on random vectors. Let us generate a sequence of random vectors $\vec{v}_1,\vec{v}_2,\ldots$ in $\mathbf{F}_2^S$ independently and $z_i$ denote the expectation of the first time $t$ such that the first $t$ vectors $\vec{v}_1,\ldots,\vec{v}_t$ have rank $i$. Then we define $x_i:=z_i-z_{i-1}$ to denote the expected number of random vectors to increase the rank from $i-1$ to $i$. It is clear that $\{x_i\}_{i\in [q]}$ are independent and $x_i$ follows a geometric distribution with parameter $\frac{2^{q}-2^{i-1}}{2^q}$.
Thus $\E[x_i]=\frac{2^q}{2^q-2^{i-1}}$ and $\Var(x_i) = \frac{2^q \cdot 2^{i-1}}{(2^q-2^{i-1})^2} \leq \frac{2^q \cdot 2^{i-1}}{(2^q - 2^{q-1})^2} \le \frac{2}{2^{q-i}}.$ 

We have
\begin{align*}
	\E\left[\sum_{i=1}^q x_i\right] & =\sum_{i=1}^q\frac{2^q}{2^q-2^{i-1}}
	=\sum_{i=1}^q\frac{1}{1-2^{i-1-q}}
	\leq \sum_{i=1}^q(1+2^{i-q})
	\leq q+2,
	\\        \Var\left(\sum_{i=1}^q x_i\right) & = \sum_{i=1}^q \Var(x_i)
	\le \sum_{i=1}^q \frac{2}{2^{q-i}}
	\leq 4.
\end{align*}
By the Chebyshev's inequality (Theorem~\ref{thm:chebyshev}), we know
$$\Pr\left[\left|\sum_{i=1}^q x_i-\E\left[\sum_{i=1}^q x_i\right]\right|\geq 20\right]\leq \frac{\Var(\sum_{i=1}^q x_i)}{20^2}\leq 0.01.$$
Which means with probability at least 0.99, $\sum_{i=1}^q x_i\leq 20+\E[\sum_{i=1}^q x_i]\leq q+25.$

\subsection{Proof of Proposition~\ref{clm:SUBSETLEARN}}\label{sec:proof_of_less_main}
We finish the proof of Proposition~\ref{clm:SUBSETLEARN} in this section. Let $\mathbf{T_c}=\{i \in [m_1] \mid \by(i)=\langle \mX(i,),\secret \rangle\}$ denote the subset of correct samples. Similarly, $\mathbf{T_i}=\{i\in [m_1] \mid \by(i) \neq \langle \mX(i,),\secret \rangle \}$ denote the subset of incorrect samples. By the definition of $\mathbf{T_c}$, Line 10 of Algorithm~\ref{alg::LSPN} finds $\mathbf{ans}=\secret$ when $rank(\mX(T,S))=|S|$ and $T \subset \mathbf{T_c}$.

Then consider the following two random events 
\begin{align*}
    A_1 & =\{\text{Line 8 samples at least one }T\subset \mathbf{T_c} \} \\ \text{ and }
    A_2 & =\{\text{The specific }T\text{ sampled in  }A_1\text{ satisfies }rank(X(T,S))=|S| \}.
\end{align*}
Since the noise is added independently to labels,
\begin{align*}
	\Pr[\text{Line 12 returns }\mathbf{ans}] & \ge\Pr[A_1A_2]                                          \\
	                               & =\Pr[A_1]\times
	\Pr[A_2\mid A_1]                                                                         \\
	                               & =\Pr[A_1]\times
	\Pr[A_2]                                                                                 \\
	                               & \ge \Pr[A_1]\times 0.99 \tag{by Proposition~\ref{cla:T&S}}.
\end{align*}

Next, we bound $\Pr[A_1]$. The following argument shows that there are enough noiseless samples in $\mX$. We still use $q':=q+25=\frac{k}{\eta}+25$ such that $m_1 \ge 100q'$. We bound $\mathbf{T_i}$ by the Chernoff bound (Property 2 of Theorem~\ref{THM:Chernoff}):

$$\Pr\left[\abs{\mathbf{T_i}}\ge 1.01 \eta m_1 \right]\le e^{- \frac{0.01^2\eta m_1}{2+0.01}} = e^{- \Omega(k)} \le 0.01.$$

Since $|\mathbf{T_c}|+|\mathbf{T_i}|=m_1$,
$|\mathbf{T_c}|\geq (1-1.01\eta)m_1$ with probability at least $0.99$ from the above line. Assuming this event holds, for each random subset $T \sim {[m_1] \choose q'}$, $\Pr[T \subseteq \mathbf{T_c}] \ge {\binom{(1-1.01\eta)m_1}{q'}}/{\binom{m_1}{q'}}$.
We bound this probability as follows.
\begin{align*}
	\left. \binom{m_1}{q'}\middle/\binom{(1-1.01\eta) m_1}{q'} \right.
	 & \leq \frac{m_1 \cdot (m_1-1)\cdots (m_1-q'+1)}{((1-1.01\eta)m_1)\cdot ((1-1.01\eta)m_1-1)\cdots ((1-1.01\eta)m_1-q'+1)}\cr
	 & \le \left(\frac{m_1-q'}{(1-1.01\eta)m_1-q'}\right)^{q'}                \\
	 & \le \left(1+\frac{1.01\eta \cdot m_1}{(1-1.01\eta)m_1-q'}\right)^{q'} \tag{recall $m_1=100 q'$}                           \\
	 & \le \left(1+\frac{1.01\eta \cdot m_1}{(0.99-1.01\eta)m_1}\right)^{q'}\tag{$\eta <0.05 $}                                   \\
	 & \le e^{1.08 \eta \cdot q'} \textit{ for } q'=q+25.
\end{align*}
Since the algorithm samples $T$ for $O(1) \cdot e^{1.08 \cdot \eta q}$ times, with probability $0.99$, at least one $T$ will be in $\mathbf{T_c}$. Thus we know $\Pr[\text{Line 12 returns }\mathbf{ans}] \ge 0.99^2$ when $|\mathbf{T_c}|\geq (1-1.01\eta)m_1$. From all discussion above, \textsc{LearnInSubset($S,\mX,\by$)} finds $\secret$ with probability at least 0.97.

\subsection{Proof of Theorem~\ref{thm:sparse_LPN}}\label{sec:proof_sparse_LPN}
    To apply Proposition~\ref{clm:decode_subset} to $\secret(I_1),\ldots,\secret(I_t)$, we only need to show $m_j \ge C \cdot q \cdot \max\{1,\frac{\log q}{k}, \frac{\eta q \log q}{k^2}\}$ for every $j$. Let us set $m=2C \cdot \max\{1, \frac{\log q}{k}, \frac{\eta q \cdot \log q}{k^2}\} \cdot q (en/q)^k$ where $q=n^{\frac{1+\delta}{2}}$ and $C$ is defined in Line 9 of function~\textsc{PartialLearn}.
    
    Now let us consider the expectation of $m_j$. For each $j \in [t]$, the probability that a random $k$-sparse vector $\bx$ has its $k$ non-zero entries in the fixed interval $I_j$ is ${q \choose k}/{n \choose k}$ (since $|I_j|=q$). This is at least
    \[
        \frac{(q/k)^k}{(en/k)^k} = (\frac{q}{en})^k
    \]    
    Hence $\E[m_j] \ge m \cdot (\frac{q}{en})^k \ge 2C \cdot \max\{1, \frac{\log q}{k}, \frac{\eta q \cdot \log q}{k^2}\} \cdot q$. By the Chernoff bound (part 2 of Theorem~\ref{THM:Chernoff}),
    \[
    \Pr\left[|m_j-\E[m_j]| \ge \frac{1}{2} \E[m_j]\right] \le 2 e^{-\Omega(\E[m_j])} \le 1/n^3.
    \]
    Given $m_j \in (\frac{1}{2},\frac{3}{2})\E[m_j]$, Proposition~\ref{clm:decode_subset} returns $\secret(I_j)$ with probability $1-1/n^2$. Hence by a union bound over all $j \in [t]$, \textsc{Main} finds $\secret$ with probability at least $1-1/n$.

    Its running time is $O(m \cdot t) + t \cdot \textbf{TIME}(\textsc{PartialLearn})=n^{O(1)} \cdot \big( e^{O(\eta \cdot n^{\frac{1+\delta}{2}} \cdot \max\{1, \frac{\log n}{k}\})} + m \big)$ by Proposition~\ref{clm:decode_subset} with $|I|=q$.

    Finally we plug $q=n^{\frac{1+\delta}{2}}$ into the sample complexity $m$: 
    \begin{align*}
        m & =2C \cdot \max\{1, \frac{\log q}{k}, \frac{\eta q \cdot \log q}{k^2}\} \cdot q (en/q)^k \\
        & = 2C \cdot \max\{1, \frac{\log n}{k}, \frac{\eta n^{\frac{1+\delta}{2}} \cdot \log n}{k^2}\} \cdot en (en/n^{\frac{1+\delta}{2}})^{k-1} \\
        &= 2eC  \max\{1, \frac{\log n}{k}, \frac{\eta n^{\frac{1+\delta}{2}} \cdot \log n}{k^2}\} \cdot n \cdot     O(n^{\frac{1-\delta}{2}})^{k-1} 
    \end{align*}


\subsection{Proof of Proposition~\ref{clm:decode_subset}}\label{sec:proof_decode_subset}
Recall that $q=|I|=n^{\frac{1+\delta}{2}}$ and $m'=100 \cdot c_1 \cdot q \cdot \max\{1,\frac{\log q}{k}\}$ such that $d=m'/100$ random samples satisfy Proposition~\ref{lem:ksparse_span} with probability $1-1/q^5$.

Since each $\bx_i$ in $\mX$ is a random $k$-sparse vector with $supp(\bx_i) \subset I$, this is equivalent to $supp(\bx_i) \sim {I \choose k}$. Hence we consider each $\bx_i$ as a $k$-sparse random vector in $I$.

Consider the first $m'$ random samples in $(\mX,\by)$. Let $\mathbf{T_c} \subset [m']$ denote the set of samples with correct labels, i.e., $\{i \in [m']: \langle \bx_i,\secret \rangle =y_i\}$. Similarly, let $\mathbf{T_i} \subset [m']$ denote the set of incorrect samples. Since 
\[
\E_{\mX,\by}[|\mathbf{T_i}|]=\eta \cdot m',
\]
we give a concentration bound on $|\mathbf{T_i}|$ based on $\eta \cdot m'$. Then we consider the probability that when $T \sim {[m'] \choose m'/100}$, $T \subseteq \mathbf{T_c}$. That is exactly ${\mathbf{|T_c|} \choose m'/100}/{m' \choose m'/100}$.

\begin{enumerate}
    \item If $\eta \cdot m' \ge 15\log q$, then with probability $1-q^{-5}$, $|\mathbf{T_i}| \le 2\eta \cdot m'$. This follows from part 2 of Theorem~\ref{THM:Chernoff} with $\epsilon=1$.
    \[
    \Pr[|\mathbf{T_i}| > 2 \eta \cdot m'] \le e^{-\frac{\eta m' \cdot 1^2}{2+1}} = e^{-\frac{\eta m'}{3}}\le 1/q^5.
    \]
    Then we simplify
    \begin{align*}
        {|\mathbf{T_c}| \choose m'/100}/{m' \choose m'/100} & \ge {(1-2\eta)m' \choose m'/100}/{m' \choose m'/100} \\ 
        & = \prod_{i=0}^{m'/100-1} \left( \frac{(1-2\eta)m' - i}{m' - i} \right) \\
        & \ge \left( \frac{(1-2\eta)m' - m'/100}{m' - m'/100} \right)^{m'/100} = \left( \frac{99/100-2\eta}{99/100} \right)^{m'/100} \ge e^{-O(\frac{2m}{99} \cdot 2\eta)}.
    \end{align*}    
    \item Otherwise $\eta \cdot m' < 15 \log q$. Then with probability $1-q^{-5}$, $|\mathbf{T_i}| \le \eta m' + 15\log q$. To apply part~2 of Theorem~\ref{THM:Chernoff}, we choose $\epsilon \E[|\mathbf{T_i}|]=15 \log q$ such that $\epsilon \ge 1$:
    \[
    \Pr[|\mathbf{T_i}| > \eta \cdot m' + 15 \log q] \le e^{-\epsilon\E[X] \cdot \frac{\epsilon}{2+\epsilon}} \le e^{-15 \log q/3}\le 1/q^5.
    \]
    Similar to the above calculation, we simplify
    \begin{align*}
        {|\mathbf{T_c}| \choose m'/100}/{m' \choose m'/100} & \ge {(1-\eta)m'-15\log q \choose m'/100}/{m' \choose m/100} \\
        & \ge \left( \frac{m'-30\log q - m'/100}{m' - m'/100} \right)^{m'/100} \\
        & \ge \left( 1 - \frac{30 \log q}{99m'/100}\right)^{m'/100}\\
        & \ge e^{-\log q} \ge 1/q.
    \end{align*}
\end{enumerate}

    Since $L:=10 q^2 + 10 \log q \cdot e^{\frac{200c_1}{99} \cdot \eta q \max\{1,\frac{\log q}{k}\}}$, with probability $1-1/q^5$, there is at least one $T \subset \mathbf{T_c}$ after sampling $L$ times. We fix such a $T$ in this analysis.

    By Proposition~\ref{lem:ksparse_span}, $rank(X(T,I))=|I|$ with probability $1-1/q^5$. Moreover, this is independent with $T \subset \mathbf{T_c}$. When $rank(X(T,I))=|I|$ and $T \subset \mathbf{T_c}$, the Gaussian elimination finds $\mathbf{ans}=\secret(I)$. Moreover, since $m \ge C \cdot \max\{1, \frac{\log |I|}{k}, \frac{\eta |I| \cdot \log |I|}{k^2}\} \cdot |I|$ is sufficiently large, $\mathbf{ans}=\secret(I)$ is the only vector satisfying $s \gets \sum_{i=m'+1}^m \mathbf{1}\big( \langle \bx_i, \mathbf{ans} \rangle \neq y_i \big)$ less than $(m-m')\cdot (\eta + c_0 \frac{k}{q})$ by Corollary~\ref{cor:verification}.

    From all discussion above, with probability $1-O(1)/q^5$, function~\textsc{PartialLearn} returns $\secret(I)$. Its running time is $L \cdot O(m + (m')^3)$.


\subsection{Proof of Proposition~\ref{lem:ksparse_span}}\label{sec:proof_sparse_span}
We finish the proof of Proposition~\ref{lem:ksparse_span} in this section. We consider the probability that $t$ random $k$-sparse vectors are of rank $\le n-1$. We first bound the probability that $t$ random vectors are in a subspace $U$ of dimension $n-1$ then apply the union bound over all linear subspaces of dimension $n-1$.

Given a linear subspace $U$ of dimension $(n-1)$, there always exists a dual vector $\bv$ such that $U=\{\bu: \langle \bu,\bv \rangle =0 \}$. Hence
\[
\Pr[\text{$t$ random $k$-sparse vectors } \in  U]=\Pr[\text{$t$ random $k$-sparse vectors } \bot \bv]=\Pr_{\bx \sim {n \choose k}}[\langle \bx,\bv\rangle=0]^t
\]
By symmetry, this probability only depends on the Hamming weight of $\bv$, say $\ell$. We simplify this probability
\[
\Pr_{\bx \sim {n \choose k}}[\langle \bx,\bv\rangle=0]=\sum_{i=0,2,4,\ldots} \frac{{\ell \choose i}\cdot{n-\ell \choose k-i}}{{n \choose k}}
\]
into three cases:
\begin{enumerate}
    \item $\ell < 8n/k$: $\Pr_\bx[\langle \bx, \bv\rangle=0] \le 1 - \Omega(k\ell/n)$. In fact, $\Pr[\langle \bx, \bv \rangle =1]=\Theta(k \ell/n)$. The upper bound follows from a union bound. We consider the following lower bound on $\Pr[\langle \bx, \bv \rangle =1]$:
    \[
    \frac{{\ell \choose 1}{n-\ell \choose k-1}}{{n \choose k}} = \ell \cdot \frac{(n-\ell)\cdots(n-\ell-(k-1)+1)/(k-1)!}{n \cdots (n-k+1)/k!} = \frac{\ell \cdot k}{n} \cdot (\frac{n-\ell}{n-1})\cdots(\frac{n-\ell-k+2}{n-k+1}).
    \]
    We lower bound these terms $(\frac{n-\ell}{n-1})\cdots(\frac{n-\ell-k+2}{n-k+1})$ as
    \[
    (\frac{n-\ell-k+2}{n-k+1})^{k-1}=(1-\frac{\ell-1}{n-k+1})^{k-1} \ge e^{-O(\frac{(\ell-1)(k-1)}{n-k+1})} \ge e^{-O(1)}.
    \]
    by our assumption $\ell \cdot k<8 n$. 
    \item $\ell \in [8n/k,n-8n/k]$: $\Pr_\bx[\langle \bx, \bv\rangle=0] \in [0.1,0.9]$. 
    Let $Z$ denote the inner product between $\bx$ and $\bv$ in $\mathbb{Z}$(not $\mathbf{F}_2$). We know $\Pr[Z=i]=\frac{{\ell \choose i}\cdot{n-\ell \choose k-i}}{{n \choose k}}$ is a hypergeometric distribution whose expectation is $\frac{k \cdot \ell}{n}$ and variance is $\frac{k \cdot \ell}{n} \cdot \frac{(n-k)(n-\ell)}{n(n-1)}$. Let us consider $\ell \le n/2$. Since $k=o(n)$, $\Var[Z] \le \frac{k \ell}{n}$. By the Chebyshev's inequality,
    \[
    \Pr_Z \left[ |Z-\E Z| \ge 2 \sqrt{\frac{k \ell}{n}} \right] \le 1/4.
    \]
    In another word, 
    \[
    \Pr_Z \left[ Z \in [\frac{k \ell}{n}- 2 \sqrt{\frac{k \ell}{n}},\frac{k \ell}{n} + 2 \sqrt{\frac{k \ell}{n}}] \right] \ge 3/4.\]
    Since $k \ell \ge 8n$, this further implies
    \begin{equation}\label{eq:hyper_concen}
        \Pr_Z\left[Z \in \left[ (1-2/\sqrt{8})\frac{k \ell}{n},(1+2/\sqrt{8})\frac{k \ell}{n}\right] \right] \ge 3/4.
    \end{equation}
    
    Next we bound the probability $\Pr_Z[Z \text{ is odd}]$ and $\Pr_Z[Z \text{ is even}]$. Let us compare the probability $\Pr[Z=i]$ and $\Pr[Z=i+1]$ for $i \in [(1-2/\sqrt{8})\frac{k \ell}{n},(1+2/\sqrt{8})\frac{k \ell}{n}]$:
    \begin{align*}
        \Pr[Z=i]/\Pr[Z=i+1] & = \frac{{\ell \choose i}\cdot {n-\ell \choose k-i}}{{\ell \choose i+1}{n-\ell \choose k-i-1}}\\
        & = \frac{(i+1)!(\ell-i-1)!}{i! (\ell-i)!} \cdot \frac{(k-i-1)!(n-\ell-k+i+1)!}{(k-i)!(n-\ell-k+i)!} \\
        & = \frac{(i+1)(n-\ell-k+i+1)}{(\ell-1)(k-i)}.
    \end{align*}
    Since $\ell \le n/2$ implies $i \in [0.14k,0.86k]$ and $n-\ell-k+i+1 \in [0.49n,n]$, we simplify 
    \[
    \frac{(i+1)(n-\ell-k+i+1)}{(\ell-1)(k-i)} \in \left[ \frac{(i+1)\cdot 0.49n}{(\ell-1)\cdot 0.86k}, \frac{(i+1)\cdot n}{(\ell-1)\cdot 0.14k} \right].
    \]
    Since $\frac{i \cdot n}{k \ell} \in [1-1/\sqrt{2},1+1/\sqrt{2}]$, this implies that \[
    \Pr[Z=i]/\Pr[Z=i+1] \in [0.3\cdot \frac{0.49}{0.86},1.71\cdot \frac{1}{0.14}+\frac{1}{8}].
    \]
    This indicates $\Pr[Z=i]/\Pr[Z=i+1] \ge 0.17$.

    Combining this with \eqref{eq:hyper_concen}, we have
    \[
    \Pr[Z \text{ is odd}]\ge \Pr\left[Z \text{ is odd and } Z \in [(1-2/\sqrt{8})\frac{k \ell}{n},(1+2/\sqrt{8})\frac{k \ell}{n}]\right] \ge \frac{0.17}{1+0.17} \cdot \frac{3}{4} \ge 0.1.
    \]
    Similarly, we have $\Pr[Z \text{ is even}] \ge 0.1$. While we assume $\ell<n/2$ in this proof, the same argument holds for $\ell \in [n/2, n-8n/k]$.
    
    \item $\ell > n-8n/k$:
    $\Pr_\bx \left [\langle \bx, \bv\rangle=0 \right] \le 1-e^{-9}$. Since $k$ is odd, we lower bound $\Pr_\bx[\langle \bx, \bv \rangle =1]$ by
    \[
    \frac{{\ell \choose k}}{{n \choose k}} = \frac{\ell \cdots (\ell-k+1)}{n \cdots (n-k+1)} \ge (\frac{\ell-k+1}{n-k+1})^k \ge (1-\frac{n-\ell}{n-k+1})^k \ge e^{-\frac{n-\ell}{n-k+1} \cdot k} \ge e^{-9}.
    \]
\end{enumerate}
Finally, we apply a union bound over all linear subspaces of dimension $(n-1)$.
\begin{align*}
&\Pr[dimension(\text{$t$ random $k$-sparse vectors})\le n-1] \\
=& \sum_{\bv \in \mathbf{F}_2^n \setminus \vec{0}} \Pr[\text{$t$ random $k$-sparse vectors } \bot~\bv]\\
=& \sum_{\bv} \Pr_{\bx \sim {n \choose k}}[\langle \bx,\bv\rangle=0]^t\\
=& \sum_{\bv:|\bv| < 8n/k} (1-\Omega(\frac{k \ell}{n}))^t + \sum_{\bv:|\bv| \in [8n/k,n-8n/k]} (1-0.1)^t + \sum_{\bv:|\bv| >n-8n/k} (1-e^{-9})^t \\
& = \sum_{\ell<8n/k} {n \choose \ell} e^{-\Omega(\frac{k \ell \cdot t}{n})} + 2^n \cdot 0.9^t + \left( {n \choose n-8n/k} +\cdots + 1 \right) \cdot (1-e^{-9})^t.
\end{align*}
When $t=O(n \cdot \max\{1,\frac{\log n}{k}\})$, the term in the first summation ${n \choose \ell} e^{-\Omega(\frac{k \ell \cdot t}{n})}<n^{\ell} e^{-\Omega(\frac{k \ell \cdot t}{n})}<n^{-10\ell}$. Hence $\Pr[dimension(\text{$t$ random $k$-sparse vectors})\le n-1] \le n^{-5}$ .

\begin{remark}
    We have the following generalization for any even number $k$: $t=O(n) \cdot \max\{1,\frac{\log n}{k}\}$ random $k$-sparse vectors in $\mathbf{F}_2^n$ will span the $(n-1)$-dimensional space $\mathbf{F}_2^{n}$ orthogonal to $\vec{1}$ with probability $1-1/n^5$.

    Moreover, the extra term $\max\{1, \frac{\log n}{k}\}$ is necessary from the coupon collector argument: we need $k \cdot t=\Omega(n \log n)$ to hit all entries.
\end{remark}

\subsection{Proof of Corollary~\ref{cor:verification}}\label{sec:proof_verification}
The proof of Corollary~\ref{cor:verification} is in the same outline as the proof of Proposition~\ref{lem:ksparse_span}. The only difference is that we use the Chernoff bound to handle the effect of the noise here.

Given $m$ random samples $(\bx_1,y_1),\ldots,(\bx_m,y_m)$, we know 
\[
\E\left[ \sum_{i=1}^m \mathbf{1}(y_i \neq \langle \bx_i, \secret \rangle) \right]=\eta m.
\]
For any hypothesis $\bv \in \mathbf{F}_2^n$, let $m_{\bv}:=\sum_{i=1}^m \mathbf{1}(y_i \neq \langle \bx_i, \bv \rangle)$ denote the number of incorrect labels. We have
\begin{align*}
\E\left[ m_{\bv} \right] & = m \cdot \left(\eta \cdot \Pr_{\bx}[\langle \bx_i, \secret \rangle = \langle \bx_i, \bv \rangle] + (1-\eta)\Pr_{\bx}[\langle \bx_i, \secret \rangle \neq \langle \bx_i, \bv \rangle]\right) \\
& = m \cdot \left(\eta  + (1-2\eta)\Pr_{\bx}[\langle \bx_i, \secret+\bv \rangle \neq 0]\right).
\end{align*}
Since $\Pr_{\bx}[\langle \bx_i, \secret+\bv \rangle \neq 0]$ could be as small as $\Theta(k/n)$ when $|\secret+\bv|=1$, we set a threshold $m(\eta  + c_0 \cdot \frac{k}{n})$ for a sufficiently small constant $c_0$ to distinguish $\secret$ from the rest hypotheses $\bv \in \mathbf{F}_2^n \setminus \{\secret\}$. Now we bound the probability that $m_{\secret}<m(\eta + c_0 \cdot \frac{k}{n})$ and $m_v \ge (\eta + c_0 \cdot \frac{k}{n})$ for any $\bv \neq \secret$.

For any fixed $v$, let us consider the probability that $v$ passes the tester:
\begin{align*}
\Pr\left[ m_{\bv}<m(\eta  + c_0 \cdot \frac{k}{n}) \right] & = \Pr\left[ m_{\bv}< \frac{\eta  + c_0 \cdot \frac{k}{n}}{\eta  + (1-2\eta) \cdot \Pr_{\bx}[\langle \bx_i, \secret+\bv \rangle \neq 0]} \cdot \E[m_{\bv}] \right]\\
& = \Pr\left[ m_{\bv}< \big( 1-\frac{(1-2\eta) \cdot \Pr_{\bx}[\langle \bx_i, \secret+\bv \rangle \neq 0] -c_0 \cdot \frac{k}{n}}{\eta  + (1-2\eta) \cdot \Pr_{\bx}[\langle \bx_i, \secret+\bv \rangle \neq 0]} \big) \cdot \E[m_{\bv}] \right] \tag{apply $\Pr_{\bx}[\langle \bx_i, \secret+\bv \rangle \neq 0]=\Omega(k/n)$}\\
& \le \Pr\left[ m_{\bv}< \big( 1-\frac{\frac{1}{2} \Pr_{\bx}[\langle \bx_i, \secret+\bv \rangle \neq 0]}{\eta  + \Pr_{\bx}[\langle \bx_i, \secret+\bv \rangle \neq 0]} \big) \cdot \E[m_{\bv}] \right].
\end{align*}
Now we upper bound $\frac{\frac{1}{2} \Pr_{\bx}[\langle \bx_i, \secret+\bv \rangle \neq 0]}{\eta  + \Pr_{\bx}[\langle \bx_i, \secret+\bv \rangle \neq 0]}$ as $\min \big\{ \frac{\Pr_{\bx}[\langle {\bx}_i, \secret+\bv \rangle \neq 0]}{4\eta},\frac{1}{4} \big\}$ and simplify the above probability to
\[
\Pr\left[ m_{\bv}< \big( 1-\min \big\{ \frac{\Pr_{\bx}[\langle {\bx}_i, \secret+\bv \rangle \neq 0]}{4\eta},\frac{1}{4} \big\} \big) \cdot \E[m_v] \right].
\]

By the Chernoff bound (part 2 of Proposition~\ref{THM:Chernoff}), this is at most
\[
e^{-\frac{\E[m_{\bv}]\cdot \min\{\frac{\Pr_{\bx}[\langle \bx_i, \secret+\bv \rangle \neq 0]}{4\eta},\frac{1}{4}\}^2}{2}}.
\]
We simplify it as follows. If $\Pr_{\bx}[\langle \bx_i, \secret+\bv \rangle \neq 0] < \eta$, this is at most $e^{-\Omega(m \cdot \Pr_{\bx}[\langle \bx_i, \secret+\bv \rangle \neq 0]^2/ \eta)}$. Otherwise, when $\Pr_{\bx}[\langle \bx_i, \secret+\bv \rangle \neq 0] \ge \eta$, this is at most $e^{-\Omega(m \cdot \Pr_{\bx}[\langle \bx_i, \secret+\bv \rangle \neq 0])}$.

Based on the discussion about $\Pr_{\bx}[\langle \bx_i, \secret+\bv \rangle \neq 0]$ in Section~\ref{sec:proof_sparse_span}, there are three cases depends on $\secret+\bv$. For convenience, let $\ell := |\secret+\bv|$ and $s_1$ be the largest index of $\ell$ such that $\Pr_{\bx}[\langle \bx_i, \secret+\bv \rangle \neq 0]\le \eta$ for any $|\secret+\bv| \le s_1$. Since $\Pr_{\bx}[\langle \bx_i, \secret+\bv \rangle \neq 0]=\Theta(\ell k/n)$ when $\ell k<n$, $s_1 = \Theta(\eta n/k)$.

\begin{enumerate}
    \item $\ell \le s_1:$ $\Pr_{\bx}[\langle \bx_i, \secret+\bv \rangle \neq 0]= \Omega(\frac{\ell \cdot k}{n})$ and $\Pr[m_{\bv}<m(\eta  + c_0 \cdot \frac{k}{n}) ]\le e^{-\Omega(m \cdot \frac{\ell^2 k^2}{n^2 \eta})}$ in this case. Since the number of such $\bv$'s is ${n \choose \ell}\le n^\ell$, we need $m = \Omega(\frac{\eta n^2 \cdot \log n}{k^2})$ such that their union bound is at most $n^{-10}$.
    
    \item $\ell \in (s_1,8n/k)$: $\Pr_{\bx}[\langle \bx_i, \secret+\bv \rangle \neq 0]= \Omega(\frac{\ell \cdot k}{n}) > \eta$ and $\Pr[m_{\bv}<m(\eta  + c_0 \cdot \frac{k}{n})] \le e^{-\Omega(m \cdot \ell k/n)}$. Since the number of such $\bv$'s is ${n \choose \ell}\le n^\ell$, we need $m = \Omega(\frac{n \log n}{k})$ such that their union bound is at most $n^{-10}$.
    
    \item $\ell>8n/k$: $\Pr_{\bx}[\langle \bx_i, \secret+\bv \rangle \neq 0]=\Omega(1)$ and $\Pr[m_{\bv}<m(\eta  + c_0 \cdot \frac{k}{n})] \le e^{-\Omega(m)}$. Since the number of such $\bv$'s is at most $2^n$, we need $m=\Omega(n)$ for the union bound.
\end{enumerate}
Finally, we bound the probability that $m_{\secret} \ge m(\eta  + c_0 \cdot \frac{k}{n})$ by the Chernoff bound (part 2 of Theorem~\ref{THM:Chernoff}) again:
\[
\Pr \left[ m_{\secret} \ge m\eta  (1 + c_0 \cdot \frac{k}{\eta n}) \right] \le e^{-\frac{m\eta \cdot (c_0 \cdot \frac{k}{\eta n})^2}{2+c_0 \cdot \frac{k}{\eta n}}}.
\]
We need $m=\Omega(\max\{\frac{\eta n^2 \log n}{k^2}, \frac{n \log n}{k}\})$ such that this probability is at most $n^{-10}$. From all discussion above, given $m=C \cdot n \cdot \max\{1,\frac{\log n}{k}, \frac{\eta n \log n}{k^2}\}$ random samples for a sufficiently large constant $C$, we could verify $\secret=\bv$ for any $\bv \in \mathbf{F}_2^n$.

\section{LSPN Algorithm against High-Noise}\label{sec:valiant}


In this section, we present our improvements upon Valiant's classical algorithm \citep{valiant2012finding} in the high-noise regime. Our algorithm learns $k$-sparse parities in $(\frac{n}{k})^{\frac{\omega + o(1)}{3} \cdot k}$ time and $\tilde{O}(k)$ samples, which matches the best known algorithm by \cite{KKK18} in terms of the time complexity and sample complexity. Different from the approach in \cite{KKK18} based on finding matchings between parties of size $k/2$ (called the light-bulb algorithm), we revisit Valiant's framework whose first step is to generate biased examples and second step is to figure out which subsets of $[n]$ are in $\secret$.


Now we state the main result of this section as following. While this does not improve the state-of-the-art \citep{KKK18}, it provides more tools to analyze the more \emph{flexible} framework by Valiant \citep[see][]{valiant2012finding, Yu21, DSGKS21}. At the same time, we believe it is of great interests to investigate the potential of Valiant's framework on LSPN. In particular, if there exists efficient approaches to generate larger-biased examples for the 1st step of Valiant's framework, one could improve the time complexity of learning sparse parities directly.


\begin{theorem}
    \label{thm:nsc_main}
    For any $\epsilon>0$ and $\eta<1/2$, there exists an algorithm running in time $(\frac{n}{k})^{\frac{\omega+o(1)}{3} \cdot k}$ and $m=O\left(\frac{k \cdot \log^{1+\epsilon} \frac{n}{k}}{\epsilon}\right)$ samples from an $(k,\eta)$-LSPN problem to recover $\secret$.
\end{theorem}


In the rest of this appendix, we finish the proof of Theorem~\ref{thm:nsc_main}. For ease of exposition, we work in $\mathbb{R}$ instead of $\mathbf{F}_2$ in this section. Hence, we consider $\secret$ as a subset of $[n]$ such that the label of sample $\bx_i \sim \{\pm 1\}^n$ is $y_i = e_i \cdot \chi_{\secret}(\bx_i)$, where $\chi_{\secret}(\bx_i) = \prod_{j \in \secret} \bx_i(j)$ and $e_i \in \{\pm 1\}$ is the random noise. Since our algorithm improves upon Valiant's ingenious algorithm, we review this framework and highlight our improvements here. Then we describe our algorithm and its analysis in Appendix~\ref{appen:alg&pf}.

The starting point of Valiant's algorithm is to test whether a coordinate $i \in \secret$ or not. However, a uniformly random sample $x_i$ does not help with this tester. So Valiant's key idea (which influenced the subsequent works by \cite{DSGKS21,Yu21}) is to generate a biased sample. We state the definition of $\alpha$-biased distributions here.

\begin{definition}[$\alpha$-biased distribution]
    $\bx \in \{\pm 1\}^n$ is drawn from the $\alpha$-biased distribution $U_{\alpha}$ if each bit $\bx(i)$ is 1 with probability $\frac{1}{2}+\alpha$ independently, and $-1$ otherwise. 
    
    For convenience, when $\alpha=0$, let $U$ denote the uniform distribution over $\{\pm 1\}^n$. For any $S \subset [n]$, we use $U(S)$ and $U_{\alpha}(S)$ to denote the \emph{marginal distributions} on $S$.
\end{definition}
If we neglect the random flip $e$, for $y=\chi_{\secret}(\bx)=\prod_{j \in \secret} \bx(j)$,
\begin{equation}\label{eq:test_simp}
\E_{\bx \sim U_\alpha}[\bx(i) \cdot y]=\E_{\bx \sim U_\alpha}[\prod_{j \in \secret\Delta \{i\}} \bx(j)]=(2\alpha)^{|\secret \Delta \{i\}|}    
\end{equation}
where $\Delta$ denotes the set difference. Note that $|\secret \Delta \{i\}|=k-1$ if $i \in \secret$; otherwise this is $k+1$. Given Equation~\eqref{eq:test_simp}, one could test $i \in \secret$ or not by sampling from biased distributions.


Our first step tightens Valiant's analysis about the rejection sampling in Algorithm 10 of \cite{valiant2012finding} such that the bias increases from $\alpha=1/\sqrt{n}$ to $\alpha=\sqrt{k/n}$. 
As mentioned earlier, if one could have a faster procedure to generate larger bias examples, say $\alpha=0.01$ instead of $\sqrt{k/n}$, this would lead to a faster algorithm of LSPN immediately.

Then we follow the idea of Valiant's algorithm: instead of testing $j \in \secret$, it applies the fast matrix multiplication to
\begin{equation}\label{eq:test_k/3}
\E_{\bx \sim U_\alpha}[\chi_{S_1}(\bx) \cdot \chi_{S_2}(\bx) \cdot y(x)]=\E_{\bx \sim U_\alpha}[\prod_{j \in \secret\Delta S_1 \Delta S_2 } \bx(j)]=(2\alpha)^{|\secret\Delta S_1 \Delta S_2|}.
\end{equation}
such that it tests whether two subsets of size $k/3$ satisfy $(S_1 \cup S_2) \subset \secret$ or not.

Our \emph{second improvement} is to reduce the sample complexity of the above tester from $\tilde{O}(k^2)$ \citep{valiant2012finding} to $\tilde{O}(k)$. In fact, we simplify Valiant's algorithm by removing its majority vote (the for-loop of $i$ in Algorithm 13 of \cite{valiant2012finding}) and use a \emph{more involved} analysis to prove its correctness. Specifically, Valiant used the \cite{Lyubashevsky05} method to reduce the sample complexity, which makes more samples by taking linear combinations of $q$-tuples of original samples. For convenience, given a matrix $\mX \in \mathbb{R}^{m \times n}$ and $T \subset [m]$, we use $\chi_{T}(\mX) \in \mathbb{R}^n$ to denote the sample generated by the linear combination of samples in $T$: $\chi_{T}(\mX)(j)=\prod_{i\in T}\mX(i,j)$ for each $j \in [n]$. Since $\chi_T(\mX)$ and $\chi_{T'}(\mX)$ are pairwise independent given a random $\mX$, Valiant applied Chebyshev's inequality on $\chi_{T_1}(\mX),\chi_{T_2}(\mX),\ldots$ to prove the tester \eqref{eq:test_k/3} based on biased distributions still works. However, Chebyshev's inequality is too weak to support the union bound for all $\frac{k}{3}$-subsets $S_1$ and $S_2$. Hence Valiant's algorithm uses a majority vote on $O(k \log n)$ parallel repetitions. The \emph{most technical} part of our proof is to prove the correctness without this majority vote.

If we take a closer look at the new tester (replacing $\bx$ by $\chi_T(\mX)$ in $\E_{\bx \sim U_\alpha}[\chi_{S_1}(\bx) \cdot \chi_{S_2}(\bx) \cdot y]$ in \eqref{eq:test_k/3})
\[
\E_{T \sim {[m] \choose q}}\left[ \chi_{S_1}\big( \chi_T(\mX) \big) \cdot \chi_{S_2}\big( \chi_T(\mX) \big) \cdot y\big(\chi_T(\mX)\big) \right] = \E_{T \sim {[m] \choose q}}[\prod_{j \in \secret\Delta S_1 \Delta S_2 } \chi_T(\mX)(j)], 
\]
it only reads $O(k)$ bits of $\chi_T(\mX)$. Thus it sounds plausible to apply the leftover hash lemma when ${m \choose q}>>2^{O(k)}$ and use the Chernoff bound again (since $T_1, T_2,\ldots$ are independent). Indeed, this was suggested by \cite{valiant2012finding}. However, a big issue is that $\chi_T(\mX)$ has to pass the rejection sampling in the first step in order to be an $\alpha$-biased example, not taking entry-wise products directly. Since the rejection sampling depends on the Hamming weight depends of all entries, one can \emph{not} use the leftover hash lemma after the rejection sampling directly. Our main technical result (Proposition~\ref{cla:GMSruntime}) shows that the statistical closeness of the leftover hash lemma still holds after the rejection sampling. This simplifies Valiant's algorithm so that it does not need the majority vote anymore. More importantly, it reduces the sample complexity to $\tilde{O}(k)$.







\paragraph{Notations.} We map $0$ in $\mathbf{F}_2$ to $1$ in $\mathbb{R}$ and $1$ in $\mathbf{F}_2$ to $-1$ in $\mathbb{R}$ separately such that we work in $\mathbb{R}$ in this section. For convenience, for a vector $\bv \in \mathbb{R}^m$ and $S \subset [m]$, we use $\chi_{S}(\bv)$ to denote $\prod_{i\in S}\bv(i)$ for any vector $\bv$. For the hidden parity $\secret$, the correct label of $\bx$ is $\langle \bx,\secret \rangle$ in $\mathbf{F}_2$; and it becomes $\chi_{\secret}(\bx)=\prod_{i \in \secret} \bx(i)$ in $\{\pm 1\}$ in $\mathbb{R}$. 

Furthermore, for a matrix $\mX \in \mathbb{R}^{m \times n}$ and $S \subset [m]$, we use $\chi_{S}(\mX) \in \mathbb{R}^n$ to denote the entry-wise product of $S$: $\chi_{S}(\mX)(j)=\prod_{i\in S}\mX(i,j)$ for each $j \in [n]$. Moreover, given any matrix $\mX$ of dimension $m \times n$, we use $\mX(,S)\in \mathbb{R}^{m \times S}$ to denote the sub-matrix of $\mX$ constituted by all columns in $S$.

For $\bx \in \{\pm 1\}^n$, We use $\textrm{Ones}(\bx)$ to denote the number of 1s in the entries of $\bx\in\{\pm 1\}^n$. Recall that $U_{\alpha}$ denotes the $\alpha$-biased distribution where each bit is 1 with probability $\frac{1}{2}+\alpha$ independently. Let $B(n,\beta):=\textrm{Ones}(\bx)$ for $\bx \sim U_{\beta-1/2}$ which represents the binomial random variable counting the number of 1s in $\bx$, where each entry is $1$ with probability $\beta$ independently.

\subsection{Algorithm and Proof of Theorem~\ref{thm:nsc_main}}\label{appen:alg&pf}
Our algorithm consists of two steps. The first step, described in Algorithm~\ref{alg::gms}, uses $m:=O\left(\frac{k \cdot \log^{1+\epsilon} \frac{n}{k}}{\epsilon}\right)$ samples to generate $(\frac{n}{k})^{O(k)}$ random samples following the approach of \cite{Lyubashevsky05}, by XORing random $q$-tuples, where $q:=O\left(\frac{k \log \frac{n}{k}}{\epsilon \log \log \frac{n}{k}}\right)$. 

The second step, described in Algorithm~\ref{alg:lwfe1}, adopts a different tester than \cite{valiant2012finding}:
\begin{equation}\label{eq:new_tester}
    \E_{\bx \sim U_{\alpha}} \left[ \chi_{S_1}(\bx) \cdot \chi_{S_2}(\bx) \cdot y_{\bx}\right].
\end{equation}
Recall that we work in $\{\pm 1\}^n$ throughout this section such that a correct label would be $\chi_{\secret}(\bx)=\prod_{i \in \secret} \bx(i)$. This tester estimates $|(S_1 \cup S_2) \Delta \secret|$ for all $S_1$ and $S_2$ by the fast matrix multiplication techniques. Here, $y_{\bx}$ denotes the label of $\bx$. 


In the rest of this section, we analyze our algorithms in more details and finish the proof of Theorem~\ref{thm:nsc_main}. Given $m = \tilde{O}(k)$ samples $(\mX, \by)$, where $\mX \in \{\pm 1\}^{m \times n}$ and $\by \in \{\pm 1\}^m$, for any subset $T \subseteq [m]$, recall that $\chi_T(\mX) \in \{\pm 1\}^n$ denotes the entry-wise product of the rows indexed by $T$, i.e., $\left(\prod_{i \in T} \mX(i, \ell)\right)_{\ell=1,\ldots,n}$, and $\chi_T(\by)$ denotes the corresponding label, $\prod_{i \in T} \by(i)$. For a fixed parameter $q = o(k \log \frac{n}{k})$, we select a uniformly random subset $T \sim \binom{[m]}{q}$ and consider $\chi_T(\mX)$ with its label $\chi_T(y)$. Due to the presence of errors in $\by$, $\chi_T(\by)$ may be either correct or incorrect. An important observation is that the correctness of $\chi_T(\by)$ is independent with the random vector $\chi_T(\mX)$.






First, we bound the probability that $\chi_T(y)$ is correct for a random $T \sim {[m] \choose q}$. Let $p$ denote the fraction such that there are $(\frac{1}{2}+p)m$ correct labels in $\by$. Without loss of generality, we assume $p\in [1-\frac{2}{3} \cdot \eta, 1-2\eta]$, and we further assume that our algorithm knows this parameter (by enumeration). The first proposition bounds the fraction of correct labels of $\chi_T(y)$.


\begin{proposition}\label{clm:gap_prob_correct_incorrect}
Let $\mathbf{T_c}:=\big\{T \in {[m] \choose q} : \chi_T(\by) = \chi_{\secret}(\chi_{T}(\mX)) \big\}$ and $\mathbf{T_i}:=\big\{ T \in {[m] \choose q}: \chi_T(\by) \neq \chi_{\secret}(\chi_T(\mX))\big\}$. Then both $\mathbf{T_c}$ and $\mathbf{T_i}$ are of size 
$[{m \choose q} \cdot\frac{q}{m^2},{m\choose q} \cdot (1-\frac{q}{m^2})]$ if there is an incorrect label in $\by$. 

    
    Moreover, let $\mathbf{gap}:=\frac{\mathbf{T_c}}{{m \choose q}}-\frac{\mathbf{T_i}}{{m \choose q}}$.
    If there are $(\frac{1}{2}+p)m$ correct labels in $\by$, $\mathbf{gap} \ge \left(\frac{2mp-q+1}{m-q+1}\right)^q$ and there exists an efficient algorithm to compute $\mathbf{gap}$ given $p$.
\end{proposition}
For completeness, we prove Proposition~\ref{clm:gap_prob_correct_incorrect} in Section~\ref{sec:proof_gap_prob}. 

Now we analyze Algorithm~\ref{alg::gms} that generates random biased samples. Function~\textsc{AddBias} is in the same routine of \cite{valiant2012finding} but with a larger bias $\alpha=\sqrt{k/n}$ instead of $\sqrt{1/n}$. Hence we provide a tighter calculation on the time complexity and sample complexity. Specifically, we state the following proposition on $r$ to bound $m'$ and the running time of \textsc{MoreSamples}. Recall that $\textrm{Ones}(\bx)$ denotes the number of 1s in $\bx$.



\begin{algorithm}
    \caption{Step 1: Generate Biased Samples}
\label{alg::gms}    
 \SetKwFunction{AddBias}{AddBias} 
 \SetKwFunction{MoreSamples}{MoreSamples} 

\SetKwProg{Fn}{Function}{:}{}
\setcounter{AlgoLine}{0}
\KwIn{parameter $q=o(k \log n)$, $\mX \in \{\pm 1\}^{m \times n}$ and $\by \in \{\pm 1\}^m$ constituted by  $m$ random samples $(\bx_1,y_1),\ldots,(\bx_m,y_m)$}    
\KwOut{$\mX' \in \{\pm 1\}^{m' \times n}$ and $\by \in \{\pm 1\}^{m'}$ where $m' = \tilde{O}((\frac{n}{k})^{\frac{k}{3}})$ is defined on line 7}   

        \Fn{\AddBias{$\bx,r,\alpha$}}{ \tcc{Generate a biased sample by rejection sampling}
        
            \If{$\textrm{Ones}(\bx) \in [n/2-t,n/2+t]$}{
            
             Return 1 with probability $r\cdot\frac{\Pr[B(n,\frac{1}{2}+\alpha)=\textrm{Ones}(\bx)]}{\Pr[B(n,\frac{1}{2})=\textrm{Ones}(\bx)]}$
            }  
             \KwRet{0}
            }
        \textbf{EndFunction}
        
        \Fn{\MoreSamples{} }{    
         $m' \gets 8\cdot (\frac{n}{k})^{\frac{k}{3}}\cdot (\frac{1-2\eta}{3})^{-2q} \cdot
k \cdot \log\frac{n}{k}$

        Initialize $X' \in \{\pm 1\}^{m' \times n}$ and $y' \in \{\pm 1\}^{m'}$ to be empty
        
         $\alpha \gets \frac{1}{2}\sqrt{\frac{k}{n}}$, $t \gets \sqrt{3nk\cdot \log \frac{n}{k}} $, and $r \gets \frac{\Pr[B(n,\frac{1}{2})>\frac{n}{2}+t]}{\Pr[B(n,\frac{1}{2}+\alpha)>\frac{n}{2}+t]}$

        \Repeat{$\mX'$ has $m'$ rows}{
         Randomly sample $T \sim {[m] \choose q}$
         
        \For{$\ell \in[ n]$}{ 
        
         $\bx(\ell)\gets \prod_{i\in T}\mX(i,\ell)$\Comment{Generate $\chi_T(\mX)$}
        }
        
         $y \gets \prod_{i\in T}\by(i)$
        \Comment{Generate $\chi_T(\by)$}
        
        \If{\textsc{AddBias($\bx,r,\alpha$)}=1}{
            Add $\bx$ to $\mX'$ and add $y$ to $\by'$
        }
        } 
        }
   \textbf{EndFunction}
\end{algorithm}

\begin{proposition}\label{cla:ADDbiasguarantee}
    Given $\alpha=\frac{1}{2}\sqrt{\frac{k}{n}}$ and $t=\sqrt{3nk\cdot \log \frac{n}{k}} $, $r := \frac{\Pr[B(n,\frac{1}{2})>\frac{n}{2}+t]}{\Pr[B(n,\frac{1}{2}+\alpha)>\frac{n}{2}+t]}$ is at least $(\frac{n}{k})^{-o(k)}$.  
    
    Moreover, $r\cdot\frac{\Pr[B(n,\frac{1}{2}+\alpha)=\textrm{Ones}(\bx)]}{\Pr[B(n,\frac{1}{2})=\textrm{Ones}(\bx)]}$ is always in $[0,1]$ for $x$ with $\textrm{Ones}(\bx) \in [n/2-t,n/2+t]$. 
\end{proposition}

We state properties of Procedure \textsc{MoreSamples} in the following proposition, which is our main technical result. This proposition simplifies Valiant's analysis (by removing the $r$ parallel repetitions in Algorithm 13 of \cite{valiant2012finding}) and reduces the sample complexity from $\tilde{O}(k^2)$ (Theorem 5.6 in \cite{valiant2012finding}) to $\tilde{O}(k)$. Basically, it shows that after passing function \textsc{AddBias}, random samples $\chi_T(\mX)$ are close to the uniform biased distribution $U_\alpha$ on $S$. Recall that $\mX'(,S)\in \mathbb{R}^{m \times S}$ denotes the sub-matrix of $\mX'$ constituted by all columns in $S$.

\begin{proposition}\label{cla:GMSruntime}
    For any $\epsilon>0$, let $m=O(\frac{1}{\epsilon}k \cdot\log^{1+\epsilon} \frac{n}{k})$ and $q:=O\left(\frac{k \log \frac{n}{k}}{\epsilon \log \log \frac{n}{k}}\right)$ such that ${m \choose q}>(n/k)^{100k}$ and $q=o(k\cdot\log \frac{n}{k})$. Then with high probability, \textsc{MoreSamples} uses $m$ samples and runs in $(\frac{n}{k})^{\frac{1+o(1)}{3}k}$ time to generate $m'=(\frac{n}{k})^{\frac{k}{3}} \cdot 2^{O(q)}$ samples such that with probability 0.99, all submatrices $\mX'(,S)$ with $|S|$ less than $2k$ columns are $(n/k)^{-2k}$-close to $U_{\alpha}^{\otimes m'}(S)$.
\end{proposition}
Since $q=o(k\log\frac{n}{k})$, we simplify $m'$ to $(\frac{n}{k})^{\frac{(1+o(1))k}{3}}$ in the rest of this section.
We defer the proof of Proposition~\ref{cla:GMSruntime} to Section~\ref{sec:proof_of_GMSruntime} and the proof of Proposition~\ref{cla:ADDbiasguarantee} to Section~\ref{sec:proof_of_ADDBIAS} separately.

By Corollary~\ref{cor::stat_dist}, we assume that every submatrix $\mX'(,S)$ is generated by $U_{\alpha}^{\otimes m'}(S)$ in the rest of this section.
Now we describe the 2nd part of our algorithm in Algorithm~\ref{alg:lwfe1}.

\begin{algorithm}
\caption{Step 2: Find out $\secret$ by testing}
\label{alg:lwfe1}
 \SetKwFunction{Tester}{Tester} 
\SetKwProg{Fn}{Function}{:}{}
      \setcounter{AlgoLine}{0}
\KwIn{$\mathbf{gap}$ calculated in Proposition~\ref{clm:gap_prob_correct_incorrect}, $\mX' \in \{\pm 1\}^{m' \times n}$ and $\by \in \{\pm 1\}^{m'}$ where $m' = \tilde{O}((\frac{n}{k})^{\frac{k}{3}})$}    
\KwOut{subset $E \subset [n]$}     
\Fn{\Tester{}}{
         $\mD \gets 0^{m' \times {n \choose k/3}}$ and identify each column of $D$ as a subset in ${[n] \choose k/3}$
         
         Initialize two empty matrices $\mD_1,\mD_{-1}$
         
        \For{each row $\mX'(i,)$ and $E\in \binom{[n]}{\frac{k}{3}}$}{
        $\mD(i,E) \gets \prod_{j\in E} \mX'(i,j)$
        }
        \For{$i\in [m']$}{
             Add row $\mD(i,)$ to $\mD_{\by'(i)}$  \Comment{recall label $\by'(i) \in \{\pm 1\}$}
       }
        Compute $\mC_1=\mD_1^{\top} \cdot \mD_1 \in \mathbb{Z}^{{n \choose k/3} \times {n \choose k/3}}$ and $\mC_{-1}=\mD_{-1}^{\top} \cdot \mD_{-1} \in \mathbb{Z}^{{n \choose k/3} \times {n \choose k/3}}$ by fast matrix multiplication.
        
        Set $\mC=\mC_1-\mC_{-1}$ and find a pair of disjoint subsets $E_1$ and $E_2$ such that $\mC(E_1,E_2) \ge \frac{3m'}{4}(2\alpha)^{\frac{k}{3}} \cdot \mathbf{gap}$
        
         Find another disjoint $E_3 \subset [n] \setminus E_1 \cup E_2$ in row $\mC(E_1,.)$ with $\mC(E_1,E_3)\geq \frac{3m'}{4}(2\alpha)^{\frac{k}{3}} \cdot \mathbf{gap}$
         
         \KwRet{$E_1\cup E_2\cup E_3$} 
       }
\textbf{EndFuntion}
\end{algorithm}


Recall that $\mX'(i,)$ denotes row $i$ of $\mX'$ and $\chi_{E}\big( \mX'(i,) \big)$ denotes $\prod_{j \in E} \mX'(i,j)$. In Algorithm~\ref{alg:lwfe1}, $\mD(i,E)$ is calculated as the product of $\mX'(i,j)$ over $j \in E$, i.e., $\chi_{E}\big( \mX'(i,) \big)$. This allows us to compute $\mC_1$ and $\mC_{-1}$ as follows:
\[
\mC_1(E_1,E_2) = \sum_{i : \by'(i) = 1} \chi_{E_1}\big( \mX'(i,) \big) \cdot \chi_{E_2}\big( \mX'(i,) \big),
\]
\[
\mC_{-1}(E_1,E_2) = \sum_{i : \by'(i) = -1} \chi_{E_1}\big( \mX'(i,) \big) \cdot \chi_{E_2}\big( \mX'(i,) \big).
\]
Which leads to:
\[
\mC(E_1,E_2) = \sum_{i \in [m']} \chi_{E_1}(\mX'(i,)) \cdot \chi_{E_2}(\mX'(i,)) \cdot \by'(i).
\]

Thus, $\mC(E_1,E_2)$ effectively computes the difference between the contributions from biased samples in $\mX'$ with positive and negative labels. The matrix $\mC$ provides an approximation of the expected value of tester \eqref{eq:new_tester} for every $E_1$ and $E_2$, i.e., $\E_{\bx \sim U_{\alpha}} \left[ \chi_{E_1}(\bx) \cdot \chi_{E_2}(\bx) \cdot y_{\bx}\right]$. Then it uses $C$ to find subsets in $\secret$.

To finish the proof of Theorem~\ref{thm:nsc_main}, we use the following proposition. Recall that $\mX'(,T) \in \mathbb{R}^{m \times T}$ denotes the sub-matrix of $\mX'$ constituted by all columns in $T$.
\begin{proposition}
\label{lem:submatrixexp}
    If all sub-matrices $\mX'(,T)$ with less than $2k$ columns are generated from $U^{\otimes m'}_{\alpha}(T)$, then for any $S\in{[n]\choose \le 2k}$
    $$\E_{\mX'}\left[\chi_{S}(\mX'(i,)\right]=(2\alpha)^{|S|}\textit{ for all } i\in m'.$$
\end{proposition}

\begin{proofof}{Proposition~\ref{lem:submatrixexp}}
    Notice $\mX'(i,j)$ for $j\in S$ are independent. Thus 
    $$\E_{\mX'}\left[\chi_S(\mX'(i,))\right]=\prod_{j\in S}\E[\mX'(i,j)]=(2\alpha)^{|S|}.$$
\end{proofof}


Finally, we prove Theorem~\ref{thm:nsc_main} by applying a union bound over all subsets $E_1 \in {n \choose k/3}$ and $E_2 \in {n \choose k/3}$ in Algorithm~\ref{alg:lwfe1}.

\begin{proofof}{Theorem~\ref{thm:nsc_main}}
To finish the proof, we show the correctness of Algorithm~\ref{alg:lwfe1}. By Proposition~\ref{cla:GMSruntime}, we assume $\mX'$ has the following property: For every $S \in {[n] \choose \le 2k}$, $\mX'(,S)$ is generated from $U_{\alpha}^{\otimes m'}(S)$.

For any disjoint $E_1,E_2\in\binom{[n]}{\frac{k}{3}}$,
since $|\textbf{Secret}|=k$ and $|E_1|=|E_2|=\frac{k}{3}$, we always have $|(E_1\cup E_2)\triangle \textbf{Secret}|\leq 5k/3$.
Then 
\begin{align*}
    \E_{\mX'}\left[\mC(E_1,E_2)\right]=&\sum_{i=1}^{m'}\E\left[\chi_{E_1}(\mX'(i,))\cdot\chi_{E_2}(\mX'(i,))\cdot\by'(i)\right]\\
    =&\sum_{i=1}^{m'}(\E\left[\chi_{(E_1\cup E_2)\triangle \textbf{Secret}}(\mX'(i,))\mid\by'(i)=\chi_{\textbf{Secret}}(\mX'(i,))\right]\cdot\Pr[\by'(i)=\chi_{\textbf{Secret}}(\mX'(i,))]\\
    & - \E[\chi_{(E_1\cup E_2)\triangle \textbf{Secret}}(\mX'(i,))\mid\by'(i)\neq \chi_{\textbf{Secret}}(\mX'(i,))]\cdot\Pr[\by'(i)\neq\chi_{\textbf{Secret}}(\mX'(i,))])\\
    =& m' \cdot (2\alpha)^{|(E_1\cup E_2)\triangle \textbf{Secret}|} \cdot \mathbf{gap}.
\end{align*}
If both $E_1$ and $E_2$ are in $\textbf{Secret}$, $|(E_1\cup E_2)\triangle \textbf{Secret}|={\frac{k}{3}}$. Otherwise $|(E_1\cup E_2)\triangle \textbf{Secret}|\geq \frac{k}{3}+2$, and the difference of the expectation of $\mC(E_1,E_2)$ between $E_1 \cup E_2\subseteq \textbf{Secret}$ and $E_1 \cup E_2\not\subseteq \textbf{Secret}$ is 
\begin{align*}
    &m' \cdot (2\alpha)^{k/3} \cdot \mathbf{gap} - m' \cdot (2\alpha)^{k/3+2} \cdot \mathbf{gap} \\
    \ge & m'(2\alpha)^{\frac{k}{3}} \cdot \mathbf{gap} \cdot (1-4\alpha^2) \geq \frac{3m'}{4}(2\alpha)^{\frac{k}{3}} \cdot \mathbf{gap}.\tag{Recall $\alpha=\frac{1}{2}\sqrt{\frac{k}{n}}$}
\end{align*}

Then by the Chernoff bound over $m'$, 
 for disjoint $E_1 \cup E_2\subseteq \textbf{Secret}$, we have
\begin{equation}\label{eq:failure1}
    \Pr\left[ \mC(E_1,E_2)-\E[\mC(E_1,E_2)]\leq -\frac{m'}{4}\cdot(2\alpha)^{\frac{k}{3}}\cdot\mathbf{gap} \right] \leq e^{-\frac{m'(2\alpha)^{\frac{2k}{3}}\cdot\mathbf{gap}^2}{8}}.
\end{equation}
For the case where $E_1\not\subseteq \textbf{Secret}$ or $E_2\not\subseteq \textbf{Secret}$, we have

\begin{equation}\label{eq:failure2}
    \Pr\left[ \mC(E_1,E_2)-\E[\mC(E_1,E_2)]\geq \frac{m'}{4}\cdot(2\alpha)^{\frac{k}{3}}\cdot\mathbf{gap} \right] \leq e^{-\frac{m'(2\alpha)^{\frac{2k}{3}}\cdot\mathbf{gap}^2}{8}}.
\end{equation}

Next, we apply the union bound over all $E_1, E_2 \in \binom{[n]}{\frac{k}{3}}$. With probability at least $1 - \binom{n}{\frac{k}{3}}^2 \cdot e^{-\frac{m'(2\alpha)^{\frac{2k}{3}} \cdot \mathbf{gap}^2}{8}}$, for all pairs $(E_1, E_2)$ such that $E_1 \cup E_2 \subseteq \textbf{Secret}$, we have $\mC(E_1, E_2) \geq \frac{3m'}{4}(2\alpha)^{\frac{k}{3}} \cdot \mathbf{gap}$, and for all pairs $(E_1, E_2)$ such that $E_1 \cup E_2 \not\subseteq \textbf{Secret}$, we have $\mC(E_1, E_2) \leq \frac{m'}{2}(2\alpha)^{\frac{k}{3}} \cdot \mathbf{gap}$.

By Proposition~\ref{clm:gap_prob_correct_incorrect}: 
\begin{align*}
    \mathbf{gap}&\geq \left( \frac{2mp-q+1}{m-q+1} \right)^q\\
    &=\left(2p-(1-2p)\frac{q-1}{m-q-1}\right)^q\\
    &\geq \left(2p-\frac{q}{m-q}\right)^q\\
    &\geq p^q \tag{Recall $m=O(k\log^{1+\epsilon}\frac{n}{k}),q=o(k\log\frac{n}{k})$}\\ 
    &\ge \left(\frac{1-2\eta}{3}\right)^{q}\tag{Recall $p\in \left[\frac{1-2\eta}{3}, 1-2\eta\right]$}.
\end{align*}

Since $m' = 8\cdot(\frac{n}{k})^{\frac{k}{3}}\cdot (\frac{1-2\eta}{3})^{-2q} \cdot
k \cdot \log\frac{n}{k}$ and $\alpha=\frac{1}{2} \sqrt{\frac{k}{n}}$, we have 
\[
\frac{m'(2\alpha)^{\frac{2k}{3}}\cdot\mathbf{gap}^2}{8}\geq \left(\frac{n}{k}\right)^{\frac{k}{3}}\cdot \left(\frac{1-2\eta}{3}\right)^{-2q}\cdot k\cdot \log\frac{n}{k}\cdot\left(\frac{n}{k}\right)^{-\frac{k}{3}}\cdot\left(\frac{1-2\eta}{3}\right)^{2q} \geq  k\log\frac{n}{k}.\]

Thus, the failure probability over all $E_1$ and $E_2$ is at most \[
\binom{n}{\frac{k}{3}}^2\cdot e^{-\frac{m'(2\alpha)^{\frac{2k}{3}}\cdot \mathbf{gap}^2}{8}}\leq e^{\frac{2k}{3}\log\frac{3n}{k}-k\log\frac{n}{k}}=o(1)
\]
by the union bound. This means with probability $1-o(1)$, line 9 in \textsc{LearningWithFewerSamples} finds disjoint $E_1$ and $E_2$ in $\secret$. Similarly, $E_3$ will be disjoint with $E_1$ and $E_2$ and be in \textbf{secret}. 

The running time mainly comes from two matrix multiplications $\mD_1^{\top} \cdot \mD_1$ and $\mD_{-1}^{\top} \cdot \mD_{-1}$. Since $\mD_1$ and $\mD_{-1}$ are of size $(\le m') \times {n \choose k/3}$ for $m'=({\frac{n}{k}})^{\frac{1+o(1)}{3} \cdot k}$, the time is at most $(\frac{n}{k})^{(\frac{\omega}{3}+o(1))k}$.

Combined with Proposition~\ref{cla:ADDbiasguarantee}, we show that the algorithm runs in a time of $(\frac{n}{k})^{\frac{\omega+o(1)}{3} \cdot k}$ and requires $m = O\left(\frac{k \cdot \log^{1+\epsilon} \frac{n}{k}}{\epsilon}\right)$ samples to recover $\secret$.
\end{proofof}

\subsection{Proof of Proposition \ref{cla:GMSruntime}}\label{sec:proof_of_GMSruntime}

The proof of Proposition~\ref{cla:GMSruntime} relies on the following fact: since \(\chi_T(\by) = \chi_{\secret}(\chi_T(\mX))\) depends only on the noise in \(\by\), this event is independent of \(\mX\) and \(\chi_T(\mX)\). We will utilize the conditional distribution of \(\chi_T(\mX)\) given that the label \(\chi_T(\by)\) is correct (or incorrect). The key property is stated in the following proposition.

\begin{proposition}\label{clm:distance_XT}
With probability 0.99 over $\mX$, for all $S \in {[n] \choose \le 2k}$, all $\sigma \in \{\pm 1\}^S$ and any $j\in[m]$,  $(\mX'(j,),\by'(j))$ satisfies
 \begin{align}
    & \Pr_{\mX'(j)}\left[\mX'(j,S)=\sigma|\by'(j)=\chi_{\secret}\big( \mX'(j,) \big) \right]=\left(1\pm O\left(\left(\frac{n}{k}\right)^{-3k}\right)\right) \cdot \Pr_{\bx\sim U_{\alpha}}[\bx(S)=\sigma] \label{eq:condition_event1}\\
    \textit{and }  & \Pr_{\mX'(j)}\left[ \mX'(j,S)=\sigma|\by'(j)\neq \chi_{\secret}\big( \mX'(j,) \big) \right]=\left(1\pm O\left(\left(\frac{n}{k}\right)^{-3k}\right)\right) \cdot \Pr_{\bx\sim U_{\alpha}}[\bx(S)=\sigma].
    \label{eq:condition_event2}
    \end{align}
    Here, the randomness of $\mX'(j)$ depends on the random choice of $T \subset [m]$ (to generate $\bx=\chi_T(\mX)$) and the randomness of function~\textsc{AddBias}($\bx$).
\end{proposition}

We defer the proof of Proposition~\ref{clm:distance_XT} to Section~\ref{sec:proof_distance_XT}.  Now we use this proposition to finish the proof of Proposition~\ref{cla:GMSruntime}.

\begin{proofof}{Proposition~\ref{cla:GMSruntime}}
For any $S\in\binom{[n]}{\leq 2k}$, $j\in[m']$ and any $\sigma\in \{\pm 1\}^S$, we have
    \begin{align*}     \Pr[\mX'(j,S)=\sigma]&=\Pr[\mX'(j,S)=\sigma|\by_j=\chi_{\secret}(\mX'(i,))]\Pr[\by_j=\chi_{\secret}(\mX'(j,))]+\\ & \qquad \Pr[\mX'(j,S)=\sigma|\by_j\neq\chi_{\secret}(\mX'(j,))]\Pr[\by_j\neq\chi_{\secret}(\mX'(j,))]\\
&=\left(1\pm O\left(\left(\frac{n}{k}\right)^{-3k}\right)\right)\cdot\Pr_{\bx\sim U_{\alpha}}[\bx(S)=\sigma] \cdot \frac{|\mathbf{T_c}|}{\binom{m}{q}}+\left(1\pm O\left(\left(\frac{n}{k}\right)^{-3k}\right)\right) \cdot\Pr_{x\sim U_{\alpha}}[\bx(S)=\sigma] \cdot \frac{|\mathbf{T_i}|}{\binom{m}{q}}\\
        &=\left(1\pm O\left(\left(\frac{n}{k}\right)^{-3k}\right)\right)\cdot\Pr_{x\sim U_{\alpha}}[\bx(S)=\sigma].
    \end{align*}

Let $V$ denote the distribution of $\mX'(j,S)$.
    Thus the total variation distance between $V$ and $U_{\alpha}(S)$ should be 
    $$\|V-U_{\alpha}(S)\|_{TV}=\frac{1}{2}\sum_{\sigma\in\{\pm 1\}^{S}}\bigg|\Pr[\mX'(j,S)=\sigma]-\Pr_{\bx\sim U_{\alpha}}[\bx(S)=\sigma]\bigg|\leq O\left(\left(\frac{n}{k}\right)^{-3k}\right).$$
     Since we have $m'=(\frac{n}{k})^{\frac{(1+o(1))k}{3}}$ rows, the whole total variation distance between $\mX'(,S)$ and $U_{\alpha}^{\otimes m'}(S)$ should be $O((\frac{n}{k})^{-2k})$ by Corollary~\ref{cor::stat_dist}.

Next, we analyze the running time.

      Since $\Pr[\frac{n}{2}-t\leq B(n,1/2)\leq \frac{n}{2}+t] \ge 0.9$, each pair $(x,y)$ satisfies IF condition of Line 2 in \textsc{AddBias} with probability at least $0.9$.  
    So the expected number of total random pairs needed to be sampled is at most$\frac{m'}{0.9r}$. By the standard concentration, with probability $1-e^{-0.64m'r}$, our algorithm needs at most $\frac{2m'}{r}$ random sets drawn from $\binom{m}{q}$.  By Proposition~\ref{cla:ADDbiasguarantee}, we know $r\geq (\frac{n}{k})^{-o(k)}.$ Thus the sampling runtime should be at most $O((\frac{n}{k})^{\frac{(1+o(1))k}{3}})$.

\end{proofof}

\subsection{Proof of Proposition~\ref{clm:distance_XT}}\label{sec:proof_distance_XT}

The proof is divided into two parts. 
The starting point is by the choice of $m,q$, we have $\binom{m}{q}\geq (\frac{n}{k})^{100k}$.
Since $\textsc{AddBias}(\bx,r,\alpha)$ depends on $\textrm{Ones}(\bx)$, the first step will show the joint distribution of $\chi_T(\mX)(S)$ and $\textrm{Ones}(\chi_T(\mX))$  close to $\big( \bx(S), \textrm{Ones}(\bx) \big)$ for $\bx \sim U_\alpha$ when $T\sim \mathbf{T_c}$ or $T\sim \mathbf{T_i}$ --- see \eqref{eq:T_c} and \eqref{eq: T_i} for the exact bound. Recall that $\chi_T(\mX)\in \mathbb{R}^n$ with each entry $\chi_T(\mX)(j)$ equal to $\prod_{i\in T}\mX(i,j)$ for $j\in [n]$, $\mathbf{T_c}:=\big\{T \in {[m] \choose q} : \chi_T(\by) = \chi_{\secret}(\chi_{T}(\mX)) \big\}$, and $\mathbf{T_i}:=\big\{ T \in {[m] \choose q}: \chi_T(\by) \neq \chi_{\secret}(\chi_T(\mX))\big\}$.

Fix $S\in\binom{[n]}{\leq 2k}$, $\sigma \in \{\pm 1\}^S$, and $\ell \in [n/2-t,n/2+t]$ for now. For each $T$, we define an indicator random variable 
\begin{equation}
    I_{T}(\sigma)=\left\{
    \begin{aligned}
        &1, \chi_T(\mX)(S)=\sigma \wedge \textrm{Ones}(\chi_T(\mX))=\ell\\
        &0, \textit{otherwise}
    \end{aligned}
    \right.
\end{equation}

Notice that for $T_1\neq T_2$, $\chi_{T_1}(\mX)$ and $\chi_{T_2}(\mX)$ are pairwise independent. Thus $\{I_T(\sigma)\}_{T\in \mathbf{T_c}}$ is a family of pairwise independent random variables and 
$$\Pr_{T \sim \mathbf{T_c}}[\chi_T(\mX)(S)=\sigma \wedge \textrm{Ones}(\chi_T(\mX))=\ell]=\frac{\sum_{T\in \mathbf{T_c}}I_T(\sigma)}{|\mathbf{T_c}|}.$$

Notice that for each \(T \in \mathbf{T_c}\), \(I_T(\sigma)\) depends on the entry-wise product \(\chi_T(\mX) := \left(\prod_{i \in T} \mX(i,j)\right)_{j=1,\ldots,n}\) and $X$ is generated from the uniform distribution \(U\). Specifically, we have

\[
\E_{\mX}[I_T(\sigma)] = \Pr_{\bx \sim U}\left[\bx(S) = \sigma \wedge \textrm{Ones}(\bx) = \ell\right].
\]

Next, we will use Chebyshev's inequality to establish a concentration bound on $\sum_{T\in \mathbf{T_c}}I_T(\sigma)$. 

Firstly, 
$$\Pr_{\bx\sim U}[\bx(S)=\sigma\wedge \textrm{Ones}(\bx)=\ell]=2^{-|S|}2^{-(n-|S|)}\binom{n-|S|}{\ell-\textrm{Ones}(\sigma)}\geq 2^{-|S|}2^{-(n-|S|)}\binom{n-|S|}{\frac{n}{2}-t-|S|}.$$
Here $t= \sqrt{3nk\log \frac{n}{k}}.$

According to the Strling's formula,
\begin{align*}
    \binom{n-|S|}{\frac{n}{2}-t-|S|}&\sim\frac{\sqrt{2\pi (n-|S|)}(\frac{n-|S|}{e})^{n-|S|}}{\sqrt{2\pi (n/2+t)}(\frac{n/2+t}{e})^{n/2+t}\sqrt{2\pi (n/2-t-|S|)}(\frac{n/2-t-|S|}{e})^{n/2-t-|S|}}\\
    &=\frac{4}{\sqrt{2\pi}} \frac{(n-|S|)^{n-|S|}}{(n/2+t)^{n/2+t}(n/2-t-|S|)^{n/2-t-|S|}}  \tag{$n$ is large}\\
    &\geq  2^{n-|S|}(\frac{(n-|S|)^2}{((n-|S|)^2-(2t+|S|)^2)})^{\frac{n}{2}-t-|S|}(\frac{n-|S|}{n+2t})^{2t+|S|}\\
    &=2^{n-|S|}(1+\frac{(2t+s)^2}{(n-|S|)^2-(2t+|S|)^2})^{\frac{n}{2}-t-|S|}(1-\frac{2t+|S|}{n+2t})^{2t+|S|}\\
    &\geq 2^{n-|S|}e^{-\frac{(2t+|S|)^2}{n+2t}}\\
    &\geq 2^{n-|S|}e^{-\frac{3t^2}{n}} \geq 2^{n-|S|}\left(\frac{n}{k}\right)^{-20k}.
\end{align*}

Then we have the lower bound $$\E_{\mX}[I_T(\sigma)]=\Pr_{\bx\sim U}[\bx(S)=\sigma\wedge \textrm{Ones}(\bx)=\ell]\geq 2^{-|S|}\cdot\left(\frac{n}{k}\right)^{-20k}\geq \left(\frac{n}{k}\right)^{-22k}.$$

Using Chebyshev's inequality ($|\mathbf{T_c}|\geq \binom{m}{q}\cdot\frac{q}{m^2}\geq \left(\frac{n}{k}\right)^{50k}$ by Proposition~\ref{clm:gap_prob_correct_incorrect}),
\begin{align*}
    \Pr_{\mX}\left[ \big|\sum_{T\in \mathbf{T_c}}I_{T}(\sigma)-\E_{\mX}[\sum_{T\in \mathbf{T_c}}I_T(\sigma)]\big| \geq  \E_{\mX}[\sum_{T\in \mathbf{T_c}}I_T(\sigma)]\cdot(n/k)^{-6k} \right] &\leq \frac{\Var(\sum_{T\in \mathbf{T_c}}I_{T}(\sigma))}{(\E_{\mX}[\sum_{T\in \mathbf{T_c}}I_T(\sigma)]\cdot(n/k)^{-6k})^2}\\
    &\leq \frac{| \mathbf{T_c}|\cdot\E_{\mX}[I_T(\sigma)]}{(| \mathbf{T_c}|\cdot\E_{\mX}[I_T(\sigma)]\cdot(n/k)^{-6k})^2}\\
    &\leq \frac{(n/k)^{12k}}{|\mathbf{T_c}|\cdot\E_{\mX}[I_T(\sigma)]} \\
    &\leq \frac{(n/k)^{12k}}{\left(\frac{n}{k}\right)^{50k}\cdot (\frac{n}{k})^{-22k}} \leq \left(\frac{n}{k}\right)^{-16k}.
\end{align*}

Replacing $\textbf{T}_c$ with $\textbf{T}_i$, we get a similar inequality.
$$\Pr_{\mX}[\big|\sum_{T\in \mathbf{T_i}}I_{T}(\sigma)-\E_{\mX}[\sum_{T\in \mathbf{T_i}}I_T(\sigma)]\big|\geq  \E_{\mX}[\sum_{T\in \mathbf{T_i}}I_T(\sigma)]\cdot(n/k)^{-6k}]\leq \left(\frac{n}{k}\right)^{-16k}.$$

Then, applying a union bound over all \(S\), \(\sigma\), and \(\ell\) shows that, with probability 0.99 over \(\mX\), for any \(\ell \in [n/2 - t, n/2 + t]\) and any \(S \in \binom{[n]}{\leq 2k}\) with any \(\sigma \in \{\pm 1\}^S\), we have:
\begin{equation}
\label{eq:T_c}
   \Pr_{T \sim \mathbf{T_c}}[\chi_T(\mX)(S)=\sigma \wedge \textrm{Ones}(\chi_T(\mX))=\ell]=(1\pm (n/k)^{-6k})) \cdot \Pr_{\bx \sim U_{\alpha}}[\bx(S)=\sigma\wedge \textrm{Ones}(\bx)=\ell]. 
\end{equation}
\begin{equation}
   \label{eq: T_i}
      \Pr_{T \sim \mathbf{T_i}}[\chi_T(\mX)(S)=\sigma \wedge \textrm{Ones}(\chi_T(\mX))=\ell]=(1\pm (n/k)^{-6k})) \cdot \Pr_{\bx \sim U_{\alpha}}[\bx(S)=\sigma\wedge \textrm{Ones}(\bx)=\ell]. 
\end{equation}

In the second step, we assume both \eqref{eq:T_c} and \eqref{eq: T_i} hold and use them to finish our proof. Since $\alpha$ and $r$ are fixed in Algorithm~\ref{alg::gms}, we use $\textsc{AddBias}(\chi_T(\mX))$ as a shorthand of $\textsc{AddBias}(\chi_T(\mX),r,\alpha).$

    The condition $\by'(j)=\chi_{\secret}(\mX'(j,))$ in \eqref{eq:condition_event1} means $T \in \mathbf{T_c}$. Thus we fix $\sigma \in \{\pm 1\}^S$ and consider the probability that a random $T \sim \mathbf{T_c}$ with $\chi_T(\mX)(S)=\sigma$ will be added to $\mX'$:
    \begin{align*}        
        & \Pr_{\mX'(j,)}\left[ \chi_T(\mX)(S)=\sigma \wedge \textsc{AddBias}\big( \chi_T(\mX) \big) \bigg| \by'(j)=\chi_{\secret}(\mX'(j,) \right]
        \\        
        = & \Pr_{T \sim \mathbf{T_c}}\left[ \chi_T(\mX)(S)=\sigma \wedge \textsc{AddBias}\big( \chi_T(\mX) \big) \right]\\
        = & \sum_{j=n/2-t}^{n/2+t} \Pr_{T \sim \mathbf{T_c}}[\chi_T(\mX(,S))=\sigma \wedge \textrm{Ones}(\chi_T(\mX))=j] \cdot r \cdot \frac{\Pr[B(n,1/2+\alpha)=j]}{\Pr[B(n,1/2)=j]}.     
    \end{align*}
    Now we plug \eqref{eq:T_c} to switch from $\chi_T(\mX)$ to $\bx \sim U_{\alpha}$:
    \begin{align*}
        & \sum_{j=n/2-t}^{n/2+t} \left( \Pr_{\bx \sim U_{\alpha}}[\bx(S)=\sigma \wedge \textrm{Ones}(\bx)=j] \pm (n/k)^{-6k} \right) \cdot r \cdot \frac{\Pr[B(n,1/2+\alpha)=j]}{\Pr[B(n,1/2)=j]}       \\
        = & \left( \sum_{j=n/2-t}^{n/2+t} \sum_{\bx \in \{\pm 1\}^n: \bx(S)=\sigma, \textrm{Ones}(\bx)=j} 2^{-n} \cdot r \cdot \frac{(1/2+\alpha)^j (1/2-\alpha)^{n-j}}{2^{-n}} \right) \pm (2t+1) \cdot  (n/k)^{-6k}  \\
        = & \sum_{j=n/2-t}^{n/2+t} \sum_{\bx \in \{\pm 1\}^n: \bx(S)=\sigma,  \textrm{Ones}(\bx)=j} r \cdot (1/2+\alpha)^j (1/2-\alpha)^{n-j} \pm n\cdot (n/k)^{-6k}  \\
        = & \sum_{j=1}^{n} \sum_{\bx \in \{\pm 1\}^n: \bx(S)=\sigma,  \textrm{Ones}(\bx)=j} r \cdot (1/2+\alpha)^j (1/2-\alpha)^{n-j} \pm n \cdot (n/k)^{-6k} \pm r\cdot\Pr_{\bx \sim U_\alpha}[\textrm{Ones}(\bx)\notin [n/2-t,n/2+t]]  \\
        = & r \cdot \Pr_{\bx\sim U_{\alpha}}[\bx(S)=\sigma] \pm n \cdot (n/k)^{-6k} \pm \Pr_{\bx \sim U_\alpha}[\textrm{Ones}(\bx)\notin [n/2-t,n/2+t]].  
    \end{align*}

    Notice that 
    $$\Pr_{\bx \sim U_\alpha}[\textrm{Ones}(\bx)\geq  n/2+t]\leq e^{-\frac{2(t-n\alpha)^2}{n}}= e^{-2k(\sqrt{3\log\frac{n}{k}}-\frac{1}{2})^2}\leq \left(\frac{n}{k}\right)^{-5k},$$
    and 
    $$\Pr_{\bx \sim U_\alpha}[\textrm{Ones}(\bx)\leq  n/2-t]\leq e^{-\frac{2(t+n\alpha)^2}{n}}= e^{-2k(\sqrt{3\log\frac{n}{k}}+\frac{1}{2})^2}\leq \left(\frac{n}{k}\right)^{-5k}.$$
Thus 

\begin{equation}
    \Pr_{T \sim \mathbf{T_c}}[\chi_T(\mX)(S)=\sigma \wedge \textsc{AddBias}(\chi_T(\mX))]=r \cdot \Pr_{\bx\sim U_{\alpha}}[\bx(S)=\sigma] \pm 3\left(\frac{n}{k}\right)^{-5k}.
\label{eq:pr_sigma_ones}
\end{equation}

Since $\textrm{Ones}(\sigma)\leq |S| \in [0,2k]$, We have $$\Pr_{\bx\sim U_{\alpha}}[\bx(S)=\sigma]=\left(\frac{1}{2}+\alpha\right)^{\textrm{Ones}(\sigma)}\cdot\left(\frac{1}{2}-\alpha\right)^{|S|-\textrm{Ones}(\sigma)}\geq \left(\frac{1}{2}-\alpha\right)^{2k}.$$ Thus $r\cdot \Pr_{\bx\sim U_{\alpha}}[\bx(S)=\sigma]\geq  (\frac{n}{k})^{-o(k)}\cdot(\frac{1}{2}-\alpha)^{2k}\geq (\frac{n}{k})^{-2k}$.
    Combining them all, the error term should be $O(r\cdot \Pr_{\bx\sim U_{\alpha}}[\bx(S)=\sigma]\cdot (\frac{n}{k})^{-3k}).$

Equation \eqref{eq:pr_sigma_ones} can also be written as
\begin{equation}
    \Pr_{T\sim \mathbf{T_c}}[\chi_T(\mX)(S)=\sigma \wedge \textsc{AddBias}(\chi_T(\mX))]=r\cdot \Pr_{\bx\sim U_{\alpha}}[\bx(S)=\sigma]\pm O\left(r \cdot \Pr_{\bx\sim U_{\alpha}}[\bx(S)=\sigma]\cdot \left(\frac{n}{k}\right)^{-3k}\right).
    \label{eq:pr_simplified}
\end{equation}
Then we define $p_0:= \Pr_{T\sim \mathbf{T_c}}[\textsc{AddBias}(\chi_T(\mX))]$.
Hence,

\begin{align*}
    p_0&=\sum_{\sigma\in\{\pm 1\}^S}\Pr_{T \sim \mathbf{T_c}}[\chi_T(\mX)(S)=\sigma \wedge \textsc{AddBias}(\chi_T(\mX))]\\
    &=\sum_{\sigma\in\{\pm 1\}^S} r \cdot \Pr_{\bx\sim U_{\alpha}}[\bx(S)=\sigma]\pm O\left(r \cdot \Pr_{\bx\sim U_{\alpha}}[\bx(S)=\sigma]\cdot \left(\frac{n}{k}\right)^{-3k}\right) =r\pm O\left(r\cdot \left(\frac{n}{k}\right)^{-3k}\right).
\end{align*}

Finally, for any row $j$ of $\mX'$, we have 
\begin{align*}
    \Pr[\mX'(j,S)=\sigma|\by'(j)=\chi_{\secret}]&=\frac{\Pr_{T\sim \mathbf{T_c}}[\chi_T(\mX)(S)=\sigma \wedge \textsc{AddBias}(\chi_T(\mX))]}{p_0}\\
    &=\frac{r\cdot \Pr_{\bx\sim U_{\alpha}}[\bx(S)=\sigma]\pm O(r \cdot \Pr_{\bx\sim U_{\alpha}}[\bx(S)=\sigma]\cdot (\frac{n}{k})^{-3k}))}{r\pm O(r\cdot (\frac{n}{k})^{-3k})}\\
    &= \Pr_{\bx\sim U_{\alpha}}[\bx(S)=\sigma]\cdot\left(1\pm O\left(\left(\frac{n}{k}\right)^{-3k}\right)\right).
\end{align*}

Replacing $\mathbf{T_c}$ with $\mathbf{T_i}$ and applying \eqref{eq: T_i} , we can get a similar conclusion for the second equation of Proposition~\ref{clm:distance_XT} .

\subsection{Proof of Proposition~\ref{clm:gap_prob_correct_incorrect}}\label{sec:proof_gap_prob}

We will use the following proposition in this proof.
\begin{proposition}[\cite{Lyubashevsky05}]
\label{lem::corn}
If a bucket contains $m$ balls, $(\frac{1}{2}+p)m$ of which are colored white, and the rest colored black, and we select $k$ balls at random without replacement, then the probability that we selected an even number of black balls is at least $\frac{1}{2}+\frac{1}{2}\left(\frac{2mp-k+1}{m-k+1}\right)^k.$    
\end{proposition}

Given $m$ examples in $(\mX,\by)$ where $\mX \in \{\pm 1\}^{m \times n}$ and $\by \in \{\pm 1\}^m$ such that each label $\by(i)$ is correct with probability $\frac{1}{2}+\frac{1-2\eta}{2}$, we have $\E\big[ \big| \{i:\by(i)\textit{ is correct}\} \big| \big]=\left(\frac{1}{2}+\frac{1-2\eta}{2}\right)m$.
By the Chernoff bound,
\begin{align*}
    \Pr \left[ |\{i : \by(i)\textit{ is correct}\}|\leq  \left(\frac{1}{2}+ \frac{1-2\eta}{3}\right)m \right] &\leq e^{-\frac{(1-2\eta)^2m}{18}} \\
 \text{ and } \Pr\left[ |\{i : \by(i)\textit{ is correct}\}|\geq  \left(\frac{1}{2}+ {1-2\eta}\right)m \right] &\leq  e^{-\frac{(1-2\eta)^2m}{2}}.
\end{align*} 
Thus with probability $1-o(\frac{1}{m})$, we have $(\frac{1}{2}+p)m$ correct labels in $\by$ for $p\in [(1-2\eta)/3,1-2\eta]$.

Note that the event $\chi_T(\by)=\chi_{\textbf{secret}}(\chi_T(\mX))$ is equivalent to having an even number of incorrect labels corresponding to the rows indexed by $T$. To bound this probability, consider selecting two rows: one with the correct label and one with the incorrect label. Regardless of how we choose $q-1$ additional rows from the remaining $m-2$ rows, we can construct a set $T$ of size $q$ such that $\chi_T(\by)$ is correct. If the $q-1$ rows contain an even number of incorrect labels, we include the correct one; if they contain an odd number, we include the incorrect one. This reasoning shows that at least $\binom{m-2}{q-1}$ sets from $\binom{[m]}{q}$ have the correct label. Thus
$$\frac{\mathbf{T_c}}{{m \choose q}} = \Pr_{T\sim \binom{[m]}{q}}[\chi_T(\by)=\chi_{\textbf{Secret}}(\chi_T(\mX))]\geq \frac{\binom{m-2}{q-1}}{\binom{m}{q}}\geq \frac{q}{m^2}.$$

By the same reason, $\Pr_{T\sim \binom{[m]}{q}}[ \chi_T(\by)\neq \chi_{\textbf{Secret}}(\chi_T(\mX))]\geq \frac{\binom{m-2}{q-1}}{\binom{m}{q}}\geq \frac{q}{m^2}.$

Then we apply Proposition~\ref{lem::corn} to get $$\Pr_{T \sim {[m] \choose q}}[\chi_T(\by) = \chi_{\secret}(\chi_T(\mX))] - \Pr_{T \sim {[m] \choose q}}[ \chi_T(\by) \neq \chi_{\secret}(\chi_T(\mX))] \ge \left(\frac{2mp-q+1}{m-q+1}\right)^q.$$

Now we present the algorithm to compute $\mathbf{gap}:=\frac{\mathbf{T_c}}{{n \choose q}} - \frac{\mathbf{T_i}}{{n \choose q}}$. We assume that the number of correct labels $c:=(\frac{1}{2}+p)m$ is known (or can be determined by enumeration).

\begin{algorithm}[H]
\label{alg:ComputeGap}
\caption{Compute the gap}
   \SetKwFunction{ComputeGap}{ComputeGap} 
\SetKwProg{Fn}{Function}{:}{}  

        \Fn{\ComputeGap{$m,c,q$}}{
        \If{$m=c$ or $q=0$}{
        \KwRet{1}
        
        \Else{
         \KwRet{ $\frac{m-c}{m}(1-\textsc{ComputeGap}(m-1,c,q-1))+\frac{c}{m}\textsc{ComputeGap}(m-1,c-1,q-1)$}
        }
        }
        }
\textbf{EndFunction}
\end{algorithm}

Define $p_{m,c,q} := \Pr_{T\sim\binom{[m]}{q}}[ \chi_T(\by)=\chi_{\secret}(\chi_T(\mX))]$ such that $\mathbf{gap}=2p_{m,c,q}-1$. Note that for $p_{m,c,q}$, we have the following recursion formula:
$$p_{m,c,q}=\frac{m-c}{m}\cdot (1-p_{m-1,c,q-1})+\frac{c}{m}\cdot p_{m-1,c-1,q-1}.$$
There are two boundary cases in function \textsc{ComputeGap}: $m=c$ or $q=0$. When $m=c$, all the labels are correct. In this case we have $p_{m,c,q}=1$. When $q=0$, we do not draw any samples. In this case we also have $p_{m,c,q}=1$. Hence The output of this algorithm is always equal to $p_{m,c,q}$.
Thus,  $\mathbf{gap}=2\cdot(\textsc{ComputeGap}(m,c,q)-1)$ where $c=(\frac{1}{2}+p)m$ is the number of rows with correct labels.

\subsection{Proof of Proposition~\ref{cla:ADDbiasguarantee}}\label{sec:proof_of_ADDBIAS}
One ingredient in the proof of Proposition~\ref{cla:ADDbiasguarantee} is Proposition 5.1 in~\cite{valiant2012finding} about the lower bound of $r$. For completeness, we state it here and provide its proof in the end of this section.
\begin{proposition}
    \label{ratl}
    For $\alpha>0$ with $\alpha=o(1)$ and $s>\sqrt{n}+\alpha n,$     $$\frac{\Pr[B(n,\frac{1}{2})\geq \frac{n}{2}+s]}{\Pr[B(n,\frac{1}{2}+\alpha)\geq \frac{n}{2}+s]}\geq \frac{1}{(1+2\alpha)^{2s}\sqrt{n}} $$
    holds for sufficiently large $n$.
\end{proposition}

\begin{proofof}{Proposition~\ref{cla:ADDbiasguarantee}}
In Algorithm~\ref{alg::gms}, we set $\alpha=\frac{1}{2}\sqrt{\frac{k}{n}}$ and $s=t=\sqrt{3nk\cdot \log \frac{n}{k}}$. Then we have 
$$
r \geq \frac{1}{(1+\sqrt{\frac{k}{n}})^{2\sqrt{3nk\cdot \log \frac{n}{k}} }\cdot\sqrt{n}}\geq \frac{1}{(\frac{n}{k})^{\frac{2\sqrt{3}k}{\sqrt{\log\frac{n}{k}}}}\cdot\sqrt{n}} = \left(\frac{n}{k}\right)^{-\frac{2\sqrt{3}k}{\sqrt{\log\frac{n}{k}}}-\frac{1}{2}}\cdot\frac{1}{\sqrt{k}}\geq \left(\frac{n}{k}\right)^{-o(k)}.
$$

We will show that \( r \cdot \frac{\Pr[B(n,\frac{1}{2}+\alpha) = \textrm{Ones}(\bx)]}{\Pr[B(n,\frac{1}{2}) = \textrm{Ones}(\bx)]} \in [0, 1] \) for \( \textrm{Ones}(\bx) \in \left[\frac{n}{2} - t, \frac{n}{2} + t\right] \), ensuring it represents a valid probability. Additionally, this value is greater than 0. We then prove that \( r \cdot \frac{\Pr[B(n,\frac{1}{2}+\alpha) = \textrm{Ones}(\bx)]}{\Pr[B(n,\frac{1}{2}) = \textrm{Ones}(\bx)]} \leq 1 \). 

    We know 
\begin{align*}
    \frac{\Pr[B(n,\frac{1}{2}+\alpha)=\textrm{Ones}(\bx)]}{\Pr[B(n,\frac{1}{2})=\textrm{Ones}(\bx)]}&=\frac{\binom{n}{\textrm{Ones}(\bx)}(\frac{1}{2}+\alpha)^{\textrm{Ones}(\bx)}(\frac{1}{2}-\alpha)^{n-\textrm{Ones}(\bx)}}{\binom{n}{\textrm{Ones}(\bx)}(\frac{1}{2})^{\textrm{Ones}(\bx)}(\frac{1}{2})^{n-\textrm{Ones}(\bx)}}\\
    &=(1+2\alpha)^{\textrm{Ones}(\bx)}(1-2\alpha)^{n-\textrm{Ones}(\bx)}.
\end{align*}
This value is monotonically increasing with $\textrm{Ones}(\bx)$. Hence $\frac{\Pr[B(n,\frac{1}{2})=\textrm{Ones}(\bx)]}{\Pr[B(n,\frac{1}{2}+\alpha)=\textrm{Ones}(\bx)]}$ is monotonically decreasing with $\textrm{Ones}(\bx)$. Thus, $$r= \frac{\Pr[B(n,\frac{1}{2})>\frac{n}{2}+t]}{\Pr[B(n,\frac{1}{2}+\alpha)>\frac{n}{2}+t]}\leq  \frac{\Pr[B(n,\frac{1}{2})=\frac{n}{2}+t]}{\Pr[B(n,\frac{1}{2}+\alpha)=\frac{n}{2}+t]}.$$
So, we have for $\frac{n}{2}-t\leq \textrm{Ones}(\bx)\leq \frac{n}{2}+t,$
$$r\cdot\frac{\Pr[B(n,\frac{1}{2}+\alpha)=\textrm{Ones}(\bx)]}{\Pr[B(n,\frac{1}{2})=\textrm{Ones}(\bx)]}\leq \frac{\Pr[B(n,\frac{1}{2})=\frac{n}{2}+t]}{\Pr[B(n,\frac{1}{2}+\alpha)=\frac{n}{2}+t]}\cdot\frac{\Pr[B(n,\frac{1}{2}+\alpha)=\textrm{Ones}(\bx)]}{\Pr[B(n,\frac{1}{2})=\textrm{Ones}(\bx)]}\leq 1.$$
\end{proofof}

\begin{proofof}{Proposition~\ref{ratl}}
We first establish a lower bound for the numerator. A straightforward bound is given by 

\[
\Pr\left[B\left(n,\frac{1}{2}\right) \geq \frac{n}{2} + s\right] > \Pr\left[B\left(n,\frac{1}{2}\right) = \frac{n}{2} + s\right] = \binom{n}{\frac{n}{2} + s} \frac{1}{2^n}.
\]

Next, we compute the upper bound for the denominator. Note that for \(s' \geq s\), we have 
\begin{align*}
    \frac{\Pr[B(n,\frac{1}{2}+\alpha)=\frac{n}{2}+s'+1]}{\Pr[B(n,\frac{1}{2}+\alpha)=\frac{n}{2}+s']}&=\frac{\binom{n}{\frac{n}{2}+s'+1}(\frac{1}{2}+\alpha)}{\binom{n}{\frac{n}{2}+s'}(\frac{1}{2}-\alpha)}\\
    &=\frac{n-2s'}{2+n+2s'}\frac{\frac{1}{2}+\alpha}{\frac{1}{2}-\alpha}\\
    &\leq \frac{n-2(\sqrt{n}+\alpha n)}{2+n+2(\sqrt{n}+\alpha n)}\frac{\frac{1}{2}+\alpha}{\frac{1}{2}-\alpha}\\
    &=1-\frac{4\sqrt{n}-4\alpha+2}{(n+2\sqrt{n}+2\alpha n+2)(1-2\alpha)} \leq 1-\frac{1}{\sqrt{n}}.
\end{align*}
Then we can bound 
\begin{align*}
    \sum_{i=\frac{n}{2}+s}^{n}\binom{n}{i}(\frac{1}{2}-\alpha)^{n-i}(\frac{1}{2}+\alpha)^i&\leq \binom{n}{\frac{n}{2}+s}(\frac{1}{2}-\alpha)^{\frac{n}{2}-s}(\frac{1}{2}+\alpha)^{\frac{n}{2}+s}\sum_{i=0}^\infty(1-\frac{1}{\sqrt{n}})^i\\
    &=\binom{n}{\frac{n}{2}+s}(\frac{1}{2}-\alpha)^{\frac{n}{2}-s}(\frac{1}{2}+\alpha)^{\frac{n}{2}+s}\sqrt{n}.
\end{align*}
Combining the two inequalities, the ratio
\begin{align*}
    \frac{\Pr[B(n,\frac{1}{2})\geq \frac{n}{2}+s]}{\Pr[B(n,\frac{1}{2}+\alpha)\geq \frac{n}{2}+\alpha]}&\geq 
    \frac{\binom{n}{\frac{n}{2}+s}\frac{1}{2^n}}{\binom{n}{\frac{n}{2}+s}(\frac{1}{2}-\alpha)^{\frac{n}{2}-s}(\frac{1}{2}+\alpha)^{\frac{n}{2}+s}\sqrt{n}} \geq \frac{1}{(1+2\alpha)^{2s}\sqrt{n}}.
\end{align*}
\end{proofof}

\section{Sparse LPN Algorithms against Adversarial Noise}\label{sec:adv_sparse_LPN}

In this section, we consider the following adversarial noise model. 
\begin{definition}[$(k,\eta)$-Adversarial Sparse LPN]
    In the adversarial setting, each sample $(\bx, y)$ is drawn from a distribution $D$ on $\binom{[n]}{k} \times \mathbf{F}_2$, where $\bx$ is uniformly distributed over $\binom{[n]}{k}$. 
    Moreover, the distribution satisfies $\Pr_{(\bx,y) \sim D}[\langle \bx, \secret \rangle = y] \ge 1 - \eta$ for some unknown $\secret \in \mathbf{F}_2^n$.
\end{definition}

Let $U_n^k$ denote the uniform distribution on $\binom{[n]}{k}$ and let $W$ denote the distribution of $\bx$ conditioned on $y = \langle\bx , \secret\rangle$. 
An observation is that, the total variation distance between $W$ and $\binom{[n]}{k}$ is bounded by $\frac{\eta}{1-\eta}=O(\eta)$.

This section is organized as follows. In Appendix~\ref{sec:adv_noise_upper}, we show a upper bound $(1-\Omega(\eta/k)) \cdot n$ on the number of bits learnable under adversarial noise of rate $\eta$. In Appendix~\ref{sec:adv_noise_Gaussian_eli}, we modify Gaussian elimination to find $ans \in \mathbf{F}_2^n$ in $\tilde{O}(n)$ samples and $e^{\tilde{O}(\eta n)}$ time with Hamming distance $O(\eta /k )\cdot n$ to $\secret$ under adversarial noise of rate $\eta$. This matches the upper bound in Appendix~\ref{sec:adv_noise_upper}. Since we will use this procedure to solve sub-problems after domain reductions, we state it for learning any subset $I \subset [n]$ in Algorithm~\ref{alg::adv_sparseLPN}. Then we state our main result of sparse LPN against adversarial noise as Theorem~\ref{thm:adv_spln_main} in Appendix~\ref{sec:adv_noise_dom_red}. This is an application of our framework of domain reduction to sparse LPN under adversarial noise, similar to Theorem~\ref{thm:infor1} and Theorem~\ref{thm:infor_sparse_LPN}.

\subsection{Upper Bound on the Learnable Bits}\label{sec:adv_noise_upper}
In this subsection, we show that under this adversarial noise model, one can only guarantee learning a $1 - \Omega\left(\frac{\eta}{k}\right)$ fraction of the correct $\secret$.

\begin{theorem}
    Any $\candidate$ with $|\candidate - \secret| \le O\left( \frac{\eta n}{k}\right)$ cannot be distinguished from $\secret$ under the $(k,\eta)$-Adversarial Sparse LPN.\label{thm:adv_slpn_upper}
\end{theorem}
\begin{proofof}{Theorem~\ref{thm:adv_slpn_upper}}
    
Assume we have a $\candidate$ with $|\candidate - \secret| = \ell$. Based on the calculations in Proposition~\ref{lem:ksparse_span}, we have the following estimates on the difference between the parity functions of $\secret$ and $\candidate$:
\begin{equation}   
 \Pr_{\bx \sim \binom{[n]}{k}}[\langle \bx, \candidate \rangle \neq \langle \bx, \secret \rangle] =
    \begin{cases}
       \Theta\left(\frac{k\ell}{n}\right)   & \text{if $\ell \leq \frac{8n}{k}$} \\
       \Theta(1) & \text{if $\ell > \frac{8n}{k}$}
    \end{cases}                
\end{equation}
This means that if $\ell \leq \frac{8n}{k}$ and the adversarial noise rate $\eta \ge \Omega\left(\frac{k\ell}{n}\right)$, the oracle can introduce noise in such a way that the parity of $\candidate$ appears to be the same as the true parity of $\secret$. Therefore, we may not be able to distinguish $\candidate$ from $\secret$.
\end{proofof}

\subsection{Partial Learning Algorithm}\label{sec:adv_noise_Gaussian_eli}

A challenge in the adversarial noise model is that the clean samples may not span the entire space $\F_2^n$, which means Proposition~\ref{lem:ksparse_span} no longer holds, and the Gaussian elimination may not find a single solution. For example, when $\eta > \frac{k}{n}$, the oracle may manipulate all the random samples with $\bx(1) = 1$, preventing the first bit of $\secret$ from being learned.

A key observation is that, even under the adversarial noise model, the clean samples will still span almost the entire space, specifically a subspace of dimension $n(1 - O(\frac{\eta}{k}))$, as proved in Proposition~\ref{prop:adv_slpn_span}. Therefore, we can enumerate the entire solution space and verify the answer without a significant loss in time complexity.

We first present the partial learning algorithm for the adversarial noise model, which is similar to Algorithm~\ref{alg::sparseLPN} with a few modifications marked in blue colors.

\begin{proposition}
\label{prop:adv_slpn}
 Given a subset \( I \subset [n] \) and \( m \ge C \cdot \max\left\{1, \frac{\log |I|}{k}\right\} \cdot |I| \) random samples (for some universal constant \( C \)) drawn from an $(k,\eta)$-Adversarial Sparse LPN oracle with a planted $\secret(I)$, the function \textsc{PartialLearn} in Algorithm~\ref{alg::adv_sparseLPN} outputs \( \candidate \) such that \( |\candidate - \secret(I)| \le O\left(\frac{\eta |I|}{k}\right) \) with probability \( 1 - \frac{O(1)}{|I|^4} \), in time \( (m + |I|^{O(1)}) \cdot e^{O\left(\eta \cdot |I| \cdot \max\left\{1, \frac{\log |I|}{k}\right\}\right)} \).

\end{proposition}

\begin{algorithm}[h]
\SetKwFunction{PartialLearn}{PartialLearn} 
\SetKwProg{Fn}{Function}{:}{}
    \caption{Algorithm for adversarial noise sparse LPN}
    \label{alg::adv_sparseLPN}

\setcounter{AlgoLine}{0}
\KwIn{$I\subset [n],m,\mX \in \mathbf{F}_2^{m \times I},\by \in \mathbf{F}_2^m$, $\eta$}    
\KwOut{$ans \in \mathbf{F}_2^I$ or $\bot$}        
\Fn{\Test{$h$}}{ \tcc{sub-function of \textsc{PartialLearn} to verify that $h \in \mathbf{F}_2^I$ equals $\secret(I)$ given $m,m',\mX$, and $\by$ in \textsc{PartialLearn}}
        
$s \gets \sum_{i=m'+1}^m \mathbf{1}\big( \langle \bx_i, h \rangle \neq y_i \big)$ 
                 
\If{$s \le (m-m')\cdot \textcolor{blue}{\max\left\{\frac{k}{2n}, 2\eta\right\}} $} {        
       \KwRet{True}
    } 
}
\textbf{EndFunction}

        \Fn{\PartialLearn{$I,m,\mX,\by,\eta$}} {
        
        \tcc{the goal is to learn $\secret_I$ given $m$ samples $(\bx_i,y_i)$ in $(\mX,\by)$ with $supp(\bx_i) \subset I$} 
        
       $q \gets |I|$
       
        $C \gets $ a sufficiently large constant
        
        
         If {$m<C \cdot q \cdot \max\{1,\frac{\log q}{k}\}$}, \KwRet{$\bot$}
        
        $d \gets c_1 \cdot q \cdot \max\{1,\frac{\log q}{k}\}$ is the number of required samples from Proposition~\ref{prop:adv_slpn_span}
        
         $m' \gets 100 \cdot d$ \qquad \Comment{Use the first $m'$ samples in $(\mX,\by)$ to try Gaussian eliminations and the rest $(m-m')$ samples for verification}

         $L \gets 10 q^2 + 10 \log q \cdot e^{\frac{200c_1}{99} \cdot \eta q \max\{1,\frac{\log q}{k}\}}$ \Comment{Rounds of Gaussian eliminations}
        
        \For{$i=1,\ldots,L$}{
            Pick a random set of samples: $T \sim {[m'] \choose d}$ 
            
            \If{\textcolor{blue}{rank$\{\mX(T,I)\}\ge |I|(1-10 \frac{\eta}{k})$ }} {
            
                Apply Gaussian elimination to \textcolor{blue}{find all possible} $\candidate \in \mathbf{F}_2^I$ such that
                    $$\mX(T,I) \cdot \candidate = \by(T)$$
                    
               \textcolor{blue}{Enumerate all possible $\candidate$ and find one (called $ans$) s.t. \textsc{Test}($ans$)=True}
               
               \KwRet{$ans$}
            }
}
        \KwRet{$\bot$}
      }
\textbf{EndFunction}

\end{algorithm}


\begin{proposition}
\label{prop:adv_slpn_span}
For any odd number \( k < n / \log n \), let \( t = c_1 \cdot n \cdot \max\left\{1, \frac{\log n}{k}\right\} \), and let the \( t \) random \( k \)-sparse vectors \( \bx_1, \bx_2, \dots, \bx_t \) be drawn i.i.d. from a distribution \( W \), satisfying \( \|W - U_n^k\|_{\text{TV}} \leq O(\eta) \). Then, these vectors span a linear space of dimension at least \( n(1 - 10\frac{\eta}{k}) \) with probability at least \( 1 - n^{-5} \) (for some constant \( c_1 \)).
\end{proposition}

The proof of Proposition~\ref{prop:adv_slpn_span} is attached to Subsection~\ref{sec:adv_slpn_span}. Next, we show in Corollary~\ref{cor:adv_slpn_validation} that any $\candidate$ far from $\secret$ cannot pass \textsc{Test}($\candidate$).


\begin{corollary}[Validation of $\candidate$]
\label{cor:adv_slpn_validation}
Assume we have $m = c_2 \cdot n \cdot \max\left\{1, \frac{\log n}{k}\right\}$ random samples drawn from a $(k,\eta)$-Adversarial Sparse LPN oracle, where $k$ is odd, and the testing threshold is $t := \max\{\frac{k}{2n}, 2\eta\}$.

Then, for sufficiently large constants $c_2$ and $c_3$, with probability $1 - 1/n^5$, any candidate $\candidate \in \F_2^n$ satisfying $|\candidate - \secret| \ge c_3 \cdot \frac{\eta n}{k}$ will fail more than $t m$ tests. Furthermore, $\secret$ will fail at most $t m$ tests.
\end{corollary}

The proof of Corollary~\ref{cor:adv_slpn_validation} is attached to Subsection~\ref{sec:adv_slpn_validation}. Now, we are ready to present the proof of Proposition~\ref{prop:adv_slpn}.

\begin{proofof}{Proposition~\ref{prop:adv_slpn}}

Consider the first $m'$ random samples in $(\mX,\by)$. Let $\mathbf{T_c} \subset [m']$ denote the set of samples with correct labels, i.e., $\{i \in [m']: \langle \bx_i,\secret \rangle =y_i\}$. Similarly, let $\mathbf{T_i} \subset [m']$ denote the set of incorrect samples. 

Similar to the calculation of  Proposition~\ref{clm:decode_subset}, we have that by randomly sampling $ L := 10 q^2 + 10 \log q \cdot e^{\frac{200c_1}{99} \cdot \eta q \max\left\{1, \frac{\log q}{k}\right\}} $ sets of samples, we can find at least one $T\subset \mathbf{T_c}$ with probability $1-\frac{1}{q^5}$.

Conditioned on \( T \subset \mathbf{T_c} \), Proposition~\ref{prop:adv_slpn_span} shows that the event \( \text{rank}\{\mX(T, I)\} \ge |I|(1 - 10 \frac{\eta}{k}) \) occurs with high probability \( 1 - q^{-5} \). Since these \( T \) samples are clean, we know that \( \secret(I) \) must be one of the candidates in the solution space \( \{\candidate \mid \mX(T, I) \cdot \candidate = \by(T)\} \). Therefore, at least one \( \candidate \) must pass the \textsc{Test}($\candidate$). Moreover, by Corollary~\ref{cor:adv_slpn_validation}, with probability \( 1 - q^{-5} \), any \( \candidate \) with \( |\candidate - \secret(I)| > c_3 \cdot \left(\frac{\eta n}{k}\right) \) may not pass \textsc{Test}($\candidate$). This proves the correctness of Algorithm~\ref{alg::adv_sparseLPN}.

Combining the time to find a clean batch of samples, enumerate the solution space of dimension \( O\left(\frac{n \eta}{k}\right) \), and verify the \( \candidate \), the total complexity is \( L \cdot 2^{O\left(\frac{n \eta}{k}\right)} \cdot (m + |I|^{O(1)}) = (m + |I|^{O(1)}) \cdot e^{O\left(\eta \cdot |I| \cdot \max\left\{1, \frac{\log |I|}{k}\right\}\right)} \). This gives the desired time complexity.
\end{proofof}

\subsection{Applying the Domain Reduction}\label{sec:adv_noise_dom_red}
In this section, we apply the domain reduction framework to obtain a learning algorithm for the $(n,k,\eta)$-Adversarial Sparse LPN. The main difference between Algorithm~\ref{alg::main_adv_sparseLPN} and Algorithm~\ref{alg::sparseLPN_main} is that, instead of using a fixed partition \( I_1, I_2, \dots I_t\), we randomly sample several subsets \( I \sim \binom{[n]}{q} \), solve the sub-problem on \( I \), and perform a majority vote to merge their answers. These modifications were made to eliminate the attack that extensively manipulates the samples contained in some set \( I_i \) of the fixed partition.

\begin{algorithm}[h]
    \caption{Main Algorithm for adversarial sparse LPN}
    \SetKwFunction{Main}{Main} 
\SetKwProg{Fn}{Function}{:}{}
    \label{alg::main_adv_sparseLPN}
\setcounter{AlgoLine}{0}
\KwIn{$k$, $\eta$, and a parameter $\delta \in (0,1)$}    
\KwOut{$\mathbf{ans} \in \mathbf{F}_2^n$}   
\Fn{\Main{}}{
$q\gets n^{\frac{1+\delta}{2}}$

$t\gets c_1\cdot\max\{\frac{n}{q},\frac{\eta\cdot q}{k^2}\}\cdot \log n$

$m\gets 2C \cdot \max\{1, \frac{\log q}{k}\} \cdot q (en/q)^k$


Randomly sample $I_1,...,I_t$ from $\binom{[n]}{q}$\Comment{Sample sub-problems}

Initialize $m_j=0$ and matrix $\mX_j\in \mathbf{F}_2^{m_j\times n}$ and $\by_j\in \mathbf{F}_2^{m_j}$ for each $j\in [t]$.\Comment{$\mX_j$ records random samples for sub-problem $I_j$}

\For{$i=1,2,...,m$}{
\tcc{Allocate $m$ random samples into $t$ sub-problems}

Take a fresh sample $(\bx,y)$

\For{all $j$ s.t. $supp(\bx)\subseteq I_j$}{
$m_j\gets m_j+1$

Store $\bx$ into a new row: $\mX_j(m_j, )\gets \bx$ and $\by_j(m_j)\gets y$
}

}

\tcc{Majority vote}
$\mathbf{vote}\gets 0^{2\times n}$\Comment{record the voting results from sub-problems}

\For{$j=1,...,t$}{
  $ans_{I_j}\gets \textsc{PartialLearn}(I_j,m_j,\mX_j,\by_j,3\eta)$

\For{$i=1,...,q$}{
$\mathbf{vote}(ans_{I_j}(i),I_j(i))\gets \mathbf{vote}(ans_{I_j}(i),I_j(i))+1$
}

  }

$\mathbf{ans}\gets 0^{n}$\Comment{Final solution for the problem}

\For{$i=1,...,n$}{
If $\mathbf{vote}(0,i)\geq \mathbf{vote}(1,i)$, $\mathbf{ans}(i)\gets 0$; otherwise $\mathbf{ans}(i)\gets 1$
}

       \KwRet{$\mathbf{ans}$}

}
\textbf{EndFunction}
\end{algorithm}

\begin{theorem}
\label{thm:adv_spln_main}
       For any $(n,k,\eta)$-Adversarial Sparse LPN, given any $\delta \in (0,1)$, there is a learning algorithm that takes $m:=\max \big\{1, \frac{\log n}{k} \big\} \cdot O(n)^{1+(1-\delta) \cdot \frac{k-1}{2}}$ samples to return $\mathbf{ans}$ in time $e^{\tilde{O}(\eta \cdot n^{\frac{1+\delta}{2}}})$ such that $|\mathbf{ans}-\secret|\le O(n) \cdot (\frac{\eta}{k} + \frac{k^2}{\eta \cdot n^{\frac{1+\delta}{2}}})$ with probability $1-o(1)$ .
\end{theorem}

\begin{remark}
By substituting the Chebyshev bound in Proposition~\ref{prop:concentrate_etaI} with the Markov bound, and allowing \( \eta_I = O\left( \sqrt{\eta k} \right) \), one can obtain an algorithm that takes $m:=\max \big\{1, \frac{\log n}{k} \big\} \cdot O(n)^{1+(1-\delta) \cdot \frac{k-1}{2}}$ samples to return $\mathbf{ans}$ in time $e^{\tilde{O}(\sqrt{\eta k} \cdot n^{\frac{1+\delta}{2}}})$ such that $|\mathbf{ans}-\secret|\le O(n) \cdot \sqrt{\frac{\eta}{k}}$ with probability $1-o(1)$ .
\end{remark}
 
To prove this theorem, we use the following propositions.

\begin{proposition}
    \label{prop:concentrate_etaI}
    Let \( D \) denote the distribution of the $(k,\eta)$-Adversarial Sparse LPN.
    For any \( I \in \binom{[n]}{q} \), let \( D_I \) be the distribution of \( (\bx, y) \) obtained by restricting \( \bx \) to \( \binom{[I]}{k} \). Define 
    \[
    \eta_I := \Pr_{(\bx, y) \sim D_I}[\langle \bx, \secret \rangle \neq y].
    \]
    Then, we have:
    \[
    \Pr_{I \sim \binom{[n]}{|I|}}\left[\eta_I  \geq 3\eta\right] \leq O\left(\frac{k^2}{\eta I}\right).
    \]
\end{proposition}

We defer the proof of Proposition~\ref{prop:concentrate_etaI} to Subsection~\ref{subsec:proof_prop_con_etaI}. Now, we begin to prove Theorem~\ref{thm:adv_spln_main}.

\begin{proofof}{Theorem~\ref{thm:adv_spln_main}}

\paragraph{Sufficient Samples for Each Subset.}
First, we show that \( m_j \geq C \cdot q \cdot \max\left\{1, \frac{\log q}{k}\right\} \) for every \( j \in [t] \) to satisfy the sample requirement for each sub-problem. Let us set \( m = 2C \cdot \max\left\{1, \frac{\log q}{k}\right\} \cdot q \left(\frac{en}{q}\right)^k \), where \( q = n^{\frac{1 + \delta}{2}} \). By a similar calculation to that in Theorem~\ref{thm:sparse_LPN}, we have, with probability at least \( 1 - \frac{1}{n^2} \), that \( m_j \geq C \cdot q \cdot \max\left\{1, \frac{\log q}{k}\right\} \) for every \( j \in [t] \).




\paragraph{Many Subsets with Low Noise.} 
Secondly, define \( \mathbf{L_1} := \{i \mid \eta_{I_i} \leq 3\eta\} \) and \( \mathbf{L_2} := [t] \setminus \mathbf{L_1} \). By applying Proposition~\ref{prop:adv_slpn} to all subsets \( I_j \) indexed by \( j \in \mathbf{L_1} \), we know that \textsc{PartialLearn}$(I_j, m_j, \mX_j, \by_j, 3\eta)$ will return \( ans_{I_j} \) such that \( |ans_{I_j} - \secret(I_j)| \leq O\left(\frac{\eta \cdot q}{k}\right) \) with probability \( 1 - \frac{O(1)}{q^4} \). By applying the union bound over \( \mathbf{L_1} \subset [t] \), we obtain that with probability \( 1 - \frac{1}{q} \), for all \( j \in \mathbf{L_1} \), \textsc{PartialLearn}$(I_j, m_j, \mX_j, \by_j, 3\eta)$ will return \( ans_{I_j} \) such that \( |ans_{I_j} - \secret(I_j)| \leq O\left(\frac{\eta \cdot q}{k}\right) \).

For \( j \in \mathbf{L_2} \), we assume \textsc{PartialLearn}$(I_j, m_j, \mX_j, \by_j, 3\eta)$ will return arbitrarily. By Proposition~\ref{prop:concentrate_etaI}, we have \( \E[|\mathbf{L_2}|] = O\left(\frac{t\cdot  k^2}{\eta \cdot q}\right) \). Now, we prove that \( |\mathbf{L_2}| \leq O\left(\frac{t\cdot  k^2}{\eta \cdot q}\right) \) with high probability.

There are two cases for $\E[\mathbf{L_2}]$:
\begin{enumerate}
    \item If $\E[|\mathbf{L_2}|]\geq 8\frac{t\cdot k^2}{\eta \cdot q}$, then by the Chernoff bound (part 2 of Theorem~\ref{THM:Chernoff}) and our choice of $t= c_1\cdot\max\{\frac{n}{q},\frac{\eta\cdot q}{k^2}\}\cdot \log n$:
    $$\Pr\left[|\mathbf{L_2}-\E[\mathbf{L_2}]|>\frac{1}{2}\E[|\mathbf{L_2}|]\right]\leq 2e^{-\frac{15t\cdot k^2}{8\eta \cdot q}}\leq \frac{1}{n^3}.$$
    \item If $\E[|\mathbf{L_2}|]< 8\frac{t\cdot k^2}{\eta \cdot q}$, then we choose $\epsilon\geq 2$ such that $\epsilon\cdot \E[|\mathbf{L_2}|]=16\frac{t\cdot k^2}{\eta\cdot q}$. By the Chernoff bound (part 2 of Theorem~\ref{THM:Chernoff}) and our choice of $t= c_1\cdot\max\{\frac{n}{q},\frac{\eta\cdot q}{k^2}\}\cdot \log n$:
    $$\Pr\left[\mathbf{L_2}>(1+\epsilon)\cdot\E[|\mathbf{L_2}|]\right]\leq e^{-\frac{16t\cdot k^2}{2\eta \cdot q}}\leq \frac{1}{n^3}.$$
\end{enumerate}

Thus, we obtain with probability at least $1-\frac{1}{n^3}$, $|\mathbf{L_2}|\leq O(\frac{t\cdot k^2}{\eta\cdot q})$.

\paragraph{Correctness of Majority Voting.} 
Thus, with probability at least \( 1 - \frac{1}{n} \), the number of incorrect votes is bounded by
\[
O\left(\frac{t \cdot k^2}{\eta \cdot q} \cdot q + \frac{\eta \cdot q}{k} \cdot t \right) = O\left(t \cdot \left(\frac{k^2}{\eta} + \frac{\eta \cdot q}{k}\right)\right).
\]

Then, for a fixed index \( i \in [n] \), it is contained in a random set \( I \) with probability \( \frac{q}{n} \). The expected number of random subsets containing \( i \) is \( \frac{tq}{n} \). 
For each \( i \in [n] \), define \( K_i := \{j \in [t] \mid i \in I_j \} \). By the Chernoff bound (part 2 of Theorem~\ref{THM:Chernoff}) and our choice of \( t = c_1 \cdot \max\left\{\frac{n}{q}, \eta \cdot q\right\} \cdot \log n \), we have
\[
\Pr\left[\left| |K_i| - \frac{tq}{n} \right| \geq \frac{1}{2} \frac{tq}{n} \right] \leq \frac{1}{n^3}.
\]
Then, by the union bound over \( [n] \), with probability at least \( 1 - \frac{1}{n^2} \), for all \( i \in [n] \), \( |K_i| \in \left(\frac{1}{2}, \frac{3}{2}\right) \frac{tq}{n} \).

To make a wrong majority vote for \( i \in [n] \), it needs to receive at least \( \frac{1}{4} \frac{tq}{n} \) incorrect votes. Thus, the total number of wrong positions is at most
\[
O\left(\frac{\frac{t \cdot k^2}{\eta} + \frac{\eta \cdot qt}{k}}{\frac{1}{4} \frac{tq}{n}}\right) = O(n) \cdot \left(\frac{k^2}{\eta \cdot q} + \frac{\eta}{k}\right).
\]

\paragraph{Time and Sample Complexity.}
The running time is $O(m \cdot t) + t \cdot \textbf{TIME}(\textsc{PartialLearn})=n^{O(1)} \cdot \big( e^{O(\eta \cdot n^{\frac{1+\delta}{2}} \cdot \max\{1, \frac{\log n}{k}\})} + m \big)$ by Proposition~\ref{prop:adv_slpn} with $|I|=q=n^{\frac{1+\delta}{2}}$.

Finally, by plugging in  $q=n^{\frac{1+\delta}{2}}$ into the sample complexity, we get  $m=2eC  \max\{1, \frac{\log n}{k}\} \cdot n \cdot     O(n^{\frac{1-\delta}{2}})^{k-1} $.
\end{proofof}

\subsection{Proof of Proposition~\ref{prop:adv_slpn_span}}
\label{sec:adv_slpn_span}
We uses the following proposition on linear subspace.
\begin{proposition}
    Assume $n\ge 3(d-1)$, and let $V\subset \mathbf{F}_2^n$ be a linear subspace of dimension $d$. Then, there exists a vector $\bx\in V$ such that $d \le |\bx| \le n-d+1$. 
    \label{prop:moderate_weight_vector}
\end{proposition}
\begin{proofof}{Proposition~\ref{prop:moderate_weight_vector}}
\label{sec:moderate_weight_vector}
Choose a basis $\bb_1,\bb_2,\ldots,\bb_d$ of $V$ in reduced row echelon form. This guarantees that for every $i$, 
\[
|\bb_i|\le n-(d-1)=n-d+1.
\]

We consider two cases.
\begin{description}
    \item[Case 1:] There is an index $i$ with $|\bb_i|\in [d, n-d+1]$. In this case, the basis vector $\bb_i$ satisfies the claim.
    \item[Case 2:] For every $i$, $|\bb_i|\in [1, d-1]$. Define the partial sums
\[
\bx_k=\sum_{j=1}^{k}\bb_j, \quad 1\le k\le d.
\]
Since each vector $\bb_i$ is of hamming weight less than $d$, we have:
\[
||\bx_i| -|\bx_{i-1}||\le |\bb_i| <d
\]
Because the basis is in reduced row echelon form, we know that:
\[
n \ge|\bx_d| \ge d, |\bx_0|=0.
\]
Thus, by the discrete intermediate value theorem, there exists an integer $k\in [d]$ where the partial sum $|\bx_{k}|\in [d,n-d+1]$.
\end{description}
\end{proofof}

We will now proceed with the proof of Proposition~\ref{prop:adv_slpn_span}.

Let \( d = 10 \frac{\eta n}{k} \) and let \( S \) denote the span of the \( t \) random vectors. We will prove that no linear subspace \( V \) of dimension \( d \) is contained in the orthogonal complement \( S^{\perp} \).

By Proposition~\ref{prop:moderate_weight_vector}, for any subspace \( V \) of dimension \( d \) there exists a vector \( \bv \in V \) with Hamming weight \( \ell := |\bv| \) satisfying \( d \le \ell \le n-d+1 \). Considering the inner product of \( \bv \) with the \( t \) random vectors, if there is some \( \bx_i \) such that \( \langle \bv, \bx_i \rangle = 1 \), then \( \bv \notin S^{\perp} \) and hence \( V \not \subseteq S^{\perp} \).

It remains to show that, with high probability, any vector \( \bv \) with moderate Hamming weight $\ell$ (i.e., \( d \le \ell \le n-d+1 \)) has a nonzero inner product with at least one of the \( t \) random vectors. Based on the calculations in Proposition~\ref{lem:ksparse_span}, we have the following estimates for a given \( \bv \):
\begin{enumerate}
    \item When \( \ell \in [d, 8n/k) \), one obtains \( \Pr_{\bx_i}[\langle \bx_i, \bv\rangle=0] \le 1 - \Omega(k\ell/n) \).
    \item When \( \ell \in [8n/k, n-8n/k] \), the probability \( \Pr_{\bx_i}[\langle \bx_i, \bv\rangle=0] \) lies in the interval \([0.1, 0.9]\).
    \item When \( \ell \in (n-8n/k, n-d+1] \), one has \( \Pr_{\bx_i}[\langle \bx_i, \bv\rangle=0] \le 1-e^{-9} \).
\end{enumerate}

These probabilities were computed under the assumption that \( \bx_i \sim \binom{[n]}{k} \). However, since \( \|\binom{[n]}{k} - W\|_{TV} \le 2\eta \) and \( \ell \ge d \) implies \( \frac{k\ell}{n} \ge 10\eta \), the probabilities remain nearly unchanged even if $\bx_i \sim W$.

Finally, applying a union bound over all vectors with Hamming weight $\ell\in [d,n-d+1]$ (as in Proposition~\ref{lem:ksparse_span}) completes the proof.

\subsection{Proof of Corollary~\ref{cor:adv_slpn_validation}}
\label{sec:adv_slpn_validation}
Recall that the testing threshold is $t := \max\left\{\frac{k}{2n}, 2\eta\right\}$, and the number of samples is $m = c_2 \cdot n \max\left\{1, \frac{\log n}{k}\right\}$.
Assume that there are exactly $\eta' m$ false samples among the $m$ samples. By the Chernoff Bound (part 2 of Theorem~\ref{THM:Chernoff}):
\begin{align}
    \Pr[\eta' > t] \le \exp\left(-\frac{\eta m \max\left\{1, \frac{k}{2n\eta} - 1\right\}^2}{2 + \max\left\{1, \frac{k}{2n\eta} - 1\right\}}\right)
    \le
    \begin{cases}
    \exp\left(-\frac{\eta m}{3}\right) & \text{if } \eta \ge \frac{k}{4n}, \\
    \exp\left(-\frac{\eta m}{3} \cdot \frac{k}{4n\eta}\right) & \text{if } \eta < \frac{k}{4n}.
    \end{cases}
\end{align}
Therefore, when $m \ge \min\left\{ 15 \frac{\log n}{\eta}, 60 \frac{n \log n}{k} \right\}$, the above probability is bounded by $n^{-5}$.

By definition, the number of failed tests is
\begin{align*}
\sum_{i=1}^m \mathbf{1}\Bigl(y_i\neq\langle \candidate,\bx_i\rangle\Bigr)
&=\sum_{i=1}^m \mathbf{1}\Bigl(\langle \secret-\candidate,\bx_i\rangle+e_i\neq 0\Bigr)\\
&\ge \sum_{i=1}^m\left(\mathbf{1}\Bigl(\langle \secret-\candidate,\bx_i\rangle\neq 0\Bigr)-\mathbf{1}(e_i=1)\right)\\
&\ge \sum_{i=1}^m \mathbf{1}\Bigl(\langle \secret-\candidate,\bx_i\rangle\neq 0\Bigr)-\eta'm.
\end{align*}

Let $\bv:= \secret-\candidate$ and $\ell:=|\bv|$. Since $\eta'\le t$ w.h.p, it remains to show that 
\[
\sum_{i=1}^m \mathbf{1}\Bigl(\langle \bv,\bx_i\rangle\neq 0\Bigr) \ge 2 t m
\]
for any $\candidate$ far from $\secret$. We discuss two cases based on the calculation in Proposition~\ref{lem:ksparse_span}:

\textbf{Case 1.} When $\ell\in [c_3\cdot (\eta n/k),\, 8n/k)$, we have $\Pr_{\bx_i}\Bigl[\langle \bx_i, \bv\rangle\neq 0\Bigr]=\Omega(k\ell/n).
$ Thus, by the Chernoff bound (second part of Theorem~\ref{THM:Chernoff}), for a fixed $\bv$ we obtain
\[
\Pr\Biggl[\sum_{i=1}^m \mathbf{1}\Bigl(\langle \bv,\bx_i\rangle\neq 0\Bigr) < 2t m\Biggr] \le \exp\Bigl(-\Omega\left(\frac{k\ell m}{n}\right)\Bigr),
\]
which is at most $\exp(-\Omega(\ell\log n))$ when $c_2,c_3$ are sufficiently large.

\textbf{Case 2.} When $\ell\in [8n/k,\, n]$, we have $\Pr_{\bx_i}[\langle \bx_i,\bv\rangle\neq 0]=\Omega(1)$. Then, for a fixed $\bv$,
\[
\Pr\Biggl[\sum_{i=1}^m \mathbf{1}\Bigl(\langle \bv,\bx_i\rangle\neq 0\Bigr) < 2t m\Biggr]
\le \binom{m}{2t m}(1-\Omega(1))^{m(1-2t)}.
\]
Using the standard bound $\binom{m}{2t m}\le \left(\frac{e}{2t}\right)^{2t m}$, we deduce that
\[
\Pr\Biggl[\sum_{i=1}^m \mathbf{1}\Bigl(\langle \bv,\bx_i\rangle\neq 0\Bigr) < 2t m\Biggr]
\le \Bigl(\left(\frac{e}{2t}\right)^{2t}(1-\Omega(1))^{1-2t}\Bigr)^m
= (1-\Omega(1))^{\Omega(m)},
\]
which is at most $\exp(-\Omega(n))$ for sufficiently large $c_2$ and $t =o(1)$.

Finally, a union bound over all possible $\bv$ with $\ell\ge c_3\cdot (\eta n/k)$ gives
\[
\sum_{\ell=c_3\cdot (\eta n/k)}^n \sum_{\bv:|\bv|=\ell} \Pr\Biggl[\sum_{i=1}^m \mathbf{1}\Bigl(\langle \bv,\bx_i\rangle\neq 0\Bigr) < 2t m\Biggr]
\le \sum_{\ell=c_3\cdot (\eta n/k)}^{8n/k-1} \binom{n}{\ell}e^{-\Omega(\ell\log n)}
+\sum_{\ell=8n/k}^{n} \binom{n}{\ell} e^{-\Omega(n)} = o(1).
\]
This completes the proof.

\subsection{Proof of Proposition~\ref{prop:concentrate_etaI}}
    \label{subsec:proof_prop_con_etaI}

Let \( e_{\bz} \) denote the noise rate over a fixed vector \( \bz \), i.e., \( \Pr_{(\bx, y) \sim D}[y \neq \langle \bx, \secret \rangle \mid \bx = \bz] \). By definition, we have
\[
\eta := \E_{\bx \sim \binom{[n]}{k}}[e_{\bx}] = \sum_{\bx \in \binom{[n]}{k}} \frac{e_{\bx}}{\binom{n}{k}},
\]
and 
\[
\eta_I := \E_{\bx \sim \binom{I}{k}}[e_{\bx}] = \sum_{\bx \in \binom{I}{k}} \frac{e_{\bx}}{\binom{|I|}{k}}.
\]

We now calculate \( \E[\eta_I] \):
\[
    \E_{I \sim \binom{[n]}{|I|}}[\eta_I] = \E_{I \sim \binom{[n]}{|I|}}\left[\E_{\bx \sim \binom{I}{k}}[e_{\bx}]\right] 
    = \E_{I \sim \binom{[n]}{|I|}, \bx \sim \binom{I}{k}}[e_{\bx}] 
    = \E_{\bx \sim \binom{[n]}{k}}[e_{\bx}] 
    = \eta.
\]

Next, we bound the variance of \( \eta_I \). 

We denote \( \textbf{idx} \) as a fixed permutation of \( \binom{[|I|]}{k} \). We then denote \( \textbf{I}_p \) as a random permutation of the set \( I \). Finally, we generate all possible vectors \( \bx_1, \bx_2, \dots, \bx_{\binom{|I|}{k}} \) whose support is restricted to \( I \) as follows:
\[
\bx_{i} := \{\textbf{I}_p(j) \mid j \in \textbf{idx}(i) \}.
\]
Note that the intersection relationships between different \( \bx_i \)'s are fixed through the permutation \( \textbf{idx} \), and are independent of the process of randomly generating \( I \) and \( \textbf{I}_p \). Additionally, the generation of \( \textbf{I}_p \) is only for the ease of analysis.

Thus we can write the variance of $\eta_I$ as:
\begin{equation}
    \Var_{I \sim \binom{[n]}{|I|}}(\eta_I) = \binom{|I|}{k}^{-2}\left( \sum_{i=1}^{\binom{|I|}{k}} \Var_{I \sim \binom{[n]}{|I|}}(e_{\bx_i}) + 2 \sum_{i=1}^{\binom{|I|}{k}} \sum_{j=i+1}^{\binom{|I|}{k}} \cov_{I \sim \binom{[n]}{|I|}}(e_{\bx_i}, e_{\bx_j}) \right). \label{eq:eta_I_var}
\end{equation}

Next, we bound the term \( \cov_{I \sim \binom{[n]}{|I|}}(e_{\bx_i}, e_{\bx_j}) \). Let \( p := \Pr_{\by, \bz \sim \binom{[n]}{k}}[\by \cap \bz \neq \emptyset] = O\left(\frac{k^2}{n}\right) \).

\begin{align}
    \left|\cov_{I \sim \binom{[n]}{|I|}}(e_{\bx_i}, e_{\bx_j})\right| &= \left|\E_{I \sim \binom{[n]}{|I|}}[e_{\bx_i} \cdot e_{\bx_j}] - \E_{I \sim \binom{[n]}{|I|}}[e_{\bx_i}] \cdot \E_{I \sim \binom{[n]}{|I|}}[e_{\bx_j}]\right| \\
    &= \left|\E_{I \sim \binom{[n]}{|I|}}[e_{\bx_i} \cdot e_{\bx_j}] - \eta^2\right| \\
    &\leq \begin{cases}
        \eta & \text{if } \bx_i \cap \bx_j \neq \emptyset, \\
        2p \cdot \eta & \text{if } \bx_i \cap \bx_j = \emptyset.
    \end{cases} \label{eq:eta_I_cov}
\end{align} 

We divide the discussion into two parts:

\begin{enumerate}
    \item \( |\bx_i \cap \bx_j| = t > 0 \). Under this condition,
    \[
    \E_{I \sim \binom{[n]}{|I|}}[e_{\bx_i} \cdot e_{\bx_j}] = \E_{\by, \bz \sim \binom{[n]}{k}}[e_{\by} \cdot e_{\bz} \mid |\by \cap \bz| = t].
    \]
    Since \( e_{\bz} \leq 1 \), we obtain
    \begin{equation}
        \E_{\by, \bz \sim \binom{[n]}{k}}[e_{\by} \cdot e_{\bz} \mid |\by \cap \bz| = t] \leq \E_{\by, \bz \sim \binom{[n]}{k}}[e_{\by} \mid |\by \cap \bz| = t] = \E_{\by \sim \binom{[n]}{k}}[e_{\by}] = \eta.\label{eq:intersec_e}
    \end{equation}
    Thus, we prove the result.
    
    \item \( |\bx_i \cap \bx_j| = 0 \). Under this condition,
    \begin{align}
        \E_{I \sim \binom{[n]}{|I|}}[e_{\bx_i} \cdot e_{\bx_j}] &= \E_{\by, \bz \sim \binom{[n]}{k}}[e_{\by} \cdot e_{\bz} \mid \by \cap \bz = \emptyset] \\
        &= \frac{\left(\E_{\by, \bz \sim \binom{[n]}{k}}[e_{\by} \cdot e_{\bz}] - p \cdot \E_{\by, \bz \sim \binom{[n]}{k}}[e_{\by} \cdot e_{\bz} \mid \by \cap \bz \neq \emptyset]\right)}{1 - p} \\
        &= \eta^2 + \frac{p}{1 - p} \cdot \left(\eta^2 - \E_{\by, \bz \sim \binom{[n]}{k}}[e_{\by} \cdot e_{\bz} \mid \by \cap \bz \neq \emptyset]\right).\label{eq:disjoint_e}
    \end{align}
    The second equality follows from the definition of conditional expectation. Combining \eqref{eq:disjoint_e} and \eqref{eq:intersec_e} with \( p \leq \frac{1}{2} \), we obtain:
    \[
    \left|\E_{I \sim \binom{[n]}{|I|}}[e_{\bx_i} \cdot e_{\bx_j}] - \eta^2\right| \leq 2p \eta.
    \]
\end{enumerate}

The term \( \Var_{I \sim \binom{[n]}{|I|}}(e_{\bx_i}) \) can be bounded by \( \E_{I \sim \binom{[n]}{|I|}}(e_{\bx_i}) \). Combining this fact with \eqref{eq:eta_I_var} and \eqref{eq:eta_I_cov}, the variance of \( \eta_I \) can be bounded by:
\begin{align*}
    \Var_{I \sim \binom{[n]}{|I|}}(\eta_I) &= \binom{|I|}{k}^{-2} \left( \sum_{i=1}^{\binom{|I|}{k}} \Var_{I \sim \binom{[n]}{|I|}}(e_{\bx_i}) + 2 \sum_{i=1}^{\binom{|I|}{k}} \sum_{j=i+1}^{\binom{|I|}{k}} \cov_{I \sim \binom{[n]}{|I|}}(e_{\bx_i}, e_{\bx_j}) \right) \\
    &\leq \binom{|I|}{k}^{-2} \left( \binom{|I|}{k} \cdot \eta + \underbrace{\binom{|I|}{k} \left( \sum_{j=1}^{k-1} \binom{|I|-k}{j} \binom{k}{k-j} \right) \cdot \eta}_{\text{Intersecting pairs}} + \underbrace{\binom{|I|}{k}^2 \cdot 2p \eta}_{\text{Non-intersecting pairs}} \right) \\
    &\leq \binom{|I|}{k}^{-1} \cdot \eta + \binom{|I|}{k}^{-1} \left( \sum_{j=1}^{k-1} \binom{|I|-k}{j} \binom{k}{k-j} \right) \cdot \eta + 2p \eta.
\end{align*}
Since the summation \( \sum_{j=1}^{k-1} \binom{|I|-k}{j} \binom{k}{k-j} \) is dominated by \( j = k-1 \), and \( p = O\left(\frac{k^2}{n}\right) \), the above can be bounded by:
\begin{align*}
    \Var_{I \sim \binom{[n]}{|I|}}(\eta_I) \leq O\left(\frac{k^2 \eta}{|I|}\right).
\end{align*}

By Chebyshev's bound (Theorem~\ref{thm:chebyshev}),
\begin{equation}
    \Pr_{I \sim \binom{[n]}{|I|}}\left[\eta_I \geq 3\eta\right] \leq \frac{\Var(\eta_I)}{(2\eta)^2} \leq O\left(\frac{k^2}{\eta |I|}\right).
\end{equation}

\section{Distinguishing Algorithm for sparse LPN}\label{sec:distinguish}
We extend the learning algorithm in Section~\ref{sec:sparse_sample} to a distinguishing algorithm in this section. We provide a definition of the distingshing problem here.
\begin{definition}[Distinguishing problem of sparse LPN]
    In a distinguishing problem of sparse LPN, the input is generated with equal probabilities from the following two cases: (1) the random samples $(\bx_i,\by_i)_{i=1,\ldots,m}$ are generated from a sparse LPN instance; (2) each $\bx_i \sim \mathbf{F}_2^n$ and its label $\by_i \sim \mathbf{F}_2$ independently. Then a distinguishing algorithm outputs 1 or 2 to indicate which case it is.
\end{definition}

Then we state our distinguishing algorithm for sparse LPN, which could be viewed as a significant extension of \cite{DJ24} (which has a similar time complexity only for $\eta \le n^{-\frac{1+\delta}{2}}$).
\begin{theorem}\label{thm:distinguish}
    Given any $n$, $k$, and $\eta$, if $\delta \in (0,1)$ satisfies $\eta<c_0 \cdot \min\{n^{-\frac{1-\delta}{2}}, n^{-\frac{1+\delta}{4}}, 1.1^{-k} \cdot n^{\frac{k}{2}(1-\delta)-\frac{1+\delta}{2}} \}$ for some universal constant $c_0$, then Algorithm~\ref{alg:distinguish} takes $m=1.1 \cdot n^{1 + \frac{(k-1)}{2} \cdot (1-\delta)}$ random samples and succeeds with probability $0.8$ in time $\big( m+n^{O(1)} \big) \cdot e^{\eta n^{\frac{1+\delta}{2}}}$.
\end{theorem}


While there are three constraints on the noise rate $\eta$, the first constraint $\eta<n^{-\frac{1+\delta}{4}}$ would be the tightest in most cases. Essentially, the second one $\eta \le n^{-\frac{1-\delta}{2}}$ is to compare with previous distinguishing algorithms \citep{Feige2008,RRS17} with $m=n^{1+(\frac{k}{2}-1)(1-\delta)}$ samples and $e^{\tilde{O}(n^\delta)}$ time. Our algorithm is only faster than previous algorithm when $e^{\eta n^{\frac{1+\delta}{2}}}<e^{\tilde{O}(n^\delta)}$ although its sample complexity $n^{1+\frac{(k-1)(1-\delta)}{2}}$ is slightly larger. The last constraint $\eta < 1.1^{-k} \cdot n^{\frac{k}{2}(1-\delta)-\frac{1+\delta}{2}}$ requires $\delta$ to be at most $1-\Omega(1/k)$. For example, if $\delta<0.98-2/k$, $1.1^{-k} \cdot n^{0.01 \cdot k}$ would be a trivial constraint on $\eta$. 

\begin{figure}[h]
    \centering
    \label{fig:example1}
    \subfigure[$k=5$ and $m=n^{1.8}$]{
    \begin{minipage}[b]{.45\linewidth}
    \centering
    \includegraphics[scale=0.5]{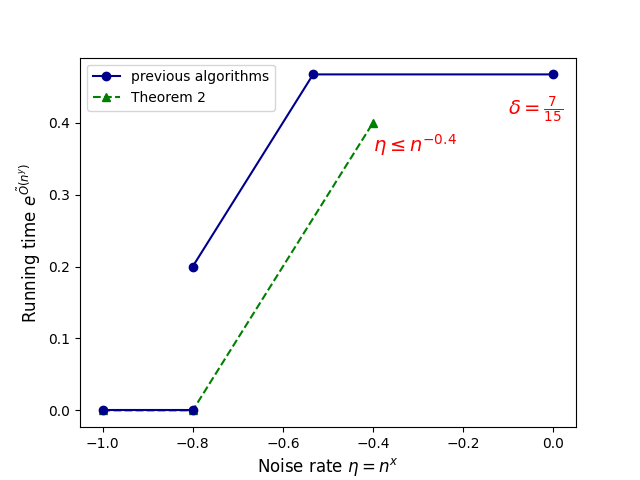}
    \end{minipage}
    }
    \subfigure[$k=8$ and $m=n^{3}$]{
    \begin{minipage}[b]{.45\linewidth}
    \centering
    \includegraphics[scale=0.5]{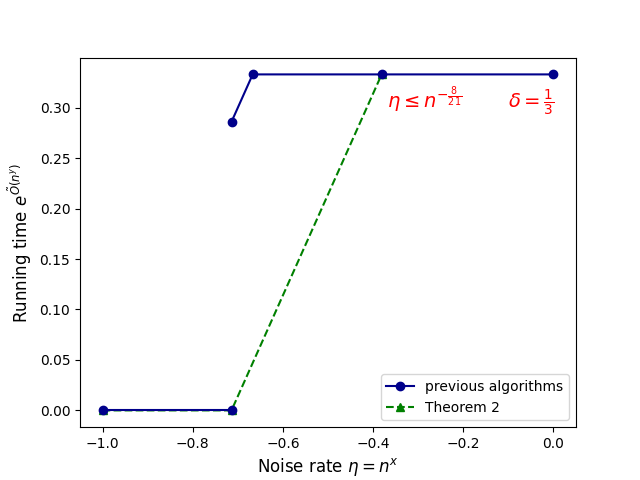}
    \end{minipage}
    }
    \caption{The running time of Theorem~\ref{thm:distinguish} compared to related art under the same number of samples for (a) $k=5$ and $m=n^{1.8}$ and (b) $k=8$ and $m=n^{3}$.}
\end{figure}

Let us compare its running time with previous results. First of all, our algorithm extends the distinguishing algorithm in \cite{DJ24} to a larger range of parameters. Essentially, \cite{DJ24} shows the time complexity of the distinguishing problem is $n^{O(1)}$ when $\eta = o(n^{-\frac{1+\delta}{2}})$ and leaves it open for larger noise rates. In respect of the distinguishing algorithm in \cite{DJ24}, Theorem~\ref{thm:distinguish} shows the time complexity is still $e^{\eta \cdot e^{\frac{1+\delta}{2}}}$ when $\eta<\min\{n^{-\frac{1+\delta}{4}}, n^{-\frac{1-\delta}{2}}\}$.

Then we compare it with previous distinguishing algorithms~\citep{Feige2008,RRS17} of running time $e^{\tilde{O}(n^{\delta})}$ under the \emph{same sample complexity}. When $k=3$ and $m=n^{1.4}$, our algorithm would have running time about $e^{\eta n^{0.8}}$ (for any $\eta<n^{-0.4}$ by choosing $\delta=0.6$). This is faster than  $e^{\tilde{O}(n^{0.2})}$ when $\eta<n^{-0.6}$. For the popular choice of $\eta=n^{-0.5}$ \citep{Alekhnovich, DMN12_PKC,KMP14_PKC,Yu19}, our algorithm would run faster than previous algorithms given $m=n^{1+\frac{(k-1)}{2} \cdot (1-\delta)}$ samples for any $\delta>\frac{2}{k}$. 
Moreover, our algorithm is always faster than the Gaussian elimination algorithm of time $e^{\eta n}$ since $\delta<1$. The final remark is that while there are several reductions \citep{ABW10, Applebaum13, BSV2019} from distinguishing algorithms to learning algorithms, plugging this distinguishing algorithm into those reductions does not provide any improvement upon our learning algorithm in Theorem~\ref{thm:sparse_LPN}.


\begin{algorithm}[h]

    \caption{Distinguishing Algorithm for sparse LPN}
    \label{alg:distinguish}
         \SetKwFunction{Distinguish}{Distinguish} 

\SetKwProg{Fn}{Function}{:}{}
\setcounter{AlgoLine}{0}
\KwIn{$k,n,\delta,\eta$}    
\KwOut{Claim ``SLPN" or ``random labels"}   

        \Fn{\Distinguish{}}{
        
       $m\gets 1.1 \cdot n^{1+\frac{k-1}{2}(1-\delta)}$ and receive $m$ random samples $(\bx_1,y_1),\ldots,(\bx_m,y_m)$
       
      $t\gets n^{\frac{1-\delta}{2}}, q\gets n^{\frac{1+\delta}{2}}, N \gets 10 \cdot e^{\eta \cdot n^{\frac{1+\delta}{2}}}$
      
 randomly sample $N$ sets $S_i$ from $\binom{[n]}{q}$.
 
      $count\gets 0$
      
        \For{$i\in [N]$}{
         $T_i\gets \{(\bx_j,y_j) : supp(\bx_j) \subset S_i\}$
         
        \If{$|T_i|>q$}{
         Find a linearly dependent set $T'_i$ in $T_i$
         
         $count \gets count+1$ when $\sum_{j \in T'_i} y_j=0$
        }
       }
         \KwRet{SLPN} if $count>\sum_{i=1}^N (\frac{1}{2}+\frac{(1-2\eta)^{|T'_i|}}{4})$; otherwise claim random labels
       }
\textbf{EndFunction}
\end{algorithm}
In the rest of this section, we finish the analysis of Algorithm~\ref{alg:distinguish}. While Algorithm~\ref{alg:distinguish} is a natural extension of the distinguishing algorithm by \cite{DJ24} from $\eta<n^{-\frac{1+\delta}{2}}$ to higher noise rates, its analysis becomes more involved. 
The basic idea is still a domain reduction via projecting random vectors $\bx_i$ to a small subset $I \subset [n]$. After collecting $|I|+1$ vectors in $I$, one could find a linearly dependent combination among $I$. Furthermore, the sum of the corresponding labels $y_i$ provides a distinguisher between noisy labels and random labels. Since the linear combination is of size $\Theta(|I|)$ w.h.p., the advantage of this distinguisher is $(1-2\eta)^{\Theta(|I|)}$. Thus, a natural idea is to repeat this for $1/(1-2\eta)^{|I|}$ times. 

However, different from the two learning algorithms, the reductions on different subsets $I_1,\ldots,I_{N}$ (say $N=1/(1-2\eta)^{|I|}$) are highly correlated. For example, when $k=3$ and $m=n^{1.4}$, $|I|=n^{0.8}$ indicates that most pairs of the linearly dependent combinations among $I_1,\ldots,I_N$ have a non-trivial intersection. Furthermore, the covariance between two linearly dependent combinations depends on the size of their intersection. Hence, the most technical part of the distinguishing algorithm is to bound the covariance of those linearly dependent combinations.  Although a direct bound on the intersection of two linearly dependent combinations is $|I_j \cap I_{j'}|$, we use the randomness of $\bx$ to show a tighter bound on intersections of linearly dependent combinations. This implies that the natural distinguishing algorithm would work for a broaden range of parameters

We start by bounding $\Pr[\sum_{j \in T'_i} y_j=0]$. Since $\sum_{j \in T'_i} \bx_j = \vec{0}$, $\sum_{j \in T'_i} y_j=0$ is equivalent to the event that the sum of the noise in these labels is 0. 
\begin{proposition}\label{prop:covariance}
    Let $e_i$ denote the Bernoulli noise of rate $\eta$ in the label $y_i$. Then $\Pr_{\eta}[\sum_{i \in S} e_i = 0 ] = \frac{1 + (1-2\eta)^{|S|}}{2}$ and $\Pr_{\eta}[\sum_{i \in S} e_i = 0 \wedge \sum_{i \in T} e_i = 0] = \frac{1 + (1-2\eta)^{|S|} + (1-2\eta)^{|T|} + (1-2\eta)^{|S \triangle T|}}{4}$.
\end{proposition}

For completeness, we provide a proof in Section~\ref{sec:correlation}. This proposition implies that $\E_y[count] =\sum_{i=1}^N \frac{1 + (1-2\eta)^{|T'_i|}}{2}$; otherwise $\E_y[count]=N/2$.

Next, we bound the variance of $count$. Because the linear dependent set $T'_i$ is of size $\Theta(q)=\Theta(n^{\frac{1+\delta}{2}})$ (in expectation) and $m=n^{O(1)}$, most random events $\sum_{j \in T'_i} y_j=0$ in $count$ are correlated. To bound the covariance of two random events $\sum_{j \in T'_i} y_j=0$ and $\sum_{j \in T'_{i'}} y_j=0$, the key is $|T'_i \cap T'_{i'}|$. Although a direct bound is $|T'_i \cap T'_{i'}| \le |S_i \cap S_{i'}|$, this would require the noise rate $\eta$ to be tiny. To prove that our algorithm works for a wider range of parameters, we consider the number of random vectors fallen into $S_i \cap S_{i'}$ to bound $|T'_i \cap T'_{i'}|$.
\begin{proposition}\label{prop:bound_sizes_S_T}
With probability $1-e^{0.1 \cdot \eta n^{\frac{1+\delta}{2}}}$ over $\bx$ and $S_i$, we have the following properties:
\begin{enumerate}
    \item $|S_i \cap S_j| \le 1.1 \cdot n^{\delta}$ for all $i \neq j$.
    \item $|T_i|>q$ for every $i \in [N]$.
    \item $|T_i \cap T_j| \le \max \big\{ 2 \cdot 1.1^{k} \cdot n^{\frac{1+\delta}{2} - \frac{k}{2}(1-\delta)} , 40 \eta \cdot n^{\frac{1+\delta}{2}} \big\}$ for all $i \neq j$.
\end{enumerate}    
\end{proposition}
\begin{proof}
    To prove Property 1, observe that $|S_i \cap S_j|$ is a hypergeometric random variable with mean $|S_i| \cdot |S_j|/n=n^{\delta}$ and variance $\le n^{\delta}$. By the concentration of hypergeometric random variables (e.g., Theorem 1 of \cite{hypergeometric_GW}), 
    \[
    \Pr \bigg[|S_i \cap S_j| \ge 1.1 \cdot n^{\delta} \bigg] \le e^{-0.01 \cdot n^{\delta}}.
    \]
    At the same time, the number of pairs $(i,j)$ is $\le N^2 = O(e^{2 \eta \cdot n^{\frac{1+\delta}{2}}})$. Since $\eta \le c_0 \cdot n^{-\frac{1-\delta}{2}}$ in the assumption, this is at most $e^{2c_0 \cdot n^{\delta}}$. By a union bound, we have $|S_i \cap S_j| \le 1.1 n^{\delta}$.

    To prove Property 2, we observe that $\Pr_{x \sim {[n] \choose k}}[supp(x) \subset S_i] = {|S_i| \choose k}/{n \choose k} \ge 0.99 \cdot (q/n)^k$ for every $S_i$. Thus $\E[|T_i|] \ge m \cdot 0.99 \cdot (q/n)^k > 1.08 \cdot n^{\frac{1+\delta}{2}}$. By the standard Chernoff bound (Theorem~\ref{THM:Chernoff}),
    \[
    \Pr\bigg[ |T_i| \le 0.93 \E|T_i| \bigg] \le e^{-0.07^2 \E |T_i|/2} = e^{-\Omega(n^{\frac{1+\delta}{2}})}.
    \]
    By a union bound over $i \in [N]$, we have $|T_i|>q$.

    To prove the last property, observe that $|T_i \cap T_j|=\{(\bx_{\ell},y_{\ell}):supp(\bx_{\ell}) \subset S_i \cap S_j\}$. Hence 
    $\Pr[(\bx_{\ell},y_{\ell}) \in T_i \cap T_j]=\frac{{|S_i \cap S_j| \choose k}}{{n \choose k}}$. Moreover, because $|S_i \cap S_j| \le 1.1 \cdot n^{\delta}$ by Property 1, 
    \[
    \E\bigg[ |T_i \cap T_j| \bigg] = m \cdot \frac{{|S_i \cap S_j| \choose k}}{{n \choose k}} \le m \cdot (1.1 \cdot n^\delta/n)^k = 1.1^{k+1} \cdot n^{\frac{1+\delta}{2} - \frac{k}{2}(1-\delta)}.
    \]
    
    To bound the deviation for all $i$ and $j$ in $[N]$, there are two cases depending on $\log N \approx \eta \cdot n^{\frac{1+\delta}{2}}$ and $\E[|T_i \cap T_j|]$.
    \begin{enumerate}   
        \item $\E[|T_i \cap T_j|] \ge 9\eta \cdot n^{\frac{1+\delta}{2}}$. Then we apply the Chernoff bound
        \[
            \Pr\bigg[ |T_i \cap T_j| \ge 1.8 \E|T_i \cap T_j| \bigg] \le e^{-0.8^2 \E |T_i \cap T_j|/2.8} \le e^{-2.2 \cdot \eta \cdot n^{\frac{1+\delta}{2}}}.
        \]

        \item $\E[|T_i \cap T_j|] < 9\eta \cdot n^{\frac{1+\delta}{2}}$. Now consider 
        \begin{align*}
        \Pr \left[ |T_i \cap T_j| \ge 40 \eta \cdot n^{\frac{1+\delta}{2}} \right] & \le {m \choose 40 \eta \cdot n^{\frac{1+\delta}{2}}} \cdot \left( (2n^\delta/n)^k \right)^{40 \eta \cdot n^{\frac{1+\delta}{2}}} \\
        & \le (\frac{em}{40 \eta \cdot n^{\frac{1+\delta}{2}}})^{40 \eta \cdot n^{\frac{1+\delta}{2}}} \cdot \left( (2n^\delta/n)^k \right)^{40 \eta \cdot n^{\frac{1+\delta}{2}}} \\
        & \le \left( \frac{em \cdot (2n^\delta/n)^k}{40 \eta \cdot n^{\frac{1+\delta}{2}}} \right)^{40 \eta \cdot n^{\frac{1+\delta}{2}}} \\
        & \le \left( \frac{em \cdot (2n^\delta/n)^k}{40/9 \cdot \E[|T_i \cap T_j|] } \right)^{40 \eta \cdot n^{\frac{1+\delta}{2}}} = (9e/40)^{40 \eta \cdot n^{\frac{1+\delta}{2}}}
        \end{align*}
    \end{enumerate}
    By a union bound, with high probability, $|T_i \cap T_j| \le \max \big\{ 1.8 \cdot \E |T_i \cap T_j| , 40 \eta \cdot n^{\frac{1+\delta}{2}} \big\}$ for all $i \neq j$.
\end{proof}

Next we bound the variance of $count$.
\begin{lemma}\label{lem:variance}
    $\Var(count) \le N/4 + 0.01 \cdot \bigg( \sum_i (1-2\eta)^{|T'_i|} \bigg)^2$ given the set of parameters in Theorem~\ref{thm:distinguish}.
\end{lemma}
\begin{proof}
    First of all, we rewrite $\Var(count)$ as
    \[
        \Var\left( \sum_i \mathbf{1} \bigg(\sum_{j \in T'_i}y_j =0 \bigg) \right) = \sum_i \Var(Z_i) + \sum_{i \neq i'} \cov(Z_i,Z_{i'}) \text{ for } Z_i :=\mathbf{1} \bigg(\sum_{j \in T'_i}y_j =0 \bigg)
    \]
    Since $Z_i$ is a $\{0,1\}$-random variable, $\Var(Z_i) \le 1/4$. Next we bound $\cov(Z_i,Z_{i'})$ by Proposition~\ref{prop:covariance}: 
    \begin{align*}
        & \Pr[Z_i=1 \wedge Z_{i'}=1] - \Pr[Z_i=1] \Pr[Z_{i'}=1]\\
        & = \frac{1 + (1-2\eta)^{|T'_i|} + (1-2\eta)^{|T'_{i'}|} + (1-2\eta)^{|T'_i \triangle T'_{i'}|}}{4} - \frac{1 + (1-2\eta)^{|T'_i|}}{2} \cdot \frac{1 + (1-2\eta)^{|T'_{i'}|}}{2} \\
        & = (1-2\eta)^{|T'_i|+|T'_{i'}|} \cdot \left( (1-2\eta)^{-2 |T'_i \cap T'_{i'}|}-1 \right)
    \end{align*}
    Now we use $T'_i \cap T'_{i'} \subset T_i \cap T_{i'}$ and $|T_i \cap T_{i'}| \le \max \big\{ 2 \cdot 1.1^{k} \cdot n^{\frac{1+\delta}{2} - \frac{k}{2}(1-\delta)} , 40 \eta \cdot n^{\frac{1+\delta}{2}} \big\}$ from Proposition~\ref{prop:bound_sizes_S_T} to simplify $\cov(Z_i,Z_{i'})$ furthermore. Since $\eta < c_0 \cdot \min\{n^{-\frac{1+\delta}{4}}, 1.1^{-k} \cdot n^{\frac{k}{2}(1-\delta)-\frac{1+\delta}{2}}\}$ for a small constant $c_0$, $(1-2\eta)^{-2 |T'_i \cap T'_{i'}|}=1+8\eta \cdot |T'_i \cap T'_{i'}| \le 1.01$.

    Thus
    \[
        \Var\left( \sum_i \mathbf{1} \bigg(\sum_{j \in T'_i}y_j =0 \bigg) \right)  \le N/4 + \sum_{i \neq i'} 0.01  (1-2\eta)^{|T'_i|+|T'_{i'}|} \le N/4 + 0.01 \cdot \bigg( \sum_i (1-2\eta)^{|T'_i|} \bigg)^2.
    \]
\end{proof}

\begin{proofof}{Theorem~\ref{thm:distinguish}}
    We apply Chebyshev's inequality to prove the correctness of our algorithm. We assume all properties in Proposition~\ref{prop:bound_sizes_S_T} and Lemma~\ref{lem:variance} hold.
    \begin{align*}
    \Pr\left[ |count-\E[count]| > \sum_i (1-2\eta)^{|T'_i|}/4 \right] & \le  \frac{\Var(count)}{(\sum_i (1-2\eta)^{|T'_i|}/4)^2} 
    \\ & \le \frac{N/4 + 0.01 \cdot \bigg( \sum_i (1-2\eta)^{|T'_i|} \bigg)^2}{(\sum_i (1-2\eta)^{|T'_i|}/4)^2}<0.18.        
    \end{align*}
    Thus our algorithm succeeds with probability 0.8.

    The running time of Algorithm~\ref{alg:distinguish} is $N \cdot (m+q^{O(1)})$.
\end{proofof}

\subsection{Proof of Proposition~\ref{prop:covariance}}\label{sec:correlation}
We finish the proof of Proposition~\ref{prop:covariance} in this section.
We prove the first part by induction. The base case $|S|=0$ is true. Then we consider $i_0 \in S$ for the inductive step:
\begin{align*}
\Pr_{\eta}\left[\sum_{i \in S} e_i = 0 \right] &=\Pr_{\eta}\left[e_{i_0}=0\right] \cdot \Pr_{\eta}\left[\sum_{i \in S\setminus \{i_0\}} e_i = 0 \right] 
+\Pr_{\eta}\left[e_{i_0}=1\right] \cdot \Pr_{\eta}\left[\sum_{i \in S\setminus \{i_0\}} e_i = 1 \right]\\
&=(1-\eta)\cdot\frac{1 + (1-2\eta)^{|S|-1}}{2}+\eta\cdot\left(1-\frac{1 + (1-2\eta)^{|S|-1}}{2}\right)=\frac{1+(1-2\eta)^{|S|}}{2}.
\end{align*}
For the second part, 
\begin{align*}
\Pr_{\eta}[\sum_{i \in S} e_i = 0 \wedge \sum_{i \in T} e_i = 0] &= \Pr_{\eta}[\sum_{i \in S \setminus T} e_i = 0 \wedge \sum_{i \in T \setminus S} e_i = 0 \wedge \sum_{i \in T \cap S} e_i = 0]\\
&+  \Pr_{\eta}[\sum_{i \in S \setminus T} e_i = 1 \wedge \sum_{i \in T \setminus S} e_i = 1 \wedge \sum_{i \in T \cap S} e_i = 1]\\
&=\frac{1}{8}(1+(1-2\eta)^{|S\setminus T|})(1+(1-2\eta)^{|T\setminus S|})(1+(1-2\eta)^{|S\cap T|})\\
&+\frac{1}{8}(1-(1-2\eta)^{|S\setminus T|})(1-(1-2\eta)^{|T\setminus S|})(1-(1-2\eta)^{|S\cap T|})\\
&=\frac{1}{4}(1+(1-2\eta)^{|T|}+(1-2\eta)^{|S|}+(1-2\eta)^{|T\triangle S|})
\end{align*}

\end{subappendices}

\end{document}